\documentclass[a4paper,11pt,oneside]{book}

\usepackage{amsmath}
\usepackage{amsfonts}

\numberwithin{equation}{section}

\def\doubleset#1#2{\bgroup%
\def\doit#1#2{%
\setbox\dblsetbox=\hbox{$\cstyle #1$}%
\raise#2\ht\dblsetbox\copy\dblsetbox%
\hskip-\wd\dblsetbox%
\raise-#2\ht\dblsetbox\box\dblsetbox}%
\mathchoice%
{\def\cstyle{\displaystyle}\doit#1#2}%
{\def\cstyle{\textstyle}\doit#1#2}%
{\def\cstyle{\scriptstyle}\doit#1#2}%
{\def\cstyle{\scriptscriptstyle}\doit#1#2}\egroup}
\newbox\dblsetbox

\newcommand{\plpl}{\doubleset{+}{.185}}
\newcommand{\mimi}{=}
\newcommand{\rd}{\mathrm{d}}
\newcommand{\intd}[1]{\int\mspace{-6mu}\rd#1\,}
\newcommand{\ints}[2]{\int_{#1}\mspace{-6mu}#2\,}
\newcommand{\Nabla}{\Tilde{\nabla}}
\newcommand{\bR}{\mathbb{R}}
\newcommand{\bC}{\mathbb{C}}
\newcommand{\du}[3]{#1_{#2}^{\phantom{#2}#3}}
\newcommand{\ud}[3]{#1^{#2}_{\phantom{#2}#3}}
\newcommand{\dud}[4]{#1_{#2\phantom{#3}#4}^{\phantom{#2}#3}}
\newcommand{\LieG}{\mathfrak{g}}
\newcommand{\bZ}{\mathbb{Z}}
\newcommand{\Dirac}{\nabla\mspace{-13mu}/\mspace{4mu}}

\begin{document}

\thispagestyle{plain}

\begin{center}
  \begin{large}
    \vspace{.7cm}

    \textbf{Supersymmetric sigma models, gauge theories and vortices}

    \vspace{2cm}

    William Machin 

    \vspace{1cm}

    \textit{Dissertation submitted for the degree of Ph.D.\\ King's
    College, London \\ September 2002}
    \vspace{2cm}
  \end{large}
\end{center}

\newpage

\chapter*{Abstract}
This thesis considers one and two dimensional supersymmetric nonlinear
sigma models. First there is a discussion of the geometries of one and
two dimensional sigma models, with rigid supersymmetry.

For the one-dimensional case, the supersymmetry is promoted to a local
one and the required gauge fields are introduced.  The most general
Lagrangian, including these gauge fields, is found. The constraints of
the system are analysed, and its Dirac quantisation is investigated.

In the next chapter we introduce equivariant cohomology which is used
later in the thesis.

Then actions are constructed for (p,0)- and (p,1)- supersymmetric, $1
\leq p \leq 4$, two-dimensional gauge theories coupled to non-linear sigma
model matter with a Wess-Zumino term.

The scalar potential for a large class of these models is derived. It
is then shown that the Euclidean actions of the (2,0) and
(4,0)-supersymmetric models without Wess-Zumino terms are bounded by
topological charges which involve the equivariant extensions of the
K\"ahler forms of the sigma model target spaces evaluated on the
two-dimensional spacetime. 

Similar bounds for Euclidean actions of appropriate gauge theories
coupled to non-linear sigma model matter in higher spacetime
dimensions are given which now involve the equivariant extensions of
the Kahler forms of the sigma model target spaces and the second Chern
character of gauge fields. It is found that the BPS configurations are
generalisations of abelian and non-abelian vortices.

\newpage

\begin{center}
\vspace*{\stretch{1}}
\textit{Dedicated to George for his infinite patience and wisdom}
\vspace*{\stretch{2}}
\end{center}

\tableofcontents

\chapter{Introduction}

This thesis is on the theory of two dimensional sigma models. Sigma
models have found a wide variety of applications in particle and
theoretical physics. We shall first outline briefly the historical
background and development of the theory, putting sigma models into
their proper context and then motivating their study by making the
reader aware of a small selection of their wide range of applications.
We shall also outline the results of this thesis.

\section{History and Motivation}

Nonlinear sigma models have been used to describe the dynamics of
relativistic and non-relativistic particles; their original motivation
was the study of the propagation of such particles in curved
space-times. Sigma models, in the modern sense, were first described
in~\cite{Gell-Mann:1960np} as a model of high energy hadron physics;
it was postulated that the pion field depended on a scalar meson
called $\sigma$ proposed in~\cite{Schwinger:1957em}. This idea was
formalised in \cite{Meetz:1969}, where chiral invariant Lagrangians
were also written down,
\begin{equation}
\label{eq:firstsigmaaction}
  S = \intd{^4x} \eta^{\mu\nu}g_{ij}(\phi) \partial_\mu \phi^i \partial_\nu \phi^j
\end{equation}
where $\phi^i$ are coordinates of $M$ and $x^\mu$ are coordinates of
$\Xi$. The coupling $g$ is a metric on the sigma model manifold $I$.

Today it is understood that strong interactions and hadron physics are
described not by nonlinear sigma models but by the Standard Model.
However, nonlinear two-dimensional ($\dim\Xi=2$) sigma models have
found new applications in string theory
\cite{Brink:1976sc,Deser:1976rb}. The Lagrangian for this is a natural
generalisation of \eqref{eq:firstsigmaaction}, with the addition of
supersymmetry.

Supersymmetry is a symmetry which rotates bosons into fermions. It
improves the short distance behaviour of quantum theories and gives an
elegant solution to the hierarchy problem. In string theory, the five
known consistent string theories are supersymmetric.

Supersymmetric sigma models have a rich geometrical structure. This
was first observed in \cite{Zumino:1979et} for four dimensional models
and later developed by \cite{Alvarez-Gaume:1981hm} for two dimensional
models. In particular it was shown that $N=1$ sigma models require the
target space to be a (pseudo-)Riemannian manifold, $N=2$ requires the manifold
to be K\"ahler and $N=4$ requires the manifold to be hyper-K\"ahler.
If the manifold has torsion, the model generalises to have chiral
supersymmetry~\cite{Hull:1985jv,Papadopoulos:1994mf,Papadopoulos:1994tn},
where the number of left handed supersymmetries differs from the
number of right handed supersymmetries. We will review this in the
next chapter.

Massive supersymmetric sigma models were first constructed in
\cite{Alvarez-Gaume:1983ab} by the addition of potentials. The
inclusion of a potential may impose additional restrictions on the
target space geometry.

\subsection{One-dimensional sigma models}

The simplest examples of nonlinear sigma models are those where the
worldspace is one-dimensional. The sigma models obtained can therefore
be thought of as particles, where the dependence on the worldsheet
signifies time dependence. Such sigma models describe the dynamics of
point particles. 

In the past the model that was mostly investigated was a relativistic
or non-relativistic particle propagating in a Riemannian manifold with
a metric $g$. These results were extended to $N=1$ locally
supersymmetric particle models in~\cite{Brink:1976sz,Brink:1977uf}.
Later the action of supersymmetric particles with extended
supersymmetry was given in~\cite{Howe:1988ft}.

One-dimensional sigma models have many applications. The simplest
example is that they form the basis for supersymmetric quantum
mechanics, when $N=4$ this can be associated
\cite{Ivanov:1991cz,Pashnev:1991cf,Donets:1999jx} with $N=1$, $d=4$
supersymmetric field theories after an appropriate dimensional
reduction.

The conditions on the target space required by supersymmetry in $d=1$
manifolds is given by \cite{Coles:1990hr, Gibbons:1997iy,
  Gibbons:1993ap, DeJonghe:1997fb, Michelson:1999zf,
  Hellerman:1999nr}.

In~\cite{Coles:1990hr} it was found that rigid supersymmetry in one
dimension allows for the construction of more general models than
those considered in the past.  In particular, the manifold that the
particle propagates in can have torsion which is not a closed 3-form.
Such models have been found to describe the effective theory of
multi-black hole systems~\cite{Maloney:1999dv}. A scalar potential and
a coupling to a magnetic field in the action of~\cite{Coles:1990hr}
was added to the sigma model formalism in~\cite{Hull:1993ct},
following the earlier work of~\cite{Barducci:1976qu,Barducci:1977xq}.

More recently, interest in sigma models has arisen in the context of
the AdS/CFT conjecture. It has been observed
\cite{Claus:1998ts,Kallosh:1999mi,deAzcarraga:1998ni} that
supersymmetric sigma models describe the dynamics of a supersymmetric
particle near the AdS horizon of an extreme Reissner-Nordstr\"om black
hole.

In this case, superconformal supersymmetry is required, which imposes
additional constraints on the target space geometry.  Superconformal
quantum mechanics may even provide a dual description of string theory
on $AdS_2$ \cite{Michelson:1999zf}.

\subsection{Two-dimensional sigma models and Gauge theories}

Two dimensional sigma models are interesting in their own right
because they have a rich mathematical structure. In addition to the
fact that supersymmetry constrains the target space, they have
improved UV behaviour and in $d=2$, they have improved UV behaviour
and power counting shows that they are renormalisable.  In fact models
with $N=4$ supersymmetry are finite to all orders in perturbation
theory.

In addition to string theory, two dimensional sigma models can be used
to describe certain properties of four-dimensional gauge theories.
For example, the magnetic monopole and dyon in the gauge theory
correspond to the kink and Q-kink solution of sigma models.  There are
also applications to the theory of integrable systems and
supersymmetric models have been used in the investigation of duality.

Two-dimensional sigma models and some two-dimensional gauge theories
have been used to model the dynamics of fundamental and D-strings,
respectively.  The small fluctuations of strings which arise as
intersections in various brane configurations are described by
two-dimensional gauge theories coupled to scalars.  Because of this,
many of the properties and the various objects that arise in gauge
theories coupled to scalars have a brane interpretation
\cite{Strominger:1996ac, Papadopoulos:1996uq, Hanany:1997ie,
Bergshoeff:1999jc}.  Supersymmetric gauge theories coupled to linear
sigma models have been constructed in \cite{Witten:1993yc} and they
have been used to illuminate the relation between Landau-Ginzburg
models and Calabi-Yau spaces.  Recently a two-dimensional gauge
theory coupled to a linear sigma model was used to investigate aspects
of the dynamics of vortices using branes \cite{Tong:2002rq}.

In two dimensions, the Wess-Zumino term has the same mass dimension as
the kinetic term of sigma model scalars. Therefore two-dimensional
supersymmetric gauged theories can couple to non-linear sigma model
matter which also has a non-vanishing Wess-Zumino coupling. Such a
theory is renormalizable.  The gauging of supersymmetric
two-dimensional non-linear sigma models with a Wess-Zumino term have
been considered in \cite{Hull:1989jk, Hull:1991uw}. However in these
papers the part of the action which involves the gauge field kinetic
terms has not been given.  It has been found in \cite{Hull:1989jk}
that the Wess-Zumino term of a non-linear sigma model cannot always be
gauged. The conditions for gauging a Wess-Zumino term have been
identified as the obstructions to the extension of the closed form
associated with the Wess-Zumino term to an element of the equivariant
cohomology \cite{Atiyah:1984} of the sigma model target space
\cite{Figueroa-O'Farrill:1994ns,Figueroa-O'Farrill:1994dj}.  Scalar
potentials for supersymmetric two-dimensional sigma models with
Wess-Zumino term have been investigated in \cite{Hull:1993ct,
  Papadopoulos:1994mf,Papadopoulos:1994tn,Papadopoulos:1995kj}.

\section{Outline}

\subsection{Supersymmetry in sigma models}
In the second chapter, we review one and two dimensional sigma model
with various amounts of supersymmetry. We begin with a discussion of
two-dimensional sigma models $\phi:\Xi\rightarrow M$ where the target
space has a torsion $H$. The torsion is introduced into the action by
a Wess-Zumino term, however this term is only defined locally on a
given coordinate patch of $M$, therefore the action $S$ is not
globally defined; in particular the exponential $\exp(iS)$ that
appears in the path integral may not be well defined. We will discuss
the global aspects of torsion, for example we will derive a
topological condition, analogous to Dirac's quantisation condition for
strings, for $\exp(iS)$ to be well defined and therefore write the
action in a globally defined way. Next we shall introduce the
($p$,$q$) supersymmetry model where $q=0,1$. The (1,0) model is
enlarging the worldspace $\Xi$ to the superspace $\Xi^{1,0}$ by adding
a Grassman (anticommuting) degree of freedom; the sigma model map will
then be a multiplet consisting of a scalar $\phi$ and a fermion field
$\lambda$.  We will write down the supersymmetric action as an action
over the superspace $\Xi^{1,0}$ and construct the Noether supercharge
$Q_+$ that generates the supersymmetry. On writing the action as an
integral over the bosonic coordinates of the superspace only, ie. the
coordinates of $\Xi$ only, we will recover the original
non-supersymmetric action plus fermion terms. Then we will introduce
($p$,0) supersymmetry for $p=2,4$. We will find that (2,0)
supersymmetry requires that the target manifold $M$ is a KT manifold
and that (4,0) supersymmetry requires that $M$ is HKT, demonstrating
the relationship between supersymmetry and target space geometry. We
will show that no other interesting cases for ($p$,0) supersymmetry
exist. Next we will introduce ($p$,1) supersymmetry for $p=1,2,4$. We
will find that these models are analogous to the ($p$,0) models; in
particular (2,1) supersymmetry requires that $M$ is KT and (4,1)
supersymmetry requires that $M$ is HKT. We will demonstrate why it is
not possible to define a (2,1) supersymmetric theory on a manifold
with vanishing torsion. Finally we will introduce one dimensional
sigma models, which are maps $\phi:\Sigma\rightarrow M$ where the
(bosonic) dimension of the superspace is $\dim\Sigma=1$. In this
section we will also introduce a fermionic superfield $\chi$, which
can be thought of as belonging to the Yang-Mills sector of the theory.
We will describe an $N=1$ supersymmetric action for $\phi$ and $\chi$
and construct the Noether supercharge of the symmetry. We will briefly
comment on the differences of the geometry of the target space between
two-dimensional and one-dimensional sigma models with extended
supersymmetry.

\subsection{Supergravity and one-dimensional sigma models}

In the third chapter, we will introduce the most general one
dimensional $N=1$ supersymmetric action with dimensionless couplings,
and describe the action in components. We will then describe the
Noether procedure, where the supersymmetry parameter $\epsilon$ is no
longer assumed to be constant, but to depend on the coordinates of the
worldspace. Such models are known as \emph{supergravity} models
because they can be shown to include gravity. We demonstrate the
Noether technique explicitly on a simple Lagrangian, introducing gauge
fields which cancel the variations proportional to derivatives of
$\epsilon$. We then describe the most general $N=1$ action after
gauging its supersymmetry and introducing the supergravity gauge
fields. We derive the Hamiltonian of the theory and calculate the
first and second class constraints; we show that these are conserved
with time and express the Hamiltonian as a linear combination of the
second class constraints. We then quantise the theory using the
Dirac-Bergmann method; we calculate the Dirac operator and show that
it squares to the Klein-Gordon operator. We will give two
representations of the Dirac operator in terms of Dirac matrices, then
finally we will comment on the existence of zero modes of the Dirac
operator on curved manifolds.

\subsection{Equivariant cohomology}
In the fourth chapter, we review the results of equivariant cohomology
that we will use in the following chapters. Equivariant cohomology
describes the cohomology of gauge invariant forms that arise when a
manifold $M$ admits a group action with gauge group $G$; we will give
a mathematician's definition in terms of the universal classifying
space $EG$ of $G$. We will then construct a model for the equivariant
cohomology as follows. We define the Weil algebra, which can be
thought of as the differential graded algebra generated by the gauge
potential $A$ and the field strength $F$. The Weil algebra is used to
generalise the space of forms to include forms which are polynomial in
$A$ and $F$; the subspace invariant under gauge transformations is
called the space of \emph{basic forms} $\Omega_\LieG^*(M)$. The
cohomology of $\Omega_\LieG^*(M)$ is the equivariant cohomology. We
will define an equivariant extension of a form, and in a final section
which can be read independently of the rest of the chapter, we will
give a more useful, physicist's way of thinking about equivariant
cohomology.

\subsection{New two dimensional gauge theories}
In the fifth chapter, we shall construct the actions of (p,0)- and
(p,1)-supersymmetric, $1\leq p\leq 4$, two-dimensional gauge theories
coupled to non-linear sigma model matter and with non-vanishing
Wess-Zumino term. In addition we shall also consider the scalar
potentials that arise in these theories.  This will generalise various
partial results that have already appeared in the literature.  To
simplify the description of the results from here on, we shall use the
term \lq sigma models' instead of the term \lq non-linear sigma
models' unless otherwise explicitly stated.  The method we shall use
to construct the various actions of supersymmetric two-dimensional
gauged theories coupled to sigma models is based on the superfields
found in the context of supersymmetric sigma models \cite{Howe:1987qv,
  Howe:1988cj} and later used in the context supersymmetric gauged
sigma models \cite{Hull:1991uw}. One advantage of this method is that
it keeps manifest the various geometric properties of the couplings
that appear in these theories. This will be used in the fourth chapter
to construct of various bounds for vortices.  Since the parts of the
actions that we shall describe involving the kinetic term of the sigma
model scalars, the Wess-Zumino term and their couplings to gauge
fields are known, we shall focus on the kinetic term of the gauge
fields and the scalar potentials of these theories. We shall allow the
gauge couplings to depend on the sigma model scalars and we shall
derive the various conditions on these couplings required by gauge
invariance and supersymmetry.  We shall find that the scalar potential
of gauge theories coupled to sigma models in two dimensions, even in
the presence of Wess-Zumino term, is the sum of a `F' term or a `D'
term or both.  The presence of a D-term, or Fayet-Iliopoulos term, may
come as a surprise. This is because in the presence of a Wess-Zumino
term the geometry of the target space, say, of (2,0)-supersymmetric
models where such a term in expected, is {\it not} K\"ahler and but
K\"ahler with torsion (KT). Thus the K\"ahler form is not closed and
there are no obvious moment maps. However, it has been shown in
\cite{Grantcharov:2002fk} that KT geometries under certain conditions
admit moment maps and are those that appear in the Fayet-Iliopoulos
term of these models. Similar results hold for the
(4,0)-supersymmetric models.  We shall observe that the gauged (p,1),
$p=1,2,4$, multiplets are associated with scalar superfields. For the
gauge theories with (2,1) and (4,1) supersymmetry, these scalar
multiplets satisfy the same supersymmetry constraints as the
associated sigma model multiplets.  Therefore these gauge theories can
be thought of as sigma models with target spaces $\LieG\otimes \bR^p$,
$p=1,2,4$.  This will allow us to combine the (p,1) gauge multiplet
and the standard sigma model $(p,1)$ multiplet to a new sigma model
multiplet. As a result, sigma model type of actions can be written for
these gauge theories coupled to matter for which the associated
couplings depend on the scalar fields of both the sigma model and
gauge multiplets. This generalizes the results of \cite{Hull:1991uw}.

\subsection{Vortices and bounds}
In the sixth chapter, we shall show that the Euclidean
actions of (2,0)- and (4,0)-supersymmetric two-dimensional gauge
theories coupled to sigma models with a Fayet-Iliopoulos term but with
{\it vanishing} Wess-Zumino term admit bounds. In particular we shall
find that the Euclidean action $S_E$ of the (2,0)-supersymmetric
theory is bounded by the absolute value of a topological charge ${\cal
  Q}$ which is the integral over the two-dimensional spacetime of the
{\it equivariant extension of the K\"ahler form} of its sigma model
target space, $S_E\geq |{\cal Q}|$.  The sigma model manifold in the
(4,0)-supersymmetric theory is hyper-K\"ahler and so there are three
K\"ahler forms each having an equivariant extension. The Euclidean
action $S_E$ of the (4,0)-supersymmetric theory is bounded by the
length of the three topological charges ${\cal Q}_1, {\cal Q}_2, {\cal
  Q}_3$ each associated with the integral over the two-dimensional
spacetime of the equivariant extensions of the three K\"ahler forms,
$S_E\geq \sqrt {{\cal Q}^2_1+ {\cal Q}^2_2+ {\cal Q}^2_3}$.  This is
another application of the equivariant cohomology in the context of
two-dimensional gauged sigma models which is distinct from that found
in \cite{Figueroa-O'Farrill:1994ns,Figueroa-O'Farrill:1994dj} and we
have mentioned above.  (For many other applications see for example
\cite{Cordes:1995fc}.)  The configurations that saturate these bounds
are vortices and include the Nielsen-Olesen type of vortices
\cite{Nielsen:1973cs} associated with gauge theories coupled to linear
sigma models.  In particular the bounds above generalise that found by
Bogomol'nyi in \cite{Bogomolny:1976de} for the abelian vortices of a
gauge theory coupled to a single linear complex scalar field.

We also find that similar bounds exist in higher dimensions for action
type of functionals that involve maps between two K\"ahler manifolds of
any dimension coupled to gauge fields or maps from a K\"ahler manifold
into a hyper-K\"ahler manifold again coupled to gauge fields. The
structure of these functionals is such that it includes the Euclidean
actions of some supersymmetric gauge theories in higher dimensions
coupled to sigma models with Fayet-Iliopoulos terms.  In particular
the first case, which involves maps between two-K\"ahler spaces,
includes the Euclidean action of a four-dimensional $N=1$
supersymmetric gauge theory coupled to a sigma model. The latter case,
which involved maps from a K\"ahler manifold into a hyper-K\"ahler
one, can be associated with the Euclidean action of a four-dimensional
$N=2$ supersymmetric gauge theory coupled to a sigma model. Note that
in $N=1$ theories in four dimensions the sigma model target space is
K\"ahler while in the $N=2$ theories in four dimensions the sigma
model target space is hyper-K\"ahler.  In all these cases the action
functionals are bounded by {\it topological charges} which involve the
{\it equivariant extensions} of the K\"ahler forms of the sigma model
target space as well as the second Chern character of the gauge
fields.  Our results are different from those of \cite{Bradlow:1989,
  Bradlow:1995vn} for non-abelian vortices which involve non-abelian
gauge theories coupled to linear sigma model matter. Note that in the
bound constructed in \cite{Bradlow:1989}, the topological term
involves the first class and second Chern character of the gauge
fields instead of the equivariant extension of the K\"ahler form and
the second Chern character of the gauge field that we find. It turns
out that in the case of gauge theories coupled to {\it linear} sigma
models of \cite{Witten:1993yc} the two different topological charges
can be related, see also \cite{Taubes:1980tm}. However this involves a
partial integration procedure in which various surface terms are taken
to vanish. We remark that in the construction of the bounds that
involve the equivariant extensions of the K\"ahler forms, and also in
\cite{Bogomolny:1976de}, the topological terms are identified with
what remains after writing the Euclidean actions of the theories as a
sum of squares without the use of partial integrations.  It is clear
that our results can also be used to construct bounds for solitons in
appropriate gauge theories coupled to sigma model matter in
odd-spacetime dimensions. This is the usual situation where instantons
in an n-dimensional theory can be thought of as static solitons of an
(n+1)-dimensional theory. In particular, there is a bound for the
energy of static configurations of a three-dimensional $N=2$
supersymmetric gauge theory coupled to sigma model matter. This new
bound is an extension to gauged theories of the of bounds found in
\cite{Abraham:1992vb} and generalizes that of \cite{Jacobs:1979ch}.

\chapter{Sigma Models}

A sigma model is defined by two manifolds: the \emph{worldspace} (or
\emph{superspace}) $(\Xi,h)$, the \emph{target space} $(M,g)$ and the
\emph{sigma model fields} which are maps $\phi:\Xi\rightarrow M$. The
dimension of the sigma model is the dimension of $\Xi$; in this
chapter we will describe the cases $\dim\Xi=2$ and $\dim\Xi=1$.

\section{Two dimensional sigma models}
\label{sec:2dsigmamodels}
In the two dimensional case, sigma model fields are maps
$\phi:\Xi\rightarrow M$ where $\dim \Xi=2$. It is usual to take $\Xi$
to be a flat Minkowski manifold with coordinates $x^\mu=(x,t)$ and
metric $\eta_{\mu\nu}$ because this represents objects with string-like
degrees of freedom; we can think of the timelike direction $t$ as
proper time and the spacelike direction $x$ as the degree of freedom
along the stringlike object.

Since the Lorentz group acts on $\Xi$ it is convenient to use the null
coordinates $(x^\plpl,x^\mimi)=(x+t,x-t)$; with this choice of
parameterisation the group action is diagonalised and the indices
denote the Lorentz weights:
\begin{align}
  [M,v_\plpl]&=+v_\plpl & [M,v_\mimi]&=-v_\mimi
\end{align}
where $v_\mu$ is a vector and $M$ is an infinitesimal Lorentz group
generator. An object is Lorentz invariant if its total weight is zero.

\subsection{The action}
\label{sec:sigmaaction}

The fields of interest satisfy an action principle with action
\begin{align}
\label{eq:sigmaaction}
  S[\phi] = \ints{\Xi}{\rd^2x} g_{ij}(\phi)\partial_\plpl\phi^i \partial_\mimi\phi^j
\end{align}
where $g_{ij}$ is the metric of the target space manifold $M$. This
action is invariant under the following transformations: (i)
coordinate changes $\phi^i\rightarrow\phi'^i(\phi)$ of the target
manifold $M$, (ii) Lorentz transformations on $\Xi$, and (iii) linear
transformations of the field $\phi$ given by 
\begin{align}
\label{eq:phitranslation}
  x^\plpl &\rightarrow x^\plpl + \alpha^\plpl \nonumber \\
  x^\mimi &\rightarrow x^\mimi + \alpha^\mimi \\
  \phi^i &\rightarrow \phi^i - \alpha^\mu \partial_\mu \phi^i \nonumber
\end{align}
where $\alpha^\mu=(\alpha^\plpl,\alpha^\mimi)$ is a real constant.
The third set of transformations are generated by the Noether charge
$P_\mu=(P_\plpl,P_\mimi)$ associated with energy and momentum, and is
realised on $\phi$ by $P_\mu=(-\partial_\plpl,-\partial_\mimi)$. In
fact the action~\eqref{eq:sigmaaction} is also invariant under
conformal transformations $\alpha^\plpl=\alpha^\plpl(x^\plpl)$,
$\alpha^\mimi=\alpha^\mimi(x^\mimi)$.

\subsection{The Wess-Zumino term}

It is natural to add torsion to the theories described above, as we
shall see this is essential in the formulation of certain
supersymmetric theories. Other applications of torsion include the
cancellation of anomaly terms, see for example~\cite{Howe:1987nw}.

The previous action with a Wess-Zumino term is
\begin{align}
\label{eq:sigmaactiontorsion}
S[\phi] = \intd{^2x} (g_{ij}+b_{ij})\partial_\plpl\phi^i
\partial_\mimi\phi^j
\end{align}
where $b_{ij}$ is a two-form; the torsion of $M$ is $H=\rd b$. 

The action with torsion \eqref{eq:sigmaactiontorsion} has the same
invariances as before, ie. under (i) coordinate changes of $M$, (ii)
Lorentz transformations of $\Xi$, and (iii) linear translations of
$\phi$. However $b_{ij}$ may change by a gauge transformation when we
move between coordinate patches on $M$.  Therefore $b_{ij}$ and the
action~\eqref{eq:sigmaactiontorsion} may only be locally defined if
the topology of $M$ is non-trivial. Note though that $H=\rd b$ is
always globally defined. We discuss this in the next section.

\subsection{Global aspects of torsion}
\label{sec:globaltorsion}

The equations of motion are globally defined because they depend on
the torsion $H$ not the Wess-Zumino term $b$, which is locally
defined. It is not possible classically for the action of the
Wess-Zumino term to be globally defined.  However in quantum theory,
$\exp (iS_E)$ must be globally defined, where $S_E$ is the Euclidean
action. 

Choose a `background' map $\phi_0:\Xi\rightarrow M$ which is homotopic
to the sigma model map $\phi$ defined before. Then we define the
Euclidean action of the Wess-Zumino term as
\begin{align}
  S_E[\Tilde\phi] =  \int_{I\times\Xi} \Tilde\phi^*H
\end{align}
where $\Tilde\phi:I\times\Xi\rightarrow M$, $I$ is the interval
$I=[0,1]$, and $\Tilde\phi$ are the interpolating maps with
$\Tilde\phi(0,x^\mu)=\phi_0$ and $\Tilde\phi(1,x^\mu)=\phi(x^\mu)$.

$S_E$ depends on the choice of $\Tilde\phi$.
To determine this dependence, take a second homotopy $\Tilde{\phi}'$
between $\phi$ and $\phi_0$ and calculate the difference,
\begin{align}
  \Delta S_E = S_E[\Tilde\phi']-S_E[\Tilde\phi]  ~.
\end{align}
This can be rewritten as
\begin{align}
  \Delta S_E = S_E[\Tilde\phi'']
  &= \int_{S^1\times\Xi} \Tilde{\phi}''^*H
\end{align}
where
\begin{align}
\label{eq:Hdifference}
  \phi''(t,x^\mu) = \begin{cases} \phi(2t,x^\mu)& 0 \leq t \leq
  \tfrac{1}{2} \\
  \phi'(-2t+2,x^\mu)& \tfrac{1}{2}\leq t \leq 1  
\end{cases}
\end{align}
and satisfies $\phi''(0,x^\mu)=\phi''(1,x^\mu)=\phi_0$.
The difference \eqref{eq:Hdifference} is the integral of a closed
three-form over a compact three-manifold  without boundary and in
general evaluates to a real number. 

Therefore, if $[H]/2\pi\in H^3(M,\bZ)$ then $\Delta S_E=2\pi n$ where
$n$ is an integer, so $\exp(iS_E)$ will be independent of the choice
of the interpolation between $\phi_0$ and $\phi$ and so well-defined.

\subsection{Supersymmetric sigma models}

We will introduce supersymmetry through its algebra. Let $Q$ be the
Noether (super)charge generating a supersymmetry, then the
\emph{supersymmetry algebra} is
\begin{align}
\label{eq:easysusyalgebra}
  [ Q, Q ]_+ &= 2iP 
\end{align}
where the brackets are anticommutators. Note that $Q$ must be
anticommuting (Grassman odd).

In components, $P=(P_\plpl,P_\mimi)$ so Lorentz consistency requires
that $Q=(Q_+,Q_-)$ with Lorentz weights $+1/2,-1/2$ respectively.
Fields with an even number of plus or minus signs are bosonic
(commuting); those with an odd number are fermionic (Grassman odd).

A $(p,q)$ supersymmetric model has $p$ left-handed supercharges $Q_+$
and $q$ right-handed supercharges $Q_-$. We will construct various
$(p,q)$ supersymmetric models in the following sections.

\subsection{(1,0) supersymmetric sigma models}
\label{sec:10susy}

In this section we will construct a two-dimensional sigma model which
realises the (1,0) supersymmetry algebra 
symmetry algebra
\begin{align}
\label{eq:10susyalgebra}
  [ Q_+, Q_+ ]_+ = 2iP_\plpl ~.
\end{align}
We begin with the superspace $\Xi^{1,0}$, which we take to be a
supermanifold parameterised by $(x^\plpl,x^\mimi,\theta^+)$ where
$(x^\plpl,x^\mimi)$ are the null coordinates defined in
section~\ref{sec:2dsigmamodels} and $\theta^+$ is a real
\emph{Grassman} coordinate, which is anticommuting and in particular
$(\theta^+)^2=0$. The index denotes the Lorentz weight, we define an
upper ($\pm$) index the same as a lower ($\mp$) index.

The Grassman derivative $\partial_{\theta+}\equiv
\tfrac{\partial}{\partial\theta^+}$ satisfies
$\partial_{\theta^+}\theta^+=1$, and we define the supercovariant
derivative $D_+$ by $(D_+)^2=i\partial_\plpl$, in components this is
\begin{align}
\label{eq:defnDplus}
  D_+ &= \frac{\partial}{\partial\theta^+} +
  i\theta^+\frac{\partial}{\partial x^\plpl} ~.
\end{align}

To construct the Noether charges, $P$ is realised exactly as before,
$P=(P_\plpl,P_\mimi)=(-\partial_\plpl,-\partial_\mimi)$. We may now
define $Q_+$ by the property that 
\begin{align}
 [Q_+,D_+]_+=0 ~,
\end{align}
in components this is
\begin{align}
\label{eq:10susycharge}
Q_+ &= \frac{\partial}{\partial\theta^+} -
  i\theta^+\frac{\partial}{\partial x^\plpl}  ~.
\end{align}
It may be checked that $P_\plpl$ and $Q_+$ indeed obey the supersymmetry
algebra \eqref{eq:10susyalgebra}.

The two-dimensional (1,0) supersymmetric sigma model is defined by the
superspace (worldspace) $\Xi^{1,0}$, the target manifold $(M,g)$ and
the sigma model fields $\phi:\Xi^{1,0}\rightarrow M$. The fields
satisfy an action principle with action
\begin{align}
\label{eq:10susyaction}
  S = -i \ints{\Xi^{1,0}}{\rd^2x\rd\theta^+} (g_{ij}+b_{ij})
  D_+\phi^i\partial_\mimi\phi^j ~.
\end{align}
The integral $\int\rd\theta^+=\partial_{\theta+}$ is the standard
Berezin integration (we remind the reader that $\rd\theta^+$ has the
\emph{opposite} Lorentz weight to $\theta^+$); $b$ is the Wess-Zumino
term of the previous section.

The action~\eqref{eq:10susyaction} is invariant under the three
transformations before: (i) coordinate changes of $M$ (up to gauge
transformations of $b_{ij}$), (ii) Lorentz transformations of
$\Xi^{1,0}$ and (iii) linear transformations of $\phi$ generated by
$P$. In addition, it is invariant under supersymmetries generated by
$Q_+$, $\delta_\epsilon\phi^i=\epsilon_-Q_+\phi^i$ where $\epsilon_-$
is a constant anticommuting parameter. To see this, set $b=0$, apply
$\delta_\epsilon$ to the fields, use $[D_+,Q_+]_+=0$ and expand $Q_+$
in components to give
\begin{align}
\label{eq:susyinvcalculation}
  S = -i \intd{^2x\rd\theta^+} \epsilon_-
\Bigl(\frac{\partial}{\partial\theta^+} -
  i\theta^+\frac{\partial}{\partial x^\plpl}\Bigr)
  g_{ij}
  D_+\phi^i\partial_\mimi\phi^j ~.
\end{align}
which evaluates to a surface term that vanishes. In general an
integral over the full superspace is manifestly supersymmetric. 

For $b\neq 0$, under the conditions of
section~\ref{sec:globaltorsion}, the action may be written globally as
\begin{align}
S[\phi] = -i\intd{^2x\rd\theta^+} g_{ij} D_+\phi^i
\partial_\mimi\phi^j
-i \intd{^2x\rd\theta^+\rd t} H_{ijk}\partial_t\Tilde\phi^i D_+\Tilde\phi^j
\partial_\mimi\Tilde\phi^k
\end{align}
which is manifestly supersymmetric. 

It remains to describe the sigma model multiplet (superfield)
$\phi=\phi(x^\mu,\theta^+)$. We calculate the Taylor expansion around
$\theta^+=0$, which quickly terminates to give
\begin{align}
  \phi^i(x^\mu,\theta^+) &= \phi^i(x^\mu)+\theta^+\lambda^i_+(x^\mu)
\end{align}
where the components are
\begin{align}
\label{eq:10components}  
  \phi^i(x^\mu)&=\phi^i(x^\mu,\theta^+)|_{\theta^+=0} 
& \lambda_+^i(x^\mu)&=D_+\phi^i(x^\mu,\theta^+)|_{\theta^+=0}  ~,
\end{align}
which are the scalar and the fermion field. The fermion can be
thought of as the superpartner of $\phi^i$. 

Under the supersymmetry transformation,
\begin{align}
\begin{split}
  \delta_{\epsilon} \phi^i(x^\mu) &= \epsilon_- \lambda_+^i(x^\mu) \\
  \delta_{\epsilon} \lambda_+^i(x^\mu) &= i \epsilon_-\partial_\plpl\phi^i(x^\mu)
\end{split}
\end{align}
we see that supersymmetry swaps a field with its superpartner. 

Evaluating the superspace integral in~\eqref{eq:10susyaction} gives the
action for the components,
\begin{align}
  S &= \intd{^2x} \bigl(
  (g_{ij}+b_{ij}) \partial_\plpl\phi^i \partial_\mimi\phi^j + i
  g_{ij}\lambda_+^i\nabla^{(+)}_\mimi\lambda_+^j \bigr) ~,
\end{align}
where $\nabla^{(+)}_\mimi \lambda_+^i = \nabla_\mimi\lambda_+^i +
\tfrac{1}{2}\ud H{i}{jk}\partial_\mimi\phi^j\lambda_+^k$, and $\nabla$
is the Levi-Civita connection; we see that $H$ is a torsion
three-form.

\subsection{(2,0) supersymmetry}

Now we will construct the (2,0) supersymmetry algebra,
\begin{align}
\label{eq:20susyalgebra}
  [Q_{p+}, Q_{q+}]_+ &= 2i\delta_{pq} P_\plpl
\end{align}
where $p,q=0,1$. We begin with the sigma model from the previous
section; recall that $\Xi^{1,0}=(x^\plpl,x^\mimi,\theta^+)$,
$\phi:\Xi^{1,0}\rightarrow M$ and that the action is
\begin{align}
\label{eq:20action}
  S &= -i \intd{^2x} (g_{ij}+b_{ij}) D_+\phi^i \partial_\mimi\phi^j ~.
\end{align}
We note that $\phi$ has the same component expansion as
before~\eqref{eq:10components} because it is defined in exactly the
same way; the second supersymmetry will not introduce any new fields.

We relabel the previous supercharge as $Q_{0+}=Q_+$ and look for a
second supercharge of the form
\begin{align}
  Q_{1+}\phi^i &= \ud Jij D_+ \phi^j~.
\end{align}
Lorentz covariance requires that $\ud Jij$ is a (1,1) tensor on $M$. We
require that $Q_{1+}$ satisfies the supersymmetry algebra; we find
\begin{align}
  \du Q{1+}{2} \phi^i &= (\ud Ji{k,l} \ud Jlj - \ud Jil 
  \ud Jl{k,j})D_+\phi^jD_+\phi^k -i \ud Jij \ud Jjk D_+ \phi^k ~,
\end{align}
and so closure of the algebra requires the constraints
\begin{subequations}
\label{eq:20algebracnds}
\begin{gather}
    \ud Jij \ud Jjk = -\ud\delta ik \\
    \ud Ni{jk} \equiv \frac{1}{4}\bigl(\ud Ji{k,l} \ud Jlj - \ud Jil 
  \ud Jl{k,j}) = 0 ~,
\end{gather}  
\end{subequations}
where $N$ is the Nijenhuis tensor. These conditions imply that $J$ is
a complex structure, therefore $(M,J)$ is a complex manifold; in
particular its dimension is even.

Invariance of the action under the second supersymmetry requires that 
\begin{align}
\label{eq:20invactioncnds}
  g_{kl}\ud Jki \ud Jlj &= g_{ij} \\
  \nabla^{(+)}_i \ud Jjk &= 0~.
\end{align}
Therefore $g$ is a Hermitian metric and $J$ is covariantly constant
with respect to $\nabla^{(+)}=\nabla+\tfrac{1}{2}H$, where $\nabla$ is
the Levi-Civita connection and $H$ is the torsion. This implies that
$M$ must be a KT (K\"ahler with Torsion) manifold, in particular
$\nabla^{(+)}$ has $U(n)$ holonomy.

We observe that the amount of supersymmetry in the model constrains
the target space geometry. The converse is also true; given a complex
structure on a Hermitian manifold satisfying the constraints above, we
can construct a second supersymmetry, and this is a general feature of
two-dimensional supersymmetric sigma models.

We present a second, equivalent, formulation of the (2,0) model by
introducing the (2,0) superspace $\Xi^{2,0}$ with coordinates
$(x^\plpl,x^\mimi,\theta_{0+},\theta_{1+})$; the sigma model fields
are now maps $\phi:\Xi^{2,0}\rightarrow M$.  Supercovariant
derivatives $D_{0+},D_{1+}$ are defined similarly to the (1,0)
case~\eqref{eq:defnDplus} and satisfy
\begin{align}
\label{eq:10Dplusalgebra}
\du D{0+}2 = \du D{1+}2&=i\partial_\plpl &
  [D_{0+},D_{1+}]_+ &=0 ~.
\end{align}

We impose the constraint
\begin{align}
\label{eq:10Dconstraint}
  D_{1+}\phi^i &= \ud Jij D_{0+}\phi^j 
\end{align}
on the superfields. The independent components of $\phi$ are then
\begin{align}
  \phi^i&=\phi^i|_{\theta_0^+=\theta_1^+=0} 
& \lambda_+^i&=D_{0+}\phi^i|_{\theta_0^+=\theta_1^+=0}  ~,
\end{align}
which correspond to those defined before in~\eqref{eq:10components}.

Consistency of the above constraint with~\eqref{eq:10Dplusalgebra}
requires precisely the conditions we had before for the closure of the
algebra~\eqref{eq:20algebracnds}; therefore $J$ is a complex structure
and $M$ is a complex manifold.

The supercharges $Q_{p+}$ are defined by $[Q_{p+},D_{q+}]_+=0$; with
the above constraint they are identical to those defined in the first
part of this section, and in particular satisfy the supersymmetry
algebra.

The action is
\begin{align}
  S &= -i \intd{^2x\rd\theta^{0+}} (g_{ij}+b_{ij}) D_{0+}\phi^i \partial_\mimi\phi^j ~,
\end{align}
which is exactly as before~\eqref{eq:20action} except that here the
$\phi^i$ are (2,0) superfields. Invariance of this action under
supersymmetry is an identical calculation to the previous case and
leads to the same constraints~\eqref{eq:20invactioncnds}; in
particular, $M$ is a KT manifold.

\subsection{(4,0) supersymmetry}

The results of the previous section are readily generalised to the
case when we have more than two left-handed supersymmetries,
\begin{align}
\label{eq:40susyalgebra}
  [ Q_{p+}, Q_{q+} ]_+ &= 2i\delta_{pq} P_\plpl,\qquad p,q=0,\ldots,n-1~.
\end{align}
Let $\Xi^{1,0}=(x^\plpl,x^\mimi,\theta^+)$ be the usual (1,0)
superspace and $\phi:\Xi^{1,0}\rightarrow M$ be the sigma model field.
We consider the supercharges
\begin{gather}
  Q_{0+}\phi^i = Q_+\phi^i \\
  Q_{r+}\phi^i = \dud Jrij D_+ \phi^j
\end{gather}
where $Q_+$ was defined in~\eqref{eq:10susycharge} and $r=1,\ldots,
n-1$. By an identical calculation~\eqref{eq:20algebracnds} to the
previous section, the requirement that the supersymmetry algebra is
satisfied implies that each $J_r$ is a complex structure.  An extra
constraint arises because the $Q_{r+}$ anticommute; this is
\begin{align}
\label{eq:40invalgebracnds}
  \dud Jrik \dud Jskj + \dud Jsik \dud Jrkj &= 0~.
\end{align}

The action of the theory is
\begin{align}
  S &= -i \intd{^2x\rd\theta^+} (g_{ij}+b_{ij}) D_+\phi^i \partial_\mimi\phi^j ~,
\end{align}
and as in the previous section, invariance of the action requires that
\begin{align}
\label{eq:40invactioncnds}
  g_{kl}\dud Jrki \dud Jrlj &= g_{ij} \\
  \nabla^{(+)}_i \dud Jrjk &= 0~.
\end{align}
Therefore the metric is Hermitian with respect to each complex
structure, and each complex structure is covariantly constant. In
particular,  $\dim M=4n$ and $\nabla^{(+)}$ has $Sp(n)$ holonomy.

From~\eqref{eq:40invalgebracnds}, if we have two complex structures
$J_1$ and $J_2$, we may form a third $J_3=J_1\cdot J_2$, so (3,0)
supersymmetry implies (4,0) supersymmetry. Furthermore, if there are
more than three complex structures, then the target manifold $M$ is
reducible~\cite{Alvarez-Gaume:1981hm}.  Thus (4,0) supersymmetry is
the only additional interesting case, and the geometry of $M$ is then
HKT (Hyper-K\"ahler with Torsion); ie. there exists a metric
connection with torsion whose holonomy is $Sp(n)$; $\dim M=4n$.

\subsection{(1,1) supersymmetry}

We will now generalise the analysis of section~\ref{sec:10susy} to have
two supercharges of opposite chirality. The supersymmetry algebra is
\begin{align}
\label{eq:11susyalgebra}
[Q_+,Q_+]_+ &= 2iP_\plpl & [Q_-,Q_-]_+ &= 2iP_\mimi \\
  [Q_+,Q_-]_+ &= 0~.    
\end{align}
We define the (1,1) superspace
$\Xi^{1,1}=(x^\plpl,x^\mimi,\theta^+,\theta^-)$ where
$(x^\plpl,x^\mimi)$ are the null coordinates defined in
section~\ref{sec:2dsigmamodels} and $\theta^+,\theta^-$ are Grassman
coordinates of opposite chirality. We define the supercovariant
derivatives $D_+,D_-$ by $\du D+2=i\partial_\plpl$, $\du
D-2=i\partial_\mimi$ and $[D_+,D_-]_+=0$; their component expansion is
\begin{align}
  D_+ &= \frac{\partial}{\partial\theta^+} +
  i\theta^+\frac{\partial}{\partial x^\plpl} ~, \\
  D_- &= \frac{\partial}{\partial\theta^-} +
  i\theta^-\frac{\partial}{\partial x^\mimi} ~.
\end{align}

The Noether supercharges are defined so that $[Q_+,D_\pm]_+=0$ and
$[Q_-,D_\pm]_+=0$; their expansions in components are
\begin{align}
  Q_+ &= \frac{\partial}{\partial\theta^+} -
  i\theta^+\frac{\partial}{\partial x^\plpl} ~, \\
  Q_- &= \frac{\partial}{\partial\theta^-} -
  i\theta^-\frac{\partial}{\partial x^\mimi} 
\end{align}
and it may be checked that $P_\plpl$, $P_\mimi$, $Q_+$ and $Q_-$ obey
the supersymmetry algebra~\eqref{eq:11susyalgebra}. 

The two dimensional (1,1) supersymmetric sigma model is defined by
the superspace $\Xi^{1,1}$, the target manifold $(M,g)$ and sigma
model maps $\phi:\Xi^{1,1}\rightarrow M$ which satisfy an action
principle with action 
\begin{align}
  \label{eq:11action}
  S &= \intd{^2x\rd\theta^+\rd\theta^-} \bigl(
    (g_{ij}+b_{ij})D_+\phi^iD_-\phi^j \bigr)~.
\end{align}
This action is invariant under (i) coordinate changes of $M$ (modulo
gauge transformations of $b$), (ii) Lorentz transformations of
$\Xi^{1,1}$ and (iii) the transformations of the fields generated by
translations and supertranslations, ie.
\begin{align}
  \delta_\alpha \phi^i &= \alpha^\plpl P_\plpl\phi^i
  +\alpha^\mimi P_\mimi\phi^i \\
  \delta_\epsilon \phi^i &= \epsilon^+Q_+ \phi^i + \epsilon^-Q_- \phi^i
\end{align}
where $\alpha^\mu$ is an infinitesimal (bosonic) constant and
$\epsilon^+,\epsilon^-$ are constant Grassman parameters. To show that
this is the case, (i) and (ii) are manifest; for part (iii),
invariance under $P_\plpl,P_\mimi$ follows similarly to
section~\ref{sec:sigmaaction} and invariance under $Q_\pm$ follows
from a similar calculation to~\eqref{eq:susyinvcalculation}.

The new right-handed supersymmetry introduces a second fermion to the
theory to the sigma model multiplet; $\phi$ now has the components
\begin{align}
  \phi^i &= \phi^i| & F^i &= \nabla^{(+)}_+ D_- \phi^i| \\
  \lambda_+^i &= D_+\phi^i| & \lambda_-^i &=D_-\phi_+^i|
\end{align}
where the vertical line denotes evaluation at $\theta^+=\theta^-=0$.

The action in components is
\begin{align}
\label{eq:11cptaction}
  \begin{split}    
    S = \intd{^2x} \bigl(& (g_{ij}+b_{ij}) \partial_\plpl\phi^i
    \partial_\mimi\phi^j + i
    g_{ij}\lambda_+^i\nabla^{(+)}_\mimi\lambda_+^j
    +i g_{ij}\lambda_-^i\nabla^{(-)}_\mimi\lambda_-^j \\
    & -g_{ij}F^iF^j
    -\frac{1}{2}R^{(+)}_{ijkl}\lambda_+^i\lambda_+^j\lambda_-^k\lambda_-^l
    \bigr) ~,
  \end{split}
\end{align}
where $\nabla^{(\pm)}=\nabla\pm\tfrac{1}{2}H$,
\begin{align}
  R^{(+)}_{ijkl} &= R_{ijkl} + \partial_i H_{jkl} - \partial_j H_{ikl}
  + H_{mjl}\ud Hm{ik} - H_{mil}\ud Hm{jk}
\end{align}
is the curvature of $\nabla^{(+)}$ and $R_{ijkl}$ is the Riemann
tensor.  We may eliminate $F$ from the above action because its
equation of motion is $F=0$.

The components transform under the supersymmetry transformations as
\begin{align}
  \delta_\epsilon \phi^i &= \epsilon^+\lambda_+^i +
  \epsilon^-\lambda_-^i \\
  \delta_\epsilon \lambda_+^i &= i\epsilon^+ \partial_\plpl\phi^i +
  \epsilon^-\bigl(\ud\Gamma{i}{jk}+\frac{1}{2}\ud
  Hi{jk}\bigr)\lambda_+^j\lambda_-^k \\
  \delta_\epsilon \lambda_-^i &= -\epsilon^+
  \bigl(\ud\Gamma{i}{jk}+\frac{1}{2}\ud
  Hi{jk}\bigr)\lambda_+^j\lambda_-^k +i\epsilon^-\partial_\mimi\phi^i ~,
\end{align}
where we have eliminated the field $F$.

\subsection{(2,1) supersymmetry}

The (2,1) supersymmetry algebra is
\begin{align}
  [Q_{p+}, Q_{q+} ]_+ &= 2i\delta_{pq}P_\plpl & [Q_-,Q_-]_+ &= 2i P_\mimi
\end{align}
where $p,q=0,1$ and all other brackets vanish. We recall the
definition of the (1,1) supersymmetric sigma model and relabel the
supercharge as $Q_{0+}=Q_+$. The second supersymmetry is defined to be
\begin{align}
  Q_{1+} \phi^i &= \ud Jij D_+ \phi^j~.
\end{align}
To satisfy the supersymmetry algebra, we must impose the same
conditions on $J$ as the (2,0) case~\eqref{eq:20algebracnds}, ie. that
$J$ is a complex structure. Invariance of the action
\begin{align}
\label{eq:21action}
  S &= \intd{^2x\rd\theta^+\rd\theta^-} (g_{ij}+b_{ij})D_+\phi^i
  D_- \phi^j
\end{align}
requires that $g$ is Hermitian with respect to $J$ and that $J$ is
covariantly constant, ie.
\begin{align}
\label{eq:21invactioncnds}
  g_{kl}\ud Jki \ud Jlj &= g_{ij} &
  \nabla^{(+)}_i \ud Jjk &= 0~.
\end{align}
Therefore $M$ is a KT manifold.

The component form of the action~\eqref{eq:21action} was given in the
previous section~\eqref{eq:11cptaction}.

A second equivalent formulation is to define the (2,1) superspace
$\Xi^{2,1}$ with coordinates
$(x^\plpl,x^\mimi,\theta_0^+,\theta_1^+,\theta^-)$. Supercovariant
derivatives $D_{p+}$, $D_-$ are defined to satisfy
\begin{align}
\label{eq:21covderivalgebra}
  \du D{0+}2=\du D{1+}2&=i\partial_\plpl &
  [D_{0+},D_{1+}]_+&=0 &
  [D_{p+},D_-]_+&=0~;
\end{align}
we constrain the sigma model maps $\phi:\Xi^{2,1}\rightarrow M$ by
\begin{align}
  D_{1+}\phi^i &= \ud Jij D_{0+}\phi^j~.
\end{align}
Consistency of this constraint with the
algebra~\eqref{eq:21covderivalgebra} of supercovariant derivatives
requires that $J$ is a complex structure; invariance of the
action~\eqref{eq:21action} requires precisely the same constraints as
before~\eqref{eq:21invactioncnds}, which imply that $M$ is a KT
manifold.

We observe how the torsion of the theory allows us to define
theories with differing numbers of right and left handed
supersymmetries. For suppose we look for an additional \emph{right}
handed supersymmetry,
\begin{align}
  Q_{1-}\phi^i &= \ud Kij D_- \phi^j~.
\end{align}
The supersymmetry algebra and invariance of the action then imply the
conditions 
\begin{gather}
\label{eq:21invcnds}
    \ud Kij \ud Kjk = -\ud\delta ik \\
    \ud Ni{jk} \equiv \frac{1}{4}\bigl(\ud Ki{k,l} \ud Klj - \ud Kil 
  \ud Kl{k,j}) = 0 \\
  g_{kl}\ud Kki \ud Klj = g_{ij} \\
  \nabla^{(-)}_i \ud Kjk = 0~.
\end{gather}
If the torsion vanishes, $H=0$, the last condition above becomes
$\nabla_i \ud Kjk=0$ which is satisfied by taking $K=J$; therefore it
is not necessary to introduce a second complex structure $K$.  Hence
on a K\"ahler manifold, (2,1) supersymmetry implies (2,2)
supersymmetry.

We note that (2,2) supersymmetry on a KT manifold $M$ requires two
commuting complex structures $J$ and $K$; in this case we can form a
\emph{product structure} $\ud\Pi ij =\ud Jik \ud Kkj$; using $\Pi$ we
may decompose $M$ to be diffeomorphic (but not isometric) to product
of two manifolds, $M_1\times M_2$ where $M_1$ and $M_2$ are both KT.
This allows the introduction of twisted multiplets, see for
example~\cite{Hull:1991uw}.

\subsection{(4,1) supersymmetry}
The results of the previous section are readily generalised to the
case when we have two or more left handed supersymmetries (and one
right handed supersymmetry). From the section on (4,0) supersymmetry,
we know that the only interesting case will be (4,1) supersymmetry.

The (4,1) supersymmetry algebra is
\begin{align}
  [Q_{p+}, Q_{q+} ]_+ &= 2i\delta_{pq}P_\plpl & [Q_-,Q_-]_+ &= 2i P_\mimi
\end{align}
where $p,q=0,\ldots,3$. We define the (4,1) superspace $\Xi^{4,1}$
with coordinates $(x^\plpl,x^\mimi,\theta_p^+,\theta_-)$; the sigma
model fields are maps $\phi:\Xi^{4,1}\rightarrow M$. We define the
supercovariant derivatives $D_{p+}$,$D_-$ to satisfy
\begin{align}
  \du D{p+}2 &= i\partial_\plpl & [D_{p+},D_{q+}]_+&=0 & [D_{p+},D_-]_+&=0
\end{align}
and impose the following constraint on the sigma model fields,
\begin{align}
  D_{r+}\phi^i &= \dud Jrij D_{0+}\phi^j \qquad r=1,2,3~.
\end{align}
The supercharges are
\begin{align}
  Q_{0-}\phi^i &= \bigl(\partial_{\theta_0-} - 
  i\theta_0^-\partial_\mimi)\phi^i \\
  Q_{0+}\phi^i &= \bigl(\partial_{\theta_0+} -
  i\theta_0^+\partial_\plpl)\phi^i \\
  Q_{r+}\phi^i &= \dud Jrij D_{0+}\phi^j
\end{align}
where $r=1,2,3$. The supercharges obey the supersymmetry algebra if
and only if the algebra of supercovariant derivatives is satisfied; for
this to hold we require
\begin{gather}
\label{eq:41invalgebracnds}
\begin{split}
    \dud Nri{jk} \equiv \frac{1}{4}\bigl(\dud Jri{k,l} \dud Jrlj - \dud Jril 
  \dud Jrl{k,j}) = 0 \\
  \dud Jrik \dud Jskj + \dud Jsik \dud Jrkj = -2\delta_{rs}\ud\delta ij ~.
\end{split}
\end{gather}
Therefore the $J_r$ are anticommuting complex structures.

The action is
\begin{align}
\label{eq:41action}
  S &= \intd{^2x\rd\theta_0^+\rd\theta^-} (g_{ij}+b_{ij})D_+\phi^i
  D_- \phi^j~;
\end{align}
invariance of the action requires that $g$ is Hermitian with respect
to each $J_r$ and that each $J_r$ is covariantly constant,
\begin{align}
  g_{kl}\ud Jki \ud Jlj &= g_{ij} &
  \nabla^{(+)}_i \ud Jjk &= 0~.
\end{align}
Therefore $M$ is a HKT manifold. We note that the torsion vanishes,
$H=0$, then we would be able to construct (4,4) supersymmetry in a
similar way to the previous section.

\section{One dimensional sigma models}

Let $\Sigma^1$ be one dimensional $N=1$ superspace with coordinates
$(t,\theta)$ where where $t$ is a real coordinate and $\theta$ is a
Grassman coordinate. A one dimensional $N=1$ sigma model superfield is
a map $q:\Sigma^1\rightarrow M$ where $(M,g)$ is the target manifold;
$q$ can be thought of as a point particle with proper time $t$.

We introduce a second fermionic superfield $\chi^a(t,\theta)$ which is
defined to be a section of the bundle $q^*\epsilon$ over $\Sigma^1$,
where $\epsilon$ is the real vector bundle over $M$. It can be thought
of as the Yang-Mills sector, and is used for example to introduce a
potential into the model, see for example~\cite{Alvarez-Gaume:1983ab}.

We define the supercovariant derivative $D$ by the property
$D^2=\partial$ where $\partial=\tfrac{\partial}{\partial t}$, in
components
\begin{align}
  D &= \frac{\partial}{\partial
  \theta}+\theta\frac{\partial}{\partial t}~.
\end{align}
Let $h_{ab}$ be the fibre metric and $\dud Biab$ the connection of the
bundle. We define the covariant derivative 
\begin{align}
  \nabla\chi^a &= D\chi^a + Dq^i \dud Biab\chi^b~;
\end{align}
without loss of generality we may assume that the fibre metric
$h_{ab}$ is compatible with the connection, $\nabla_i h_{ab}=0$.

The two multiplets satisfy an action principle with action
\begin{align}
   S = -\intd{t\rd\theta}\bigl(
     \frac{1}{2}g_{ij}Dq^i\partial q^j +
     \frac{1}{3!}c_{ijk}Dq^iDq^jDq^k 
       -\frac{1}{2} h_{ab}\chi^a \nabla \chi^b \bigr)~.
\end{align}
if $c$ is closed then it may be interpreted as the torsion of $M$,
however we will make no such restriction. 

The action is invariant under (i) reparameterisations of the
superspace $\Sigma^1$, (ii) reparameterisations of the target manifold
$M$ and (iii) linear translations of the field given by
\begin{align}
  t &\rightarrow t+\alpha \\
  q &\rightarrow q-\alpha\partial_t q
\end{align}
where $\alpha$ is real and constant; these transformations generate
the Noether charge $P=\partial$ associated with energy. The action is
also invariant under supersymmetry generated by the supercharge $Q$
defined by $[Q,D]_+=0$, in components this is
\begin{align}
  Q &= \frac{\partial}{\partial
  \theta}+\theta\frac{\partial}{\partial t}
\end{align}
and acts on the fields as $\delta_\epsilon \phi^i=\epsilon Q\phi^i$,
$\delta_\epsilon \chi^a=\epsilon Q\chi^a$, where $\epsilon$ is an
constant Grassman parameter.

Taking the commutator of two such transformations,
\begin{align}
  [ \delta_\zeta, \delta_\eta ] q^{\mu} = 2 \zeta\eta \Dot{q}^{\mu}
 = 2\zeta\eta P q^{\mu}~.
\end{align}
which shows that $Q$ and $P$ satisfy the supersymmetry algebra,
\begin{align}
  [ Q, Q]_+ &= 2iP~.
\end{align}

We expand the action in components defined by
\begin{align}
  q^{i} &= q^{i}| & \chi^{a} &= \chi^{a}| \\
  \lambda^{i} &= Dq^{i}| & y^{a} &= \nabla \chi^{a}|
\end{align}
where the line means evaluation at $\theta=0$. We are following the standard
convention of using the same letter for a superfield and its lowest
component. The field $y^{a}$ is auxiliary and we will eliminate it later.
Expanding the action~(\ref{eq:rigidactionsf}) into component form, we
find that
\begin{equation}
  \label{eq:rigidactioncom}
  \begin{split}
    S=\int\mspace{-5mu} \mathrm{d}t
    \bigl( 
    & \frac{1}{2} g_{ij} \Dot{q}^i\Dot{q}^j
    + \frac{1}{2} g_{ij} \lambda^i \nabla^{(+)}_t \lambda^j 
    - \frac{1}{2}h_{ab} y^ay^b 
    - \frac{1}{2} h_{ab} \chi^a \nabla_t \chi^b \\
    & +\frac{1}{4}G_{ijab}\lambda^i\lambda^j\chi^a\chi^b
    + \frac{1}{3!}\nabla_{[i} c_{jkl]}\lambda^i\lambda^j\lambda^k\lambda^l
    \bigr)
  \end{split}
\end{equation}
where $G_{ijab}$ is the curvature of the vector bundle
connection $B_{i\phantom{a}b}^{\phantom{i}a}$,
\begin{align}
\dud G{ij}ab = \partial_i \dud Bjab - \partial_j \dud Biab
+ \dud Biac \dud Bjcb - \dud Bjac \dud Bicb
\end{align}
and $\nabla^{(+)}$ is the covariant derivative including the $c$ term,
\begin{equation}
  \nabla_t^{(+)}\lambda^i = \nabla_t\lambda^i + \frac{1}{2} \ud ci{jk}
  \Dot{q}^j\lambda^k
\end{equation}
The $N=1$ supersymmetry transformations of the component fields are
\begin{gather}
  \delta_\epsilon q^{i} = \epsilon \lambda^{i} \\
  \delta_\epsilon \lambda^{i} = -\epsilon \Dot{q}^{i} \\
   \delta_\epsilon \chi^{a} = \epsilon (y^{a} - \lambda^{i}B^{\phantom{i}a}_{i\phantom{a}b}\chi^{b})\\
  \delta_\epsilon y^{a} = -\epsilon(\nabla_t
  \chi^{a}+\lambda^{i}B^{\phantom{i}a}_{i\phantom{a}b}y^{b})
  +\frac{1}{2}\epsilon\lambda^{i}\lambda^{j}\chi^{b} G^{\phantom{ij}a}_{ij\phantom{a}b}
\end{gather}

Extended supersymmetries, where there are $N>1$ supercharges may be
constructed in the same way as for two dimensional sigma models. We
comment that the resulting models are broadly similar to the two
dimensional case, except that the resulting geometry of the target
manifold $M$ is not so tightly constrained. For example,
in~\cite{Coles:1990hr} an $N=3$ multiplet is constructed which does
not imply $N=4$, and $N=2$ models are defined on manifolds with
foliations.

\chapter{Spinning particles and Supergravity}

\section{The $N=1$ supersymmetric action}
\label{sec:review}
The most general $N=1$ spinning particle action with rigid
supersymmetry is~\cite{Papadopoulos:2000ka},
\begin{equation} 
  \label{eq:rigidactionsf}
  \begin{split}
    S=-\int\mspace{-5mu} \mathrm{d}t\,\mathrm{d}\theta 
    \biggl[ 
    &\frac{1}{2} g_{\mu\nu} Dq^{\mu}\partial_t q^{\nu} 
    - \frac{1}{2} h_{\alpha\beta} \chi^{\alpha} \nabla \chi^{\beta}
    + \frac{1}{3!} c_{\mu\nu\rho} Dq^{\mu} Dq^{\nu} Dq^{\rho} \\
    &+ \frac{1}{2} m_{\mu\alpha\beta} Dq^{\mu} \chi^{\alpha} \chi^{\beta}             
    + \frac{1}{2} n_{\mu\nu\alpha}Dq^{\mu}Dq^{\nu}\chi^{\alpha}
    + \frac{1}{3!} l_{\alpha\beta\gamma}\chi^{\alpha}\chi^{\beta}\chi^{\gamma} \\
    &- f_{\mu\alpha}\partial_t q^{\mu}\chi^{\alpha} 
    + A_{\mu}Dq^{\mu}
    +ms_{\alpha}\chi^{\alpha}
    \biggr]
  \end{split}
\end{equation}
We recall the definitions from the previous chapter;
$D=\partial_\theta-\theta\partial_t$ and
\begin{equation}
  \nabla\chi^{\alpha} = D\chi^{\alpha} + Dq^{\mu} B_{\mu\phantom{\alpha}\beta}^{\phantom{\mu}\alpha} \chi^{\beta}
\end{equation}
where $B_{\mu\phantom{\alpha}\beta}^{\phantom{\mu}\alpha}$ is a
connection of the bundle $\epsilon$ with fibre metric
$h_{\alpha\beta}$. We may assume that the fibre metric
$h_{\alpha\beta}$ is compatible with the connection,
$\nabla_{\mu}h_{\alpha\beta}=0$.

The action includes the $c$-term $c_{\mu\nu\rho}$ and an
electromagnetic potential $A_{\mu}$. The potential of the theory is
described in terms of $s_{\alpha}$~\cite{Hull:1993ct},
$V(q)=ms_{\alpha}s^{\alpha}/2$. Also present are the Yukawa couplings
$m_{\mu\alpha\beta},n_{\mu\nu\alpha},l_{\alpha\beta\gamma}$ and
$f_{\mu\alpha}$.


The action is invariant under worldline translations generated by
$H=\partial_t$,
\begin{align}
  \delta^{(H)}_{\epsilon} q^{\mu}&=\epsilon\Dot{q}^{\mu} 
& \delta^{(H)}_{\epsilon} \chi^{\mu}&=\epsilon\Dot{\chi}^{\mu} 
\end{align}
and supersymmetry transformations generated by $Q=\partial_\theta+\theta\partial_t$,
\begin{align}
\label{eq:rst}
  \delta_\zeta q^{\mu} &= \zeta Q q^{\mu} 
  & \delta_\zeta \chi^{\alpha} &= \zeta Q \chi^{\alpha}
\end{align}
where $\zeta$ is anticommuting. These two transformations satisfy the
$N=1$ supersymmetry algebra $\{Q,Q\}=2H$.

We expand the action in components defined by
\begin{align}
  q^{\mu} &= q^{\mu}| & \chi^{\alpha} &= \chi^{\alpha}| \\
  \lambda^{\mu} &= Dq^{\mu}| & y^{\alpha} &= \nabla \chi^{\alpha}|
\end{align}
Expanding the action~(\ref{eq:rigidactionsf}) into component form, we
find that
\begin{equation}
  \label{eq:rigidactioncomfull}
  \begin{split}
    S=\int\mspace{-5mu} \mathrm{d}t
    \biggl[ 
    & \frac{1}{2} g_{\mu\nu} \Dot{q}^{\mu}\Dot{q}^{\nu}
    + \frac{1}{2} g_{\mu\nu} \lambda^{\mu} \nabla^{(+)}_t \lambda^{\nu} 
    - \frac{1}{2}h_{\alpha\beta} y^{\alpha}y^{\beta} - \frac{1}{2} h_{\alpha\beta} \chi^{\alpha} \nabla_t \chi^{\beta} \\
    & +\frac{1}{4}G_{\mu\nu\alpha\beta}\lambda^{\mu}\lambda^{\nu}\chi^{\alpha}\chi^{\beta}
    - \frac{1}{3!}\nabla_{[\mu} c_{\nu\rho\sigma]}\lambda^{\mu}\lambda^{\nu}\lambda^{\rho}\lambda^{\sigma}\\
    & + \frac{1}{2}m_{\mu\alpha\beta}\Dot{q}^{\mu}\chi^{\alpha}\chi^{\beta} + m_{\mu\alpha\beta}\lambda^{\mu}y^{\alpha}\chi^{\beta} 
    -\frac{1}{2} \nabla_{[\mu} m_{\nu]\alpha\beta} \lambda^{\mu}\lambda^{\nu}\chi^{\alpha}\chi^{\beta} \\
    & + n_{\mu\nu\alpha} \Dot{q}^{\mu} \lambda^{\nu} \chi^{\alpha} 
    - \frac{1}{2} n_{\mu\nu\alpha} \lambda^{\mu} \lambda^{\nu} y^{\alpha} 
    - \frac{1}{2} \nabla_{[\mu} n_{\nu\rho]\alpha}\lambda^{\mu}\lambda^{\nu}\lambda^{\rho}\chi^{\alpha}  \\
    & - \frac{1}{2} l_{\alpha\beta\gamma}\chi^{\alpha}\chi^{\beta} y^{\gamma} 
    - \frac{1}{3!} \nabla_{\mu} l_{\alpha\beta\gamma} \lambda^{\mu}\chi^{\alpha}\chi^{\beta}\chi^{\gamma}\\
    &+ f_{\mu\alpha}\Dot{q}^{\mu}y^{\alpha} 
    + f_{\mu\alpha}\Dot{\lambda}^{\mu} \chi^{\alpha} 
    + \nabla_{\mu} f_{\nu\alpha} \lambda^{\mu}\Dot{q}^{\nu} \chi^{\alpha} \\
    & + A_{\mu}\Dot{q}^{\mu} -  \nabla_{[\mu} A_{\nu]} \lambda^{\mu}\lambda^{\nu} 
    - ms_{\alpha} y^{\alpha} - m \nabla_{\mu} s_{\alpha} \lambda^{\mu}\chi^{\alpha}
    \smash[t]{\biggr]}
  \end{split}
\end{equation}
where $G_{\mu\nu\alpha\beta}$ is the curvature of the vector bundle
connection $B_{\mu\phantom{\alpha}\beta}^{\phantom{\mu}\alpha}$,
\begin{equation}
  G^{\phantom{\mu\nu}\alpha}_{\mu\nu\phantom{\alpha}\beta}
  =\partial_{\mu}B_{\nu\phantom{\alpha}\beta}^{\phantom{\nu}\alpha}
  - \partial_{\nu}B_{\mu\phantom{\alpha}\beta}^{\phantom{\mu}\alpha}
  +B^{\phantom{\mu}\alpha}_{\mu\phantom{\alpha}\gamma}B^{\phantom{\nu}\gamma}_{\nu\phantom{\gamma}\beta}
  -B^{\phantom{\nu}\alpha}_{\nu\phantom{\alpha}\gamma}B^{\phantom{\mu}\gamma}_{\mu\phantom{\gamma}\beta}
\end{equation}
and $\nabla^{(+)}$ is the covariant derivative including the $c$-term,
\begin{equation}
  \nabla_t^{(+)}\lambda^{\mu} = \nabla_t\lambda^{\mu} - \frac{1}{2} c^{\mu}_{\phantom{\mu}\nu\rho}\Dot{q}^{\nu}\lambda^{\rho}
\end{equation}
The $N=1$ supersymmetry transformations of the component fields are
\begin{gather}
  \delta_\zeta q^{\mu} = \zeta \lambda^{\mu} \\
  \delta_\zeta \lambda^{\mu} = -\zeta \Dot{q}^{\mu} \\
   \delta_\zeta \chi^{\alpha} = \zeta (y^{\alpha} - \lambda^{\mu}B^{\phantom{\mu}\alpha}_{\mu\phantom{\alpha}\beta}\chi^{\beta})\\
  \delta_\zeta y^{\alpha} = -\zeta(\nabla_t
  \chi^{\alpha}+\lambda^{\mu}B^{\phantom{\mu}\alpha}_{\mu\phantom{\alpha}\beta}y^{\beta})
  +\frac{1}{2}\zeta\lambda^{\mu}\lambda^{\nu}\chi^{\beta} G^{\phantom{\mu\nu}\alpha}_{\mu\nu\phantom{\alpha}\beta}
\end{gather}

\section{Supergravity in one dimension}
\label{sec:local}
To construct the $N=1$ supergravity action, we gauge the rigid
supersymmetry by promoting the supersymmetry parameter $\zeta$ to a
local parameter, $\zeta=\zeta(t)$. In general this will destroy
invariance of (\ref{eq:rigidactioncom}) under supersymmetry, because we will get terms proportional
to $\Dot{\zeta}$. Therefore it is necessary to introduce gauge fields 
whose transformations will cancel with the $\Dot{\zeta}$ terms arising
from varying (\ref{eq:rigidactioncom}).

The method to find how these fields appear in the action, and their
transformations, is the Noether technique. 

To illustrate this technique we will gauge the supersymmetry for a
simple Lagrangian in flat space, reproducing the results
of~\cite{Brink:1976sz,Brink:1977uf}.
\begin{equation}
  \mathcal{L}_0=\frac{1}{2}\eta_{\mu\nu}\Dot{q}^{\mu}\Dot{q}^{\nu} + \frac{1}{2}\eta_{\mu\nu}\lambda^{\mu}\Dot{\lambda}^{\nu}
\end{equation}
Taking the supersymmetry transformation, with $\zeta=\zeta(t)$,
\begin{equation}
\label{eq:noetherexample}
  \delta_\zeta\mathcal{L}_0=\Dot{\zeta}\eta_{\mu\nu}\lambda^{\mu}\Dot{q}^{\nu}
\end{equation}
up to surface terms, which vanish in the action. To cancel
(\ref{eq:noetherexample}), consider the Lagrangian to first order in a
parameter $g$,
\begin{equation}
  \mathcal{L}_1=\mathcal{L}_0+g\psi\eta_{\mu\nu}\lambda^{\mu}\Dot{q}^{\nu}
\end{equation}
where $\psi$ is a gauge field with
$\delta_\zeta\psi=-g^{-1}\Dot{\zeta}$. Then the variation of the
new Lagrangian vanishes to zeroth order in $g$.

Continuing this process for all orders of $g$, the Noether technique gives
\begin{equation}
  \mathcal{L}=\frac{1}{2}e^{-1}\eta_{\mu\nu}\Dot{q}^{\mu}\Dot{q}^{\nu} +
    \frac{1}{2}\eta_{\mu\nu}\lambda^{\mu}\Dot{\lambda}^{\nu}
    +ge^{-1}\psi\eta_{\mu\nu}\Dot{q}^{\mu}\lambda^{\nu}
\end{equation}
which is precisely the result of~\cite{Brink:1976sz}.
It proves necessary to introduce a second gauge field $e$ which
transforms under local supersymmetry as $\delta_\zeta
e=2\zeta\psi$. It is an einbein which is the gauge field associated
with the diffeomorphisms of the worldline.

It is also necessary to modify
the transformation for $\lambda^{\mu}$ to
\begin{equation}
  \delta_\zeta \lambda^{\mu} = -e^{-1}\zeta(\Dot{q}^{\mu}+g\psi\lambda^{\mu})
\end{equation}
Applying the Noether method to (\ref{eq:rigidactioncomfull}), the $N=1$
supergravity action is
\begin{equation}
  \label{eq:locsc}
  \begin{split}
    S=\int\mspace{-5mu} \mathrm{d}t
    \biggl[ 
    & \frac{1}{2} e^{-1} g_{\mu\nu} (\Dot{q}^{\mu} + \psi\lambda^{\mu}) (\Dot{q}^{\nu} + \psi\lambda^{\nu})
    + \frac{1}{2} g_{\mu\nu} \lambda^{\mu} \nabla_t \lambda^{\nu} \\
    & - \frac{1}{2}eh_{\alpha\beta}y^{\alpha}y^{\beta} - \frac{1}{2} h_{\alpha\beta} \chi^{\alpha} \nabla_t \chi^{\beta} \\
    & +\frac{1}{4}e G_{\mu\nu\alpha\beta}\lambda^{\mu}\lambda^{\nu}\chi^{\alpha}\chi^{\beta}\\
    & + \frac{1}{2} c_{\mu\nu\rho}(\Dot{q}^{\mu}+\frac{2}{3}\psi\lambda^{\mu})\lambda^{\nu}\lambda^{\rho}
    - \frac{1}{3!}e\nabla_{[\mu} c_{\nu\rho\sigma]}\lambda^{\mu}\lambda^{\nu}\lambda^{\rho}\lambda^{\sigma} \\
    & + \frac{1}{2}m_{\mu\alpha\beta}\Dot{q}^{\mu}\chi^{\alpha}\chi^{\beta} + em_{\mu\alpha\beta}\lambda^{\mu}y^{\alpha}\chi^{\beta}
    -\frac{1}{2} e\nabla_{[\mu} m_{\nu]\alpha\beta} \lambda^{\mu}\lambda^{\nu}\chi^{\alpha}\chi^{\beta} \\
    & + n_{\mu\nu\alpha} (\Dot{q}^{\mu}+\frac{1}{2}\psi\lambda^{\mu})\lambda^{\nu} \chi^{\alpha}  
    - \frac{1}{2} e n_{\mu\nu\alpha} \lambda^{\mu} \lambda^{\nu} y^{\alpha}     
    - \frac{1}{2} e\nabla_{[\mu} n_{\nu\rho]\alpha}\lambda^{\mu}\lambda^{\nu}\lambda^{\rho}\chi^{\alpha}  \\
    & - \frac{1}{2} el_{\alpha\beta\gamma}\chi^{\alpha}\chi^{\beta} y^{\gamma} 
    - \frac{1}{3!} \psi l_{\alpha\beta\gamma}\chi^{\alpha}\chi^{\beta}\chi^{\gamma} 
    - \frac{1}{3!} e\nabla_{\mu} l_{\alpha\beta\gamma} \lambda^{\mu}\chi^{\alpha}\chi^{\beta}\chi^{\gamma}\\
    &+ f_{\mu\alpha}(\Dot{q}^{\mu}+\psi\lambda^{\mu})y^{\alpha} 
    + f_{\mu\alpha}\Dot{\lambda}^{\mu} \chi^{\alpha} 
    + \nabla_{\mu} f_{\nu\alpha} \lambda^{\mu}(\Dot{q}^{\nu}+\psi\lambda^{\nu}) \chi^{\alpha} \\
    & + A_{\mu}\Dot{q}^{\mu} - e \nabla_{[\mu} A_{\nu]} \lambda^{\mu}\lambda^{\nu} \\
    & - ms_{\alpha}(ey^{\alpha} + \psi\chi^{\alpha}) - em \nabla_{\mu} s_{\alpha} \lambda^{\mu}\chi^{\alpha}
    \smash[t]{\biggr]}
  \end{split}
\end{equation}
We observe that the Lagrangian is not obtained by the simple minimal
coupling rule $\Dot{q}^{\mu} \rightarrow \Dot{q}^{\mu} + \psi\lambda^{\mu}$.

The geometric interpretation of the couplings with the above action is
manifest. The above action is similar to that of models with rigid
supersymmetry in~\cite{Coles:1990hr}. 

After various redefinitions of the couplings and the fields, one can
recover the action constructed in~\cite{vanHolten:1995qt} in the
special case where $f_{\mu\alpha}=0$. However, if $f_{\mu\alpha}\neq
0$ then~(\ref{eq:locsc}) is more general.

The $N=1$ supersymmetry transformations for the components fields become
\begin{gather}
  \label{eq:sgtra}
  \delta_\zeta q^{\mu} = \zeta\lambda^{\mu} \\ 
  \delta_\zeta \lambda^{\mu} = -\zeta e^{-1}(\Dot{q}^{\mu}+\psi\lambda^{\mu}) \\
   \delta_\zeta \chi^{\alpha} = \zeta (y^{\alpha} - \lambda^{\mu}B^{\phantom{\mu}\alpha}_{\mu\phantom{\alpha}\beta}\chi^{\beta})\\
  \delta_\zeta y^{\alpha} = -e^{-1}\zeta(\nabla_t
  \chi^{\alpha}+\psi y^{\alpha}) -\zeta\lambda^{\mu}B^{\phantom{\mu}\alpha}_{\mu\phantom{\alpha}\beta}y^{\beta}
  +\frac{1}{2}\zeta\lambda^{\mu}\lambda^{\nu}\chi^{\beta} G^{\phantom{\mu\nu}\alpha}_{\mu\nu\phantom{\alpha}\beta}
\end{gather}
The einbein $e$ and gravitino $\psi$ transform as
\begin{align}
  \delta_\zeta e &= 2\zeta\psi, & \delta_\zeta\psi &= -\Dot{\zeta}
\end{align}
Checking the algebra of the new transformations~(\ref{eq:sgtra}), 
\begin{align}
  [\delta_\zeta, \delta_\eta]q^{\mu} &= 2e^{-1}\zeta\eta (\Dot{q}^{\mu}+\psi\lambda^{\mu})
  & [\delta_\zeta, \delta_\eta]\lambda^{\mu} &= 2e^{-1}\zeta\eta (\Dot{\lambda}^{\mu} -e^{-1} \psi\Dot{q})\\
  &= 2e^{-1} ( \delta^{(H)}_{\zeta\eta} + \delta_{\zeta\eta\psi}) q^{\mu}  
  & &= 2e^{-1} ( \delta^{(H)}_{\zeta\eta} + \delta_{\zeta\eta\psi}) \lambda^{\mu} 
\end{align}
from which we obtain
\begin{equation}
  [ \delta_\zeta, \delta_\eta ] = 2e^{-1} ( \delta^{(H)}_{\zeta\eta} + \delta_{\zeta\eta\psi})
\end{equation}
and the same is true on the components $\chi^{\alpha},y^{\alpha}$. 

For invariance of (\ref{eq:locsc}) under worldline diffeomorphisms,
we need to specify the action of $\delta^{(H)}_{\epsilon}$ on the
einbein and the gravitino,
\begin{align}
  \delta_\epsilon^{(H)} e&=\partial_t(\epsilon e)
  & \delta_\epsilon^{(H)} \psi &= \partial_t(\epsilon\psi)
\end{align}
and those for $q^{\mu},\lambda^{\mu},\chi^{\alpha}$ and $y^{\alpha}$ are unchanged.

\section{Hamiltonian Analysis}
\label{sec:constraints}
To investigate the Hamiltonian dynamics of the system described by
(\ref{eq:elimaction}), we follow the Dirac-Bergman
procedure~\cite{Sundermeyer:1982gv} to analyse the constraints. This
will be important when we come to quantise the system in the next
section.

At this stage we introduce vielbeins $\mathrm{e}_{\mu}^{\phantom{\mu}i}$,
$\mathrm{f}_{\alpha}^{\phantom{\alpha}a}$ so that
$g_{\mu\nu}=\eta_{ij}\mathrm{e}_{\mu}^{\phantom{\mu}i}\mathrm{e}_{\nu}^{\phantom{\nu}j}$
and
$h_{\alpha\beta}=\eta_{ab}\mathrm{f}_{\alpha}^{\phantom{\alpha}a}\mathrm{f}_{\beta}^{\phantom{\beta}b}$
where $\eta_{ij}$ and $\eta_{ab}$ are the flat metrics on the
manifold and vector bundle respectively. We will use latin letters for
vielbein indices and greek letters otherwise. We take
\begin{align}
  \lambda^i &= \mathrm{e}_{\mu}^{\phantom{\mu}i}\lambda^{\mu},
  &\chi^a &= \mathrm{f}_{\alpha}^{\phantom{\alpha}a}\chi^{\alpha}
\end{align}
as our new fermion fields. This ensures that in the next section, the
Dirac brackets, hence commutation relations, between $p$ and the
fermions are zero. In the following analysis it is also necessary to
set the Yukawa coupling $f_{\mu\alpha}=0$.

Adopting this notation, and eliminating the auxiliary field
$y^{\alpha}$ from~(\ref{eq:rigidactioncomfull}) using its equation of
motion, gives the action
\begin{equation}
\label{eq:elimaction}
  \begin{split}
    S=\int\mspace{-5mu} \mathrm{d}t
    \biggl[ 
    & \frac{1}{2} g_{\mu\nu} e^{-1}(\Dot{q}^{\mu}+\psi
    \mathrm{e}^{\mu}_{\phantom{\mu}i}\lambda^{i})(\Dot{q}^{\nu}+\psi \mathrm{e}^{\nu}_{\phantom{\nu}j}\lambda^{j})
    + \frac{1}{2} \eta_{ij} \lambda^{i} \nabla_t \lambda^{j} 
    - \frac{1}{2} \eta_{ab} \chi^{a} \nabla_t \chi^{b} \\
    & +\frac{1}{2}eh_{ab}Y^{a}Y^{b} +\frac{1}{4}eG_{ijab}\lambda^{i}\lambda^{j}\chi^{a}\chi^{b}\\
    & + \frac{1}{2} c_{ijk}(\mathrm{e}_{\mu}^{\phantom{\mu}i}\Dot{q}^{\mu}+\frac{2}{3}\psi\lambda^i)\lambda^j\lambda^k
    - \frac{1}{3!}e \mathrm{e}^{\mu}_{\phantom{\mu}[i|}\nabla_{\mu} c_{|jkl]}\lambda^{i}\lambda^{j}\lambda^{k}\lambda^{l} \\
    & + \frac{1}{2}m_{\mu ab}\Dot{q}^{\mu}\chi^{a}\chi^{b} 
    -\frac{1}{2}e \mathrm{e}^{\mu}_{\phantom{\mu}[i|}\nabla_{\mu} m_{|j]ab} \lambda^{i}\lambda^{j}\chi^{a}\chi^{b} \\
    & + n_{ija} (\mathrm{e}_{\mu}^{\phantom{\mu}i}\Dot{q}^{\mu}+\frac{1}{2}\psi\lambda^{i}) \lambda^{j} \chi^{a} 
    - \frac{1}{2} e\mathrm{e}^{\mu}_{\phantom{\mu}[i|}\nabla_{\mu} n_{|jk]a}\lambda^{i}\lambda^{j}\lambda^{k}\chi^{a}  \\
    & -\frac{1}{3!}\psi l_{abc}\chi^{a}\chi^{b}\chi^{c}
    - \frac{1}{3!} e\mathrm{e}^{\mu}_{\phantom{\mu}i}\nabla_{\mu} l_{abc} \lambda^{i}\chi^{a}\chi^{b}\chi^{c}\\
    & + A_{\mu}\Dot{q}^{\mu} - e\mathrm{e}^{\mu}_{\phantom{\mu}[i|}\nabla_{\mu} A_{|j]} \lambda^{i}\lambda^{j} 
    - \psi ms_{a}\chi^{a}
    - em \mathrm{e}^{\mu}_{\phantom{\mu}i}\nabla_{\mu} s_{a} \lambda^{i}\chi^{a}
    \smash[t]{\biggr]}
  \end{split}
\end{equation}
where for convenience we define
\begin{equation}
\label{eq:defofya}
  Y_{a}=m_{iab}\lambda^{i}\chi^{b} -\frac{1}{2}n_{ija}\lambda^{i}\lambda^{j}
  -\frac{1}{2}l_{abc}\chi^{b}\chi^{c}  -ms_{a} 
\end{equation}
and the two covariant derivatives are with respect to the spin
connection $\omega_{\mu\phantom{k}l}^{\phantom{\mu}k}$ on the manifold
and the spin connection$\Omega_{\mu\phantom{a}b}^{\phantom{\mu}a}$ on
the vector bundle,
\begin{gather}
  \nabla_t \lambda^i = \partial_t \lambda^i + \Dot{q}^{\mu}\omega_{\mu\phantom{i}k}^{\phantom{\mu}i}\lambda^k\\
  \nabla_t \chi^a = \partial_t \chi^a + \Dot{q}^{\mu}\Omega_{\mu\phantom{a}b}^{\phantom{\mu}a}\chi^b
\end{gather}

The canonical momenta for $\lambda^i$ and $q^i$ are
\begin{gather}
  \pi_i = -\frac{1}{2}\eta_{ij}\lambda^j \\
  \begin{split}
    p_{\mu}  ={} &  e^{-1} g_{\mu\nu} (\Dot{q}^{\nu}+\psi \mathrm{e}^{\nu}_{\phantom{\nu}i}\lambda^i) 
    + \frac{1}{2}\omega_{\mu jk}\lambda^j\lambda^k
    -\frac{1}{2}\Omega_{\mu ab}\chi^a\chi^b\\
    &+ \frac{1}{2} \mathrm{e}_{\mu}^{\phantom{\mu}i} c_{ijk} \lambda^j\lambda^k 
    + \frac{1}{2}  m_{\mu ab}\chi^a\chi^b
    + \mathrm{e}_{\mu}^{\phantom{\mu}i} n_{ija}\lambda^j\chi^a 
    + A_{\mu} 
  \end{split}
\end{gather}
respectively. Similarly the canonical momenta for  $\chi^a$, $e$ and $\psi$ are
\begin{align}
  \pi_{\chi a} &= \frac{1}{2}\eta_{ab}\chi^b ,
  & \pi_e &= 0 ,
  & \pi_\psi &= 0
\end{align}
Clearly the system is constrained, as would be expected. The explicit
constraints are
\begin{align}
\label{eq:pc}
  \phi_i &= \pi_i+\frac{1}{2}\eta_{ij}\lambda^j\approx 0 
& \phi_{\chi a} &= \pi_{\chi a} - \frac{1}{2}\eta_{ab}\chi^b \approx 0 \\
  \phi_e &= \pi_e \approx 0
& \phi_\psi &= \pi_\psi \approx 0
\end{align}
where the $\approx$ denotes weak equality, in other words equality up
to polynomial combinations of the other
constraints.

The constrained Hamiltonian can then be found to be
\begin{equation}
  \begin{split}
    \mathcal{H}_c ={} & 
    \frac{1}{2}e \eta_{ij} P^i P^j
    - \psi \lambda^i  (P_i + \frac{1}{3}c_{ijk}\lambda^j\lambda^k 
    + \frac{1}{2}n_{ija}\lambda^j\chi^a)\\
    &- \frac{1}{2} e \eta_{ab} Y^a Y^b 
     - \frac{1}{4}eG_{ijab}\lambda^i\lambda^j\chi^a\chi^b\\
    &+ \frac{1}{3!} e \mathrm{e}^{\mu}_{\phantom{\mu}[i|}\nabla_{\mu}c_{|jkl]}\lambda^i\lambda^j\lambda^k\lambda^l 
    + \frac{1}{2} e \mathrm{e}^{\mu}_{\phantom{\mu}[i|}\nabla_{\mu} m_{|j]ab} \lambda^i\lambda^j\chi^a\chi^b\\
    &+ \frac{1}{2} e \mathrm{e}^{\mu}_{\phantom{\mu}[i|}\nabla_{\mu}n_{|jk]a}\lambda^i\lambda^j\lambda^k\chi^a
    + \frac{1}{3!} e \mathrm{e}^{\mu}_{\phantom{\mu}i}\nabla_{\mu} l_{abc} \lambda^i\chi^a\chi^b\chi^c \\
    &+ \frac{1}{3!} \psi l_{abc}\chi^a\chi^b\chi^c
    + e \mathrm{e}^{\mu}_{\phantom{\mu}[i|}\nabla_{\mu} A_{|j]} \lambda^i\lambda^j \\
    &+ m \psi s_a\chi^a 
    + e \mathrm{e}^{\mu}_{\phantom{\mu}i}\nabla_{\mu} m s_a \lambda^i\chi^a 
  \end{split}
\end{equation}  
where $Y^a$ was defined in (\ref{eq:defofya}) and 
\begin{equation}
  \begin{split}
    P_i ={} &  \mathrm{e}^{\mu}_{\phantom{\mu}i} p_{\mu} - \frac{1}{2}\omega_{ijk}\lambda^j\lambda^k
    +\frac{1}{2}\Omega_{iab}\chi^a\chi^b\\
    &- \frac{1}{2} c_{ijk} \lambda^j\lambda^k 
    - \frac{1}{2} m_{iab}\chi^a\chi^b
    - n_{ija}\lambda^j\chi^a 
    - A_i 
  \end{split}
\end{equation}
The primary Hamiltonian is defined to be
\begin{equation}
  \mathcal{H}_p = \mathcal{H}_c + \phi_i u^i +
                  \phi_{\chi a} u_{\chi}^{\phantom{\chi}a}
                  + \phi_e u_e + \phi_\psi u_\psi
\end{equation}
where the $u$ are all Lagrange multipliers (and $u^i$,
$u_{\chi}^{\phantom{\chi}a}$, $u_{\psi}$ are anticommuting).

We require that the constraints~(\ref{eq:pc}) hold for all time, 
\begin{align}
\label{eq:sccon}
  \Dot{\phi}_i &= \{\phi_i,\mathcal{H}_p\} \approx 0
&  \Dot{\phi}_{\chi a} &= \{\phi_{\chi a},\mathcal{H}_p\} \approx 0\\
\label{eq:fccon}
  \Dot{\phi}_e &= \{\phi_e,\mathcal{H}_p\} \approx 0
&  \Dot{\phi}_\psi &= \{\phi_\psi,\mathcal{H}_p\} \approx 0
\end{align}
Each condition either determines a multiplier or leads to a new
constraint. We assume the canonical Poisson brackets
\begin{align}
  \{q^\mu, p_\nu\} &= \delta^\mu_{\phantom{\mu}\nu} \\
  \{\chi^a,\pi_{\chi b} \} &= -\delta^a_{\phantom{a}b} \\
  \{\lambda^i, \pi_j \} &= -\delta^i_{\phantom{i}j} \\
  \{e,\pi_e\} &=1 \\
  \{\psi,\pi_\psi\} &= -1
\end{align}
Imposing (\ref{eq:fccon}) requires the secondary constraints
\begin{gather}
\label{eq:Hconstraint}
  \begin{split}
    \varphi_e ={} & \frac{1}{2}\eta_{ij}P^iP^j - \frac{1}{2}\eta_{ab}Y^aY^b
     -\frac{1}{4}G_{ijab}\lambda^i\lambda^j\chi^a\chi^b \\
    & + \frac{1}{3!} \mathrm{e}^{\mu}_{\phantom{\mu}[i|}\nabla_{\mu} c_{|jkl]}\lambda^i\lambda^j\lambda^k\lambda^l 
     + \mathrm{e}^{\mu}_{\phantom{\mu}[i|}\nabla_{\mu} A_{|j]}\lambda^i\lambda^j\\
   &  + m \mathrm{e}^{\mu}_{\phantom{\mu}i}\nabla_{\mu} s_a\lambda^i\chi^a 
    + \frac{1}{2} \mathrm{e}^{\mu}_{\phantom{\mu}[i|}\nabla_{\mu} m_{|j]ab}\lambda^i\lambda^j\chi^a\chi^b\\
   &  + \frac{1}{2} \mathrm{e}^{\mu}_{\phantom{\mu}[i|}\nabla_{\mu} n_{|jk]a} \lambda^i\lambda^j\lambda^k\chi^a
     + \frac{1}{3!} \mathrm{e}^{\mu}_{\phantom{\mu}i}\nabla_{\mu} l_{abc}\lambda^i\chi^a\chi^b\chi^c
  \end{split}\\
\label{eq:Qconstraint}
    \varphi_\psi =  
    \lambda^i (P_i+\frac{1}{3}c_{ijk}\lambda^j\lambda^k +\frac{1}{2}n_{ija}\lambda^j\chi^a)
    - \frac{1}{3!}l_{abc}\chi^a\chi^b\chi^c
    - ms_a \chi^a 
\end{gather}
In fact these are the charges of $H,Q$ respectively. It can be checked
that both of these are conserved over time,
\begin{align}
  \{ \varphi_e, \mathcal{H}_p \} &\approx 0 &   \{ \varphi_\psi, \mathcal{H}_p \} &\approx 0 
\end{align}
so they give rise to no new constraints.

The remaining conditions (\ref{eq:sccon}) determine
\begin{align}
\label{eq:multipliers}
  u^i &= e \eta^{ij}\{ \varphi_e,\phi_j \} + \psi \eta^{ij} \{ \varphi_\psi, \phi_j \} \\
  u_{\chi}^{\phantom{\chi}a} &= 
  -e \eta^{ab} \{ \varphi_e,\phi_{\chi b} \} 
  - \psi \eta^{ab}\{ \varphi_\psi, \phi_{\chi b} \}
\end{align}

Observe that the constrained Hamiltonian $\mathcal{H}_c$ can be
written in terms of the secondary constraints
\begin{equation}
\label{eq:cansch}
  \mathcal{H}_c = \varphi_e e + \varphi_\psi\psi
\end{equation}
Essentially this is because $\varphi_e=\{\pi_e,\mathcal{H}_c\}$ and
$\varphi_\psi=\{\pi_\psi,\mathcal{H}_c\}$ and the Hamiltonian
$\mathcal{H}_c$ is linear in the gauge fields $e$ and $\psi$.

\section{Quantisation}
\label{sec:quantise}
We observe that the constraints $\phi_i$ and $\phi_{\chi a}$ are both
second class, whereas
\begin{gather}
  \phi_e \\
  \phi_\psi \\
  \varphi'_e = \varphi_e + \eta^{ij}\{ \varphi_e,\phi_i \}\phi_j
             - \eta^{ab}\{ \varphi_e,\phi_{\chi a}\}\phi_{\chi b} \\
  \varphi'_\psi = \varphi_\psi + \eta^{ij}\{ \varphi_\psi,\phi_i \}\phi_j
             - \eta^{ab}\{ \varphi_\psi,\phi_{\chi a}\}\phi_{\chi b} 
\end{gather}
are all first class. Defining the Dirac bracket as
\begin{equation}
\label{eq:defdirac}
  \{ A, B \}_D = \{A,B\} + \eta^{ij}\{ A,\phi_i\}\{\phi_j, B\} 
  - \eta^{ab}\{ A,\phi_{\chi a}\}\{\phi_{\chi b}, B\}
\end{equation}
then Poisson brackets between the primed constraints are weakly equal
to Dirac brackets between the original unprimed constraints. 

The extra terms on the right of~(\ref{eq:defdirac}) give rise to new
relations
\begin{align}
\label{eq:diracbrackets}
  \{\lambda^i, \lambda^j \}_D &= \eta^{ij} \\
  \{\chi^a,\chi^b\}_D &=-\eta^{ab}
\end{align}

We can check that the $N=1$ supersymmetry algebra is still obeyed,
\begin{equation}
  \{\varphi'_\psi,\varphi'_\psi\} \approx \{\varphi_\psi,\varphi_\psi\}_D = 2\varphi_e \approx 0 
\end{equation}
and that all the brackets between secondary constraints vanish,
\begin{align}
  \{\phi_e,\varphi'_e\} &\approx \{\phi_e,\varphi_e\}_D = 0
&  \{\phi_{\psi},\varphi'_\psi\} &\approx \{\phi_{\psi},\varphi_\psi\}_D = 0 \\
  \{\varphi'_e,\varphi'_e\} &\approx \{\varphi_e,\varphi_e\}_D = 0 
&  \{\varphi'_e,\varphi'_\psi\} &\approx \{\varphi_e,\varphi_\psi\}_D = 0 
\end{align}

The Hamiltonian can be written in terms of these first
class constraints. From~(\ref{eq:multipliers}) and~(\ref{eq:cansch}),
\begin{equation}
  \mathcal{H}_p = \varphi_e e + \varphi_\psi \psi + \phi_eu_e + \phi_\psi u_\psi
\end{equation}
so it is vanishing weakly, as is expected for a gravitational system.

In Dirac's process of quantisation, fields become operators acting on some
Hilbert space and second class constraints are imposed as operator conditions
on the states. First class constraints generate unphysical degrees of
freedom, so it is necessary to fix a gauge,
\begin{align}
  e &= 1 & \psi &= 0
\end{align}

Moving over to the quantised system, Dirac brackets become
(anti) commutation relations. One realisation of this algebra is
using the standard Clifford algebra generators
\begin{align}
  \{ \gamma^i, \gamma^j \} &= 2\eta^{ij} & \{ \gamma^a, \gamma^b \} &= 2\eta^{ab}
\end{align}
If the dimension of the manifold is even, for example $d=4$, then we
have an element of the algebra, $\gamma^{d+1}$, satisfying
$(\gamma^{d+1})^2=-1$ and $\{ \gamma^{d+1}, \gamma^i \}=0$ so the
following realisation exists,
\begin{align}
  \Hat{\lambda}^i &= \frac{1}{\sqrt{2}} \gamma^i \otimes 1
  & \Hat{\chi}^a &= \frac{1}{\sqrt{2}} \gamma^{d+1} \otimes \gamma^a
\end{align}
The $\gamma^{d+1}$ ensures that $\Hat{\lambda}^i$ and $\Hat{\chi}^a$
anticommute.  

If the manifold $\mathcal{M}$ is not even dimensional then the
following realisation may be used,
\begin{align}
    \Hat{\lambda}^i &= \frac{1}{2} \gamma^i \otimes 1 \otimes \sigma_1
  & \Hat{\chi}^a &= \frac{1}{2} 1 \otimes \gamma^a \otimes \sigma_2
\end{align}
where $\sigma_i$ are the Pauli spin matrices which satisfy
$\{\sigma_i,\sigma_j\}=2\delta_{ij}$. For the $N$-extended case, see
for example~\cite{Gates:1995ch}.

In the following we will assume that the dimension of $\mathcal{M}$ is
even and use the first realisation given. Then the second class
constraints are imposed as conditions on physical states,
\begin{align}
  \Hat{Q}|\mathrm{phys}\rangle&=0
  & \Hat{H}|\mathrm{phys}\rangle&=0
\end{align}
where 
\begin{multline}
  \label{eq:dirac}
     \Hat{Q}=-\frac{1}{\sqrt{2}}(\gamma^i \otimes 1)\mathrm{e}^{\mu}_{\phantom{\mu}i}\frac{\partial}{\partial q^{\mu}}
     - \frac{1}{4\sqrt{2}}\omega_{ijk}(\gamma^{i} \gamma^{jk} \otimes 1)
    - \frac{1}{4\sqrt{2}}\Omega_{iab} (\gamma^i \otimes \gamma^{ab})\\
    + \frac{1}{12\sqrt{2}} c_{ijk} (\gamma^{ijk} \otimes 1)
    - \frac{1}{4\sqrt{2}} m_{iab}(\gamma^i \otimes \gamma^{ab})
    - \frac{1}{4\sqrt{2}}n_{ija}(\gamma^{ij}\gamma^{d+1} \otimes \gamma^a)\\
    - \frac{1}{\sqrt{2}}A_i (\gamma^i \otimes 1)
    + \frac{1}{12\sqrt{2}}l_{abc}(\gamma^{d+1}\otimes\gamma^{abc})
    - \frac{1}{\sqrt{2}}ms_a (\gamma^{d+1}\otimes\gamma^a)
\end{multline}
\begin{multline}
  \label{eq:kleingordon}
    \Hat{H}=\frac{1}{2}\eta_{ij}\Hat{P}^i\Hat{P}^j - \frac{1}{2}\eta_{ab}\Hat{Y}^a\Hat{Y}^b
    +\frac{1}{16}R_{ijkl}(\gamma^{ij}\gamma^{kl}\otimes 1) 
    +\frac{1}{16}G_{ijab}(\gamma^{ij}\otimes\gamma^{ab})\\
     + \frac{1}{24} \mathrm{e}^{\mu}_{\phantom{\mu}[i|} \nabla_{\mu} c_{|jkl]}(\gamma^{ijkl}\otimes 1)
    - \frac{1}{8} \mathrm{e}^{\mu}_{\phantom{\mu}[i|} \nabla_{\mu} m_{|j]ab}(\gamma^{ij}\otimes\gamma^{ab})\\
     + \frac{1}{8} \mathrm{e}^{\mu}_{\phantom{\mu}[i|}\nabla_{\mu} n_{|jk]a} (\gamma^{ijk}\gamma^{d+1}\otimes\gamma^a)
     - \frac{1}{24} \mathrm{e}^{\mu}_{\phantom{\mu}i} \nabla_{\mu} l_{abc}(\gamma^i\gamma^{d+1}\otimes\gamma^{abc})\\
     + \frac{1}{2}\mathrm{e}^{\mu}_{\phantom{\mu}[i|}\nabla_{\mu} A_{|j]}(\gamma^{ij}\otimes 1)
     + \frac{1}{2}m \mathrm{e}^{\mu}_{\phantom{\mu}i} \nabla_{\mu} s_a(\gamma^i\gamma^{d+1}\otimes \gamma^a) \\
     - \omega_{\phantom{mi}m}^{mi}\mathrm{e}^\mu_{\phantom{\mu}i}\frac{\partial}{\partial
     q^\mu} 
    -\frac{1}{8}\omega^{mi}_{\phantom{mi}m}\omega_{ikl}(\gamma^{kl}\otimes 1)\\
    - \frac{1}{2}\Gamma^{\mu}_{\phantom{\mu}\nu\rho}g^{\nu\rho}\frac{\partial}{\partial q^\mu}
    + \frac{1}{24}c^{ijk}c_{ijk} - \frac{1}{24} l^{abc} l_{abc}\\
    -\frac{1}{16}c_{ijk}\omega^{mi}_{\phantom{mi}m}(\gamma^{jk}\otimes 1)
    -\frac{1}{8}n_{ija}\omega^{mi}_{\phantom{mi}m}(\gamma^j\gamma^{d+1}\otimes\gamma^a)
    +\frac{1}{8}m_{iab}\omega^{mi}_{\phantom{mi}m}(1\otimes \gamma^{ab})    
\end{multline}
The new terms which appear here include a Riemann tensor which vanishes
classically because it is contracted with four fermions, and several
terms involving a trace of the connection which arise because of the
way we have chosen to order the equation.

In~(\ref{eq:dirac}) and~(\ref{eq:kleingordon}) we have defined as usual
\begin{equation}
  \begin{split}
      \Hat{P}_i ={} &
    -\mathrm{e}^{\mu}_{\phantom{\mu}i}\frac{\partial}{\partial q^{\mu}} 
    - \frac{1}{4}\omega_{ijk}(\gamma^{jk}\otimes1)
    -\frac{1}{4}\Omega_{iab}(1\otimes\gamma^{ab})\\
    &- \frac{1}{4} c_{ijk} (\gamma^{jk}\otimes 1)
    + \frac{1}{4} m_{iab}(1\otimes\gamma^{ab})
    - \frac{1}{2}n_{ija}(\gamma^j\gamma^{d+1}\otimes\gamma^a)
    - A_i 
  \end{split}
\end{equation}
and
\begin{equation}
    \Hat{Y}_a={} 
    \frac{1}{2}m_{iab}(\gamma^i\gamma^{d+1}\otimes\gamma^b)
  -\frac{1}{4}n_{ija}(\gamma^{ij}\otimes 1)
  +\frac{1}{4}l_{abc}(1\otimes\gamma^{ab}) 
  -ms_a(1\otimes 1)
\end{equation}
It can be checked that 
\begin{align}
  \{\Hat{Q},\Hat{Q}\}=2\Hat{H}
\end{align}
the familiar result that the square of the Dirac operator gives the
Klein Gordon equation.

Finally we note that it is not always the case that the system will
have physical states. This is because manifolds exist for which the
Dirac-like operators do not have zero modes. However it is expected
that most models will have a physical Hilbert space which is non-empty.

To see this, consider the case where the supersymmetry charge is the
Dirac operator $\Dirac$. Then
\begin{align}
  \Dirac^2 = - \nabla^2 + \frac{1}{4} R
\end{align}
where $R$ is the Ricci scalar. A physical state satisfies $\Dirac
\psi=0$. Such a state cannot exist if $R>0$, because the following
partial integration argument.
\begin{align}
  \int (\psi,(-\nabla^2+\frac{1}{4}R\psi) = \int
  [(\nabla\psi,\nabla\psi)+\frac{1}{4}R(\psi,\psi)] >0  
\end{align}
Thus $(-\nabla^2+\tfrac{1}{4}R)\psi\neq 0$ and so $\psi$ cannot solve
the Dirac equation, on the assumption that $M$ is compact and without
boundary.

However solutions exist if $R$ is negative or indefinite.

\chapter{Equivariant Cohomology}

\section{Equivariant cohomology}

Equivariant cohomology arises when the sigma model manifold $M$ admits
a Lie group action with gauge group $G$. We shall see that it
generalises the de Rham complex $\Omega^*(M)$ to the complex
$\Omega^*_\LieG(M)$ which can be thought of as the space of gauge
invariant forms generated by $\Omega^*(M)$, the gauge field $A$ and
field strength $F$, and therefore provides a natural framework for
understanding the conditions needed in the gauging of forms, see for
example \cite{Figueroa-O'Farrill:1994dj,Figueroa-O'Farrill:1994ns}. In
this chapter we will summarise some of the facts about equivariant
cohomology that we shall use later.

\subsection{Definition}

Let $M$ be a manifold which admits an action by a group $G$. The
equivariant cohomology $H_G^*$ of $M$ is defined as the de Rham
cohomology of $M_G = (EG\times M)/G$ where $EG$ is the universal
bundle over the classifying space of the group $G$, ie.
\begin{align}
  H_G^*(M) \equiv H^*(M_G)~.
\end{align}
If $G$ acts freely on $M$, then
\begin{align}
  EG \hookrightarrow M_G \rightarrow M/G
\end{align}
is a fibration. Since $EG$ is contractible, it follows that
\begin{align}
  H_G^*(M) = H^*(M/G) \ ,
\end{align}
ie. the equivariant cohomology of $M$ is identified with the standard
cohomology of $M/G$.

In the other extreme, $M$ is a point $M=\{ p \}$, and we have $M_G=BG$
where $BG$ is the classifying space $G \hookrightarrow EG
\rightarrow BG$. Thus 
\begin{align}
  H_G^*(M) = H^*(BG) \ .
\end{align}

In general there are many ways to compute the equivariant cohomology.
For example, one such method is the use of spectral sequences for the
fibrations
\begin{align}
  M \hookrightarrow M_G \rightarrow BG ~.
\end{align}
However for many applications a de Rham version of the equivariant
cohomology is needed. Such a method was proposed in \cite{Atiyah:1984}
using a model for the de Rham equivariant cohomology based on Weil's
model of the cohomology of the universal classifying space
\begin{align}
  G \hookrightarrow EG \rightarrow BG
\end{align}
First we will give a mathematical definition and then (more usefully)
we will give a physicist's way of thinking about it.

\section{The Weil Model}
First we need to define the Weil algebra $W(\LieG)$, which is used to
capture the algebraic properties satisfied by the gauge field $A$ and
field strength $F$. Let $\LieG$ be the Lie algebra of $G$. We define
\begin{align}
  W(\LieG) = \Lambda(\LieG^*) \otimes S(\LieG^*)
\end{align}
where $\Lambda$ is the exterior algebra of the dual $\LieG^*$ of
$\LieG$ and $S$ is the symmetric algebra. We attach degree 1 to the
element $\alpha\in\LieG^*$ in the exterior algebra and degree 2 the
corresponding element in the symmetric algebra.  

Note that $\alpha$ and $\theta$ can be thought of as generators
obeying the same algebraic relations as the gauge field $A$ and field
strength $F$ respectively.  $W(\LieG)$ is a graded commutative algebra
freely generated by the basis
$\{\alpha^a,\theta^a:a=1\ldots\dim\LieG\}$.

Next we define the following differential operator
\begin{align}
  \begin{split}
    d\alpha^a &= \theta^a + \frac{1}{2} \ud{f}a{bc}\alpha^b\alpha^c  \\
    d\theta^a &= - \ud{f}a{bc}\alpha^b\theta^c
  \end{split}
\end{align}
where $\ud{f}a{bc}$ are the structure constants of $\LieG$. Because of
the Jacobi identities, we can show that $d^2=0$ and that $d$ defines a
cohomology on the complex $W(\LieG)$.  It can be shown that
\begin{align}
  H_d^*(W(\LieG))= \bR
\end{align}
The general proof is given in \cite{Atiyah:1984}; in fact
$H^*_d(W(\LieG))$ should be thought of as a model for the cohomology
of $EG$. This is why it is trivial, but here we shall show this in
some special cases only. For example, take $G=T^\ell$. Then
$\LieG=\bR^\ell$, and since $T^\ell$ is abelian,
\begin{align}
  d\alpha^a - \theta^a &= 0 \\
  d\theta^a &= 0
\end{align}
A degree zero element is clearly any constant $c\in\bR$. Moreover,
$dc=0$. Thus $H^0(W(\bR^\ell))=\bR$. A general element of degree 1 is
$v=v_a\alpha^a$. $d v=0$ implies $v_a\theta^a=0$ and therefore
$v_a=0$. This implies that $v=0$, so we have $H^1(W(\bR^\ell))=0$. 

A general element of degree two is
\begin{align}
  w = \frac{1}{2}w_{ab}\alpha^a\alpha^b + u_a\theta^a
\end{align}
$d w=0$ implies that $w_{ab}d\alpha^a\alpha^b + u_a
d\theta^a=w_{ab}\theta^a\alpha^b=0$ and this implies that $w_{ab}=0$.
Therefore the closed elements are
\begin{align}
   w=u_a\theta^a \ .
 \end{align}
However these elements are exact because $w=d u$ when
$u =u_a\alpha^a$.  Therefore $H^2(W(\bR^\ell))=0$, again. In
general $H^n(W(\bR^\ell))=0$ for all $n>0$.

\subsection{The cohomology of $BG$}

It is taken that $H_D^*(W(\LieG))$ models the cohomology of $EG$. Now
a model should be constructed for the cohomology of $BG$. We should
identify the elements of $W(\LieG)$ which are associated with the
cohomology $BG$. Thinking about the fibration $G\hookrightarrow
EG\rightarrow BG$, the elements of $EG$ which are forms associated with
those that are pullbacks of $BG$ should have two properties: (i) they
should vanish when evaluated along the fibre directions and (ii) they
should be invariant under the action of $G$. To model these two
properties we define the inner derivation,
\begin{align}
\label{eq:weilcontractions}
  \iota_a\alpha^b &= \du\delta{a}{b}
  & \iota_a\theta^b &= 0 ~.
\end{align}
We can also define a Lie derivative in the usual way,
\begin{align}
  \mathcal{L}_a= \iota_a d + d\iota_a \ .
\end{align}
Let $B\LieG$ be all the elements of $W(\LieG)$ which have the
property that
\begin{align}
\label{eq:basicforms}
\iota_a\phi &= 0 & \mathcal{L}_a\phi &= 0 \ ,
\end{align}
such elements are called \emph{basic}.

To interpret this condition, we note that a gauge transformation can
be defined on ordinary forms $w\in\Omega^*(M)$ by
\begin{align}
  \delta_\lambda w = \rd\lambda^a\iota_a w + \lambda^a\mathcal{L}_a w ~.
\end{align}
With the definitions \eqref{eq:weilcontractions}, this formula extends to
$\alpha$ and $\theta$ as
\begin{align}
  \begin{split}
  \delta_\lambda \alpha^a &= \rd\lambda^a - \ud fa{bc} \lambda^b \alpha^c \\
  \delta_\lambda \theta^a &= -\ud fa{bc} \lambda^b \theta^c
  \end{split}
\end{align}
which are the familiar transformation laws for the gauge potential and
field strength. Therefore \eqref{eq:basicforms} is equivalent to the
condition for gauge invariance.

It is easy to see that in fact $B\LieG$ can be identified with the
polynomials in $S(\LieG^*)$ which are invariant under the co-adjoint
action of $\LieG$ on $\LieG^*$, ie.
\begin{align}
  B\LieG \cong \mathrm{Inv}_\LieG S(\LieG^*) \ .
\end{align}
Moreover, in \cite{Atiyah:1984} it was shown that
\begin{align}
\label{eq:atiyahresult}
  B\LieG \cong H^*(BG) \ .
\end{align}
To illustrate these, consider the special case where $G=T^\ell$ as
before. The co-adjoint action of $\LieG=\bR^\ell$ on $\bR^{*l}$ is
trivial because $T^\ell$ is abelian. Thus any $\phi\in W(\bR^\ell)$
satisfies the condition $\mathcal{L}_a\phi=0$.

It remains to find the restrictions imposed by the other condition
$\iota_a\phi=0$. Any $\phi$ can be written as
\begin{align}
  \begin{split}
  \phi={}&\sum_{k\geq 1} v_{a_1\ldots a_k b_1\ldots
  b_l}\alpha^{a_1}\cdots \alpha^{a_k}\theta^{b_1}\cdots \theta^{b_l} \\
&+ w_{b_1\ldots b_m}\theta^{b_1}\cdots \theta^{b_m}  
  \end{split}
\end{align}
Now
\begin{align}
  \iota_a\phi &= \sum_{k\geq 1} kv_{a a_1\ldots a_{k-1} b_1\ldots
  b_l}\alpha^{a_1}\cdots \alpha^{a_{k-1}}\theta^{b_1}\cdots \theta^{b_l}
\end{align}
and so $\iota_a\phi=0$ implies that $v_{a_1\ldots a_k b_1\ldots
  b_l}=0$.  Therefore $\iota_a\phi=0$ implies that $\phi\in
S(\bR^{*\ell})$.

The result \eqref{eq:atiyahresult} implies that the cohomology of
$BT^\ell$ is generated by $\theta^1,\ldots,\theta^l$ as a polynomial, ie.
\begin{align}
  H^0(BT^\ell) &&& 1 \nonumber\\
  H^1(BT^\ell) &&& 0 \nonumber\\
  H^2(BT^\ell) &&& \theta^1,\theta^2,\ldots,\theta^\ell \\
  H^3(BT^\ell) &&& 0 \nonumber\\
  H^4(BT^\ell) &&& (\theta^1)^2,(\theta^2)^2,\ldots,(\theta^\ell)^2, \nonumber\\ 
  &&& \theta^1\theta^2,\theta^1\theta^3,\ldots\nonumber
\end{align}
This can be confirmed by an independent calculation using, say, the
spectral sequence of the fibration $T^\ell\hookrightarrow
ET^\ell\rightarrow BT^\ell$.

This concludes the discussion of the Weil model for the cohomology of
$BG$.

\section{Weil model for equivariant cohomology}

A model for the equivariant cohomology of $M$ can be defined as
follows.

First we consider the complex 
\begin{align}
  \Omega^*(M) \otimes W(\LieG) \ .
\end{align}
For example, a degree one element of $\Omega^*(M)\otimes W(\LieG)$ is
\begin{align}
  \phi = r_i \rd x^i + r_a \alpha^a
\end{align}
where $r_i$ and $r_a$ are functions on $M$. A degree two element is
\begin{align}
\label{eq:deg2element}
  \phi &= \frac{1}{2}v_{ij}\rd x^i \rd x^j + v_{ai}\alpha^a \rd x^i +
  \frac{1}{2} w_{ab}\alpha^a\alpha^b + u_a \theta^a
\end{align}
and so on.

One can define the inner derivation with respect to an element of
$\LieG$ by contracting $\phi\in\Omega_\LieG^*(M)$ with the associated
vector field on $M$ and using the
contractions~\eqref{eq:weilcontractions} defined in the Weil model.
For example, for the degree two element above, we have
\begin{align}
  \iota_b \phi = \du\xi{b}{i}v_{ij}\rd x^j + v_{bi}\rd x^i +
  v_{ai}\alpha^a\du\xi{b}{i} + w_{ba}\alpha^a
\end{align}
We define an exterior derivative $\rd$ by a combinations of the
exterior derivative $\rd$ of $\Omega^*(M)$ and that defined in the
previous section on $W(\LieG)$. Denote this exterior derivative by
$\rd$. We can also define a Lie derivative as
\begin{align}
  \mathcal{L}_a = \iota_a \rd + \rd \iota_a \ .
\end{align}
Now the forms of $M_G$ in $\Omega^*(M)\otimes W(\LieG)$ are those for
which
\begin{align}
\iota_a\phi&=0 & \mathcal{L}_a\phi &= 0 \ ,
\end{align}
ie. the \emph{basic} (gauge invariant) forms, and we write
$\Omega_\LieG^*(M)$ for the space that they generate.

Let us illustrate what this means for the case $G=T^\ell$ and degree
2. Since we have done the calculation of $\iota_b\phi$ above, we set
$\iota_b\phi=0$ to find
\begin{align}
  \du\xi{b}{i} v_{ij} + v_{bj} &= 0 \\
 v_{ai}\du\xi{b}{i} + w_{ba} &= 0 \ .
\end{align}
Substituting into \eqref{eq:deg2element}, 
\begin{align}
\label{eq:basicform}
  \phi &= \frac{1}{2} v_{ij}\rd x^i \rd x^j -
  \du\xi{a}{i}v_{ij}\alpha^a\rd x^j + \frac{1}{2}
  v_{ij}\du\xi{a}{i}\du\xi{b}{j}\alpha^a\alpha^b + u_a \theta^a \\
\label{eq:basic2form}%
&= \frac{1}{2}v_{ij}(\rd x^i - \du\xi{a}{i}\alpha^a)(\rd x^j -
  \du\xi{b}{j}\alpha^b) + u_a\theta^a \ .
\end{align}
The condition $\mathcal{L}_a\phi=0$ implies that 
\begin{align}
  \mathcal{L}_a\bigl(\frac{1}{2}v_{ij}\rd x^i\rd x^j\bigr) = \mathcal{L}_a
  (v_{bi}\rd x^i) = \mathcal{L}_a w_{cd} = \mathcal{L}_a u_b = 0
\end{align}
where $\mathcal{L}_a$ is the usual Lie derivative on forms in
$\Omega^*(M)$. Thus a basic degree-two from is given by
\eqref{eq:basic2form} provided that all the components are invariant
under the action of $T^\ell$ on $M$.

Similarly a basic degree-one form is given by
\begin{align}
\label{eq:basic1form}
  \omega = r_i(\rd x^i - \du \xi{a}{i}\alpha^a)
\end{align}
where the one-form $r_i\rd x^i$ satifies $\mathcal{L}_a(r_i\rd x^i)=0$.

The cohomology of $M_G$ is given by $H^*_\rd(\Omega_\LieG(M))$. It can
be shown that under certain conditions, this is the equivariant
cohomology,
\begin{align}
  H_\rd^*(\Omega^*_\LieG(M)) = H^*(M_G)\ .
\end{align}
To understand some of the elements that represent
$H_\rd^*(\Omega^*_\LieG(M))$, let us find the closed form in
$\Omega^*_\LieG(M)$ of degree two for the case $G=T^\ell$.
Computing the exterior derivative of $\phi$ in \eqref{eq:basicform} we
get
\begin{align}
\rd\phi=0  \quad\Rightarrow\quad&  \rd(v_{ij}\rd x^i\rd x^j)=0 \nonumber\\
 \text{and}\quad& \du\xi{a}{i}v_{ij} + \partial_ju_a = 0
\end{align}
This form is exact if $\phi=\rd \omega$ where $\omega$ is a basic form
of degree one, so from \eqref{eq:basic1form} we obtain
\begin{align}
  \rd \omega = \partial_{[i} r_{j]}(\rd x^i - \du\xi{a}{i}\alpha^a)(\rd
  x^j - \du\xi{b}{j}\alpha^b) - r_i \du\xi{a}{i} \theta^a
\end{align}
Thus $\phi$ is exact \emph{if and only if} the two-form
$\frac{1}{2}v_{ij}\rd x^i \rd x^j$ is exact and $u_a=-r_i\du\xi{a}{i}$.

\subsection{Equivariant extensions}
\label{sec:equivarext}
Given a closed form $v$ in $M$ which is invariant under the
$G$-action, we say that $v$ has an \emph{equivariant extension} if
there exists a closed basic form $\phi \in \Omega_\LieG^*(M)$ satisfying
\begin{align}
  v = \phi|_{\alpha=0,\theta=0}\ .
\end{align}
Note that $\iota_a v=0$ is not a necessary condition for the existence
of an equivariant extension as we have seen.

\section{A physicist's approach to equivariant cohomology}
\label{sec:physicsequivar}
As we shall see later, we need to construct the equivariant extensions
of closed forms on a manifold $M$. Suppose that $M$ has a $G$-action,
with associated vector fields $\du\xi{a}i\partial_i$.

Let $r$ be a closed form and write $r=\tfrac{1}{k!}r_{i_1\ldots
  i_k}\rd x^{i_1}\wedge\ldots\wedge\rd x^{i_k}$. We introduce a
connection $A$ and gauge the form by performing a `minimal'
substitution $\rd x^i\rightarrow \rd x^i - \du\xi{a}{i}A^a$.

Thus we construct 
\begin{align}
  \Tilde{r} = \frac{1}{k!} r_{i_1\ldots i_k}(\rd
  x^{i_1}+\du\xi{a}{i_1}A^a)\wedge\ldots\wedge(\rd
  x^{i_k}+\du\xi{b}{i_k}A^b)\ .
\end{align}
This form is invariant under gauge transformations, but we have lost
the property that $\Tilde{r}$ is closed. To `repair' this, we add an
extra piece as follows,
\begin{align}
\label{eq:repairgauge}
  \begin{split}
  \Tilde{r} ={}& \frac{1}{k!} r_{i_1\ldots i_k}\nabla x^{i_1}\wedge\ldots\wedge\nabla
  x^{i_k} \\
  &+ \frac{1}{(k-2)!} r_{a i_1\ldots i_{k-2}} F^a\wedge\nabla
  x^{i_1}\wedge\ldots\wedge\nabla x^{i_{k-2}}\\
  &+ \frac{1}{(k-4)!} r_{a_1a_2 i_1\ldots i_{k-4}} F^{a_1}\wedge
  F^{a_2}\wedge\nabla
  x^{i_1}\wedge\ldots\wedge\nabla x^{i_{k-4}}\\
  &+ \ldots   
  \end{split}
\end{align}
where $\nabla x^i=\rd x^i - \du\xi{a}{i}A^a$.

Gauge invariance of this form implies that
\begin{align}
  \begin{split}
  \mathcal{L}_b r_{a_1\ldots a_l i_1\ldots i_k} ={}&
  \ud{f}c{ba_1}r_{ca_2\ldots a_l i_1\ldots i_k} \\
  & \text{$+$ cyclic in $a_1a_2\ldots a_l$.}  
  \end{split}
\end{align}
Since $r$ is gauge invariant, we can write $\rd r=\nabla r$ and
use $\nabla^2 x^i = -F^a \du\xi{a}{i}$.

Now, if $r$ is closed then $\nabla r=0$ which implies that 
\begin{align}
\label{eq:equivarclosed}
  \begin{split}
  \du\xi{a}{i} r_{i j_1\ldots j_{k-1}} + (k-1)
   \nabla_{[j_1}r_{|a|j_2\ldots j_{k-1}]} &= 0 \\
  \du\xi{b}{i} r_{a ij_1\ldots j_{k-3}} + (k-3)
   \nabla_{[j_1}r_{|ab|j_2\ldots j_{k-3}]} &= 0 \\
&   \hdots  
  \end{split}
\end{align}
and so on. These are the conditions for an \emph{equivariant
  extension}.

We shall apply these results to sigma models.

\chapter{New two dimensional gauge theories}

\section{Two-dimensional gauged sigma models with Wess-Zumino term}

\subsection{Geometric Data and Action}
\label{sub:data}

To describe two-dimensional supersymmetric gauge theories coupled to
sigma model matter with a Wess-Zumino term, it is instructive to begin
with a non-supersymmetric system as a toy example.  Let $\Xi$ be the
two-dimensional Minkowski spacetime with light-cone coordinates
$(x^\plpl,x^\mimi)$.  The fields of the model that we shall consider
here are the following: a gauge potential $A$ with gauge group $G$,
sigma model matter fields $\phi$ which are locally maps from $\Xi$
into a sigma model manifold or target space $M$ and real fermions
$\psi_-$ and $\lambda_+$ on $\Xi$ of opposite chirality.

The couplings of two-dimensional gauged sigma model are described by a
Riemannian metric $g$ on $M$ and the Wess-Zumino term which is a
locally-defined two-form $b$ on $M$; $H=\rd b$ is a globally defined
closed three-form on $M$. In addition, the gauge group $G$ acts on $M$
leaving invariant both the metric and the Wess-Zumino term, ie
\begin{align}
\label{eq:ghinvariant}
  \mathcal{L}_a g &=0
  & \mathcal{L}_a H&=0
\end{align}
where $\mathcal{L}_a$ is the Lie-derivative with respect to the vector
fields $\{\xi_a: a=1,\dots, \dim \LieG\}$ generated by the action of
the gauge group $G$ on $M$ and $\LieG$ is the Lie algebra of $G$.
Therefore, we have
\begin{align}
  [ \xi_a, \xi_b ]^i = f_{ab}^{\phantom{ab}c}\xi_c^{\phantom{c}i}\ ,
\end{align}
where $f$ are the structure constants of $G$; $a,b,c=1,\dots, \dim
\LieG$ are gauge indices.  The first condition in
\eqref{eq:ghinvariant} implies that $\xi_a$ are Killing vectors, ie
\begin{align}
\label{eq:killingcnd}
  \nabla_i \xi_{aj}+\nabla_j\xi_{ai}=0
\end{align}
where $\nabla$ is the Levi-Civita connection of $g$ and $i,j=1,\dots,
\dim M$.  The second condition in \eqref{eq:ghinvariant} together with
$\rd H=0$ imply that $i_a H$ is closed and so
\begin{align}
\label{eq:u1form}
  \xi_a^{\phantom{a}i}H_{ijk} &= 2\partial_{\smash{[j}}w_{k]a}
\end{align}
for some locally defined one-form $w_a$.

A sigma model with rigid transformation symmetries can be thought of
as a trivial principal $G$-bundle $P=\Xi\times G$.  The sigma model
fields $\phi$ are sections of an associated bundle $P\times_G
M\rightarrow\Xi$, and since $P$ is trivial, these are precisely the
maps $\Xi\rightarrow M$.

Promoting the gauge transformation parameters $\epsilon^a$ to be
\emph{local}, ie. functions $\epsilon^a=\epsilon^a(x)$ defined on
$\Xi$, then $P$ becomes an arbitary $G$-bundle, $P=P(\Xi,G)$. The
sigma model fields, which are sections of an associated bundle, can be
represented locally as maps $\phi:\Xi\rightarrow M$. The gauge
potential $A$ can be represented locally as the pull back of the
connection one-form of $P$ onto an open neighbourhood of $\Xi$.

To describe the couplings of the fermions, we consider two vector
bundles $E$ and $F$ over $M$ equipped with connections $B$ and $C$ and
with fibre metrics $h$ and $k$, respectively. The fermions $\psi_-$
and $\lambda_+$ can be thought of as sections of $S_-\otimes E$ and
$S_+\otimes F$, respectively, where $S_-$ and $S_+$ are spin bundles
over $\Xi$ associated to the two inequivalent real representation of
$Spin(1,1)$. Note that in two dimensions there are Majorana-Weyl
fermions and so two inequivalent one-dimensional real spinor
representations of $Spin(1,1)$.  In addition we shall assume that the
connections $B$ and $C$ as well as the fibre metrics $h$ and $k$ are
invariant under the action of the gauge group on $M$. These conditions
imply that
\begin{align}
\label{eq:bgaugecnd}
 \mathcal{L}_a B_{i\phantom{A}B}^{\phantom{i}A}&=
 -\nabla_i U_{a\phantom{A}B}^{\phantom{a}A}\\
\label{eq:hgaugecnd}
  \mathcal{L}_c h_{AB}&=-U_a{}^C{}_A h_{CB}-U_a{}^C{}_B h_{AC}
  \ ,
\end{align}
where
\begin{align}
  \nabla_i U_{a\phantom{A}B}^{\phantom{a}A} &
  =\partial_i U_{a\phantom{A}B}^{\phantom{a}A}
   +  B_i{}^A{}_C U_a{}^C{}_B
- U_a{}^A{}_C B_i{}^C{}_B \ ,
\end{align}
and $U_a$ are infinitesimal gauge transformations, $A,B,C=1,\dots, {\rm rank}\,E$,
and similarly for the connection $C$ and fibre metric $k$. The
above conditions on the connection $B$ have appeared in \cite{Hull:1994hm}.
An action for the  fields $A$, $\phi$, $\psi_-$ and $\lambda_+$ is
\begin{align}
\label{eq:gauge00action}
\begin{split}
  S ={}& \intd{^2x} \bigl(u_{ab} F^a_{\plpl\mimi} F^b_{\plpl\mimi}+
       g_{ij} \nabla_\plpl \phi^i \nabla_\mimi \phi^j
       - V(\phi)
       \bigr)\\
    & + \intd{^2x\rd t}\bigl(
        H_{ijk}\partial_t \phi^i \nabla_\plpl \phi^j
    \nabla_\mimi\phi^k - w_{ia}\partial_t\phi^i F_{\plpl\mimi}^a
    \bigr)\\
    & +\intd{^2x}(ih_{AB}\psi_-^A\Nabla_\plpl\psi_-^B-ik_{A'B'}
     \lambda_+^{A'}\Nabla_\mimi \lambda_+^{B'})
\end{split}
\end{align}
where $u_{ab}=u_{ab}(\phi)$ are the gauge couplings which in general
depend on $\phi$, $V$ is a scalar potential
and
\begin{align}
  F_{\plpl\mimi}=[\nabla_\plpl,\nabla_\mimi] ~,
\end{align}
where we have suppressed gauged indices.
The covariant derivatives in the action above are defined as follows,
 \begin{align}
  \nabla_\mu \phi^i=\partial_\mu \phi^i+ A^a_\mu \xi_a^i
\end{align}
and
\begin{align}
 \Nabla_\mu\psi_-^A &= \partial_\mu\psi_-^A
  + \nabla_\mu\phi^iB_{i\phantom{A}B}^{\phantom{i}A}\psi_-^B
  + A_\mu^a U_{a\phantom{A}B}^{\phantom{a}A}\psi^B  \ ,
\end{align}
$\mu=\plpl, \mimi$, and similarly for $\Nabla \lambda_+^{A'}$.
The latter can be rewritten as
\begin{align}\label{eq:covdev}
 \Nabla_\mu\psi_-^A &= \partial_\mu\psi_-^A
  + \partial_\mu\phi^iB_{i\phantom{A}B}^{\phantom{i}A}\psi_-^B
  + A_\mu^a \mu_{a\phantom{A}B}^{\phantom{a}A}\psi^B  ~.
\end{align}
where
\begin{align}
\mu_a{}^A{}_B=U_a{}^A{}_B+\xi_a{}^i B_i{}^A{}_B ~.
\end{align}

Observe that the part of the action involving the Wess-Zumino term
has been written as an integral over a three-dimensional space.
The conditions for this term to be written in a two-dimensional form
as well as the conditions for the gauge invariance of the action
will be investigated in the next section, see also \cite{Hull:1989jk}.

\subsection{Conditions for gauge invariance}
\label{sub:gaugecon}

The gauge transformations of the fields are
\begin{align}\begin{split}
  \delta_\epsilon A_\mu^a &=- \nabla_\mu \epsilon^a \\
  \delta_\epsilon \phi^i &= \epsilon^a \xi_a^i \\
  \delta_\epsilon \psi_-^A &= \epsilon^a U_{a\phantom{A}B}^{\phantom{a}A} \psi^B_- \\
  \delta_\epsilon \lambda_+^{A'}&=\epsilon^a V_{a\phantom{A'}B'}^{\phantom{a}A'}
   \lambda_+^{B'}
\end{split}\end{align}
where $\epsilon$ is the parameter of infinitesimal gauge
transformations.  Some of the conditions required for the
invariance of the action~\eqref{eq:gauge00action} have been
incorporated as part of the geometric data of the sigma model
in the previous section. In particular, in addition to the conditions
\eqref{eq:ghinvariant} and \eqref{eq:bgaugecnd}, we require that (i) $w_a$
is a globally defined one-form on $M$ which (ii) satisfies
\begin{align}
{\cal L}_a w_b=-f_{ab}{}^c w_c\ .
\label{giw}
\end{align}

To write the Wess-Zumino part of the action in a two-dimensional form, it
is necessary for the relevant three form to be closed. This in addition
requires that
\begin{align}
\xi^i_a w_{ib}+\xi^i_b w_{ia}=0\ .
\label{weq}
\end{align}
Then  the three-dimensional part of the action \eqref{eq:gauge00action}
can be written locally \cite{Hull:1989jk} as
\begin{align}
  \begin{split}
    S' = \intd{^2x} \bigl(&
     b_{ij} \partial_\plpl\phi^i \partial_\mimi\phi^j
   - A_\plpl^a w_{ia} \nabla_\mimi\phi^i\\
&    + A_\mimi^a w_{ia} \nabla_\plpl\phi^i
    + A_\plpl^a A_\mimi^b \xi_b^{\phantom{b}i}w_{ai}
    \bigr)\ .
  \end{split}
\end{align}
The conditions above that require (i) $H$ to be invariant under the
group action, (ii) $w_a$ to be globally defined on $M$, (iii)
\eqref{giw} and (iv) \eqref{weq} are those for the closed form $H$ to
have an extension, i.e.\  equivariant extension \cite{Atiyah:1984}, as a closed
form in $M\times_G EG$ \cite{Figueroa-O'Farrill:1994ns,Figueroa-O'Farrill:1994dj}.

Next let us consider the conditions for gauge invariance
of part of the action~\eqref{eq:gauge00action} involving the fermions.
We find that this requires that
 \begin{align}
  \mathcal{L}_c u_{ab} + u_{db}f^d_{\phantom{d}ca}+u_{ad}f^d_{\phantom{d}cb}=0
\end{align}
for the gauge couplings $u_{ab}$,
\begin{align}
\label{eq:psigaugecnd}
 \mathcal{L}_a\mu_b - [U_a,\mu_b] =
  -f_{ab}^{\phantom{ab}c}\mu_c
\end{align}
 and
\begin{align}
  \mathcal{L}_a V&=0 ~.
\end{align}
We have not assumed that $\nabla_i h = 0$. However given a connection
on a vector bundle with a fibre metric $h$, there always exist another
connection $\nabla'$ such that $\nabla_i'h=0$. Suppose that the
$\nabla'$ is used for the fermionic couplings. If this is the case,
the gauge group is $O(N)$ and therefore the right-hand-side
of~\eqref{eq:hgaugecnd} vanishes.

Observe that the equation~\eqref{eq:bgaugecnd} can be written in a
more covariant form as
\begin{align}
\xi_a^j G_{ij\phantom{A}B}^{\phantom{ij}A}
= \nabla_i \mu_{a\phantom{A}B}^{\phantom{a}A} ~.
\end{align}

In what follows we shall assume that the conditions stated in this
section by requiring gauge invariance of the non-supersymmetric model
described by the action \eqref{eq:gauge00action} hold. We shall see
that for supersymmetric sigma models more conditions are necessary.

\section{$(1,0)$ supersymmetric gauge theory}

The (1,0)-supersymmetric gauged sigma model involves the coupling of
three different (1,0)-multiplets. To simplify the construction of this
model we shall describe each multiplet and the conditions for
supersymmetry and gauge invariance separately. There are different
ways of approaching this problem. Here we shall use `standard'
(1,0)-superfields. The action will be constructed using
(1,0)-superspace methods.

\subsection{The gauge multiplet}

The (1,0)-superspace $\Xi^{1,0}$ has coordinates
$(x^\plpl,x^\mimi,\theta^+)$, where $(x^\plpl,x^\mimi)$ are bosonic
light-cone coordinates and $\theta^+$ is a Grassmann odd-coordinate.
The (1,0)-supersymmetric Yang-Mills multiplet with gauge group $G$ is
described by a connection $A$ in superspace which has components
$(A_\plpl, A_\mimi, A_+)$.  In addition, it is required that these
satisfy the supersymmetry constraints \cite{Hull:1991uw}
\begin{gather}
  [ \nabla_+, \nabla_+ ] = 2i\nabla_\plpl
\mspace{100mu} [ \nabla_\plpl, \nabla_\mimi ] = F_{\plpl\mimi}     
\nonumber\\
[ \nabla_+, \nabla_\mimi ] = W_- \ ,
\end{gather}
where $F_{\plpl\mimi}$, $W_-$ are the components of the curvature of the
superspace connection $A$. (The gauge indices have been suppressed.)
Jacobi identities imply that
\begin{align}
  \nabla_+ W_- &= iF_{\plpl\mimi}
\end{align}
Therefore the independent components of the gauge multiplet are
\begin{align}
  \chi_-^a &= W_-|_{\theta^+=0} & F^a_{\plpl\mimi} &=
  -i\nabla_+W^a_-|_{\theta^+=0}
\end{align}
where $\chi_-^a$ is the gaugino and $F^a$ is the two-form gauge field
strength.

\subsection{Sigma model multiplets}

To describe the sigma model  multiplet that couples to the
above gauge field, we introduce a Riemannian manifold $M$ with metric
$g$ and a locally defined two form $b$. In addition we assume that $M$
admits a vector bundle $E$ with fibre metric $h$, connection $B$ and a section $s$.
The data required for the description of the sigma model multiplet are the same
as those given in section \ref{sub:data}. In addition
we take the section $s$ to satisfy
 \begin{align}
 \label{eq:sgaugecnd}
  \mathcal{L}_a s_A = -U^{\phantom{a}B}_{a\phantom{B}A} s_B\ .
  \end{align}

The $(1,0)$-supersymmetric sigma model multiplet is described by a
real scalar superfield $\phi$ and a fermionic {\it superfield} $\psi_-$. The
superfield $\phi$ is a map from the superspace $\Xi^{1,0}$ into a
sigma model manifold $M$ and the fermionic superfield $\psi_-$ is a
section of the bundle $\phi^*E\otimes S_-$; $S_-$ is a spin bundle
over $\Xi^{1,0}$.

The components of the superfields $\phi$, $\psi_-$ are
\begin{align}
  \phi^i &= \phi^i|_{\theta^+=0}
  & \lambda^i_+ &= \nabla_+\phi^i|_{\theta^+=0}  \nonumber\\
  \psi_-^A &= \psi_-^A|_{\theta^+=0} & \ell^A &=
  \Nabla_+\psi_-^A|_{\theta^+=0} \ ,
\end{align}
where the covariant derivatives are defined as
\begin{align}
  \nabla_+\phi^i &= D_+\phi^i+A_+^a \xi_a^i \nonumber\\
  \Nabla_+\psi_-^A &= D_+\psi_-^A + \nabla_+\phi^i
  B_{i\phantom{A}B}^{\phantom{i}A} \psi_-^B + A_+^a
  U_{a\phantom{A}B}^{\phantom{a}A} \psi_-^B ~,
\end{align}
where $D_+$ is the usual flat superspace derivative, $D_+^2=i\partial_\plpl$.

\subsection{Supersymmetric action}

It is straightforward to couple the gauge multiplet to
$(1,0)$-supersymmetric sigma model matter. The full action is
\begin{align}
\label{eq:01ta}
  S &= S_g +S_\sigma+ S_f + S_p
\end{align}
where
\begin{align}
  S_g &=- \intd{^2x\rd\theta^+}\bigl( u_{ab}W_-^a\nabla_+W_-^b -i z_a W^a_-\bigr) \ ,
\end{align}
where $u_{ab}=u_{ab}(\phi)$ and $u_{ab}$ is not necessarily symmetric in
the gauge indices and $z_a$ is a $\theta$-type of term which may depend
on the scalar field $\phi$.

The gauge covariant supersymmetric action for the fields $\phi$
is~\cite{Hull:1991uw}
\begin{align}
\begin{split}
  S_\sigma ={}&-i \intd{^2x\rd\theta^+}
  g_{ij}\nabla_+\phi^i\nabla_\mimi\phi^j \\
  &-i \intd{^2x\rd t \rd\theta^+}\bigl(
  H_{ijk}\partial_t\phi^i\nabla_+\phi^j\nabla_{\mimi}\phi^k -
  w_{ia}\partial_t\phi^i W_-^a \bigr)
\end{split}
\end{align}
which as in section \ref{sub:data} can also be written as an integral
over $\Xi^{1,0}$ superspace provided
that $H$ admits an equivariant extension. In particular we have
\begin{align}
\label{eq:gauge10action}
\begin{split}
  S_\sigma = -i\intd{^2x\rd\theta^+}\bigl(&
    g_{ij}\nabla_+\phi^i\nabla_\mimi\phi^j
    + b_{ij} D_+\phi^i \partial_{\mimi}\phi^j \\
&   - A_+^a w_{ia} \partial_\mimi\phi^i
    +A_\mimi^a w_{ia} D_+\phi^i
    + A_+^a A_\mimi^b \xi^i_{[b} w_{a]i}
    \bigr)~.
\end{split}
\end{align}

The action of the gauged fermionic multiplet  is~\cite{Hull:1994hm}
\begin{align}
  S_f &= \intd{^2x\rd\theta^+}h_{AB}\psi_-^A\Nabla_+\psi_-^B~.
\end{align}
The definition of the covariant derivative $\Nabla_+$ is similar to the
one given in section \eqref{sub:data} for the covariant derivative
$\Nabla_\plpl$.

The action for the potential term is
\begin{align}
  S_p &= \intd{^2x\rd\theta^+} m s_A \psi_-^A
\end{align}
which is similar to that of the ungauged model in~\cite{Hull:1993ct}.

The superfields transform under the gauge group $G$ as
\begin{align}
\delta A_\mu&=-\nabla_\mu\epsilon^a
  & \delta\phi^i &= \epsilon^a\xi_a^{\phantom{a}i}(\phi)
& \delta\psi_-^A &= \epsilon^a U_{a\phantom{A}B}^{\phantom{a}A} \psi_-^B ~.
\end{align}
where $\mu=\plpl, \mimi, +$ is a $\Xi^{1,0}$ superspace index and $\epsilon^a$ is an infinitesimal
gauge transformation parameter.
 Gauge invariance of the action \eqref{eq:01ta} requires, in addition
to the conditions given in section \eqref{sub:gaugecon}, the condition
\eqref{eq:sgaugecnd} and
\begin{align}
{\cal L}_a z_b=-f_{ab}{}^c z_c~.
\end{align}

\subsection{The action of (1,0)-model in components and scalar potential}

The action of (1,0)-supersymmetric two-dimensional gauged sigma model
described by the action \eqref{eq:01ta} can be easily written in
components by performing the $\theta^+$ integration and using the
definition of the various component fields of the (1,0)-multiplets
which we have described in the previous sections.  In particular we
find for the part of the action involving the kinetic term of the
gauge multiplet that
\begin{align}
\begin{split}
  S_{g}=\intd{^2x}\bigl(&
  u_{ab}F_{\plpl\mimi}^aF_{\plpl\mimi}^b+z_a F_{\plpl\mimi}^a \\
  &-i
  u_{ab}\chi_-^a\nabla_{\plpl}\chi_-^b
  +i\partial_iu_{ab} \lambda^i_+ \chi_-^a F_{\plpl\mimi}^b +i\partial_i z_a \lambda_+^i \chi_-^a
   \bigr)\ .   
\end{split}
\end{align}
Next we find that
\begin{align}
  \begin{split}
    S_\sigma = \intd{^2x} \bigl(&
    g_{ij} \nabla_{\plpl}\phi^i \nabla_\mimi\phi^j
    + b_{ij} \partial_\plpl\phi^i \partial_\mimi\phi^j
      +i \lambda_+^i \Nabla^{(+)}_\mimi \lambda_+^j
    \\
&   +i w_{ia} \lambda_+^i \chi_-^a
    -  A_\plpl^a w_{ia} \partial_\mimi\phi^i
    +  A_\mimi^a w_{ia} \partial_\plpl\phi^i \\
&   -  A_\plpl^a A_\mimi^b \xi_{[b}^{\phantom{[b}i} w_{a]i}
    \bigr)~,
  \end{split}
\end{align}
where $\nabla^{(\pm)}$ are the usual metric connections with torsion
$\pm H$ and $\Nabla^{(\pm)}$ are the associated connections
involving also the gauge connection $A$.  For the fermionic multiplet
we have
\begin{align}
  S_f = \intd{^2x} \bigl(
      -i h_{AB} \psi_-^A \Nabla_\plpl \psi_-^B
      + h_{AB} \ell^A \ell^B
      - \frac{1}{2} G_{ijab} \psi_-^A \psi_-^B \lambda_+^i \lambda_+^j
      \bigr)
\end{align}
and
\begin{align}
S_p = \intd{^2x} \bigl(
      \nabla_i s_A \lambda_+^i
      + s_A \ell^A
      \bigr)\ .
\end{align}

The scalar potential in these models is precisely that of  the ungauged
(1,0) sigma models in \cite{Hull:1993ct}, ie.
\begin{align}
  V=\frac{1}{4}m^2 h^{AB} s_A s_B~ .
\end{align}
So we find that only `$F$-terms' contribute to the potential. This is
because the gauge multiplet does not have a auxiliary field.
Therefore the classical vacua of the theory are the points of the
sigma model manifold $M$ for which the section $s$ vanishes.  If there
is such a point, then $s$ will vanish at the orbit of the gauge group
$G$ in $M$ which contains that point. This is because the section $s$
is invariant under the action of the gauge group. So in general the
theory will not have isolated vacua unless they are fixed points of
the group action of $G$ on $M$.

\section{$(2,0)$ supersymmetry}

\subsection{The gauge multiplet}

The $(2,0)$ superspace $\Xi^{2,0}$ has coordinates
$(x^\plpl,x^\mimi,\theta_{0}^+,\theta_{1}^+)$ where
$(x^\plpl,x^\mimi)$ are the usual light-cone coordinates and
$\{\theta_{p}^+: p=0,1\}$ are anticommuting coordinates. The
(2,0)-supersymmetric Yang-Mills multiplet is described by a connection
$A$ in $\Xi^{2,0}$ superspace with components
$(A_\plpl,A_\mimi,A_{p+})$, $p=0,1$. In addition it is required that
these satisfy the supersymmetry constraints \cite{Hull:1991uw}
\begin{align}
  [ \nabla_{p+},  \nabla_{q+} ] &= 2i\delta_{pq}\nabla_\plpl
& [ \nabla_\plpl, \nabla_\mimi ] &= F_{\plpl\mimi}
& [ \nabla_{p+},  \nabla_\mimi ] &= W_{p-} \ ,
\end{align}
where $p,q=0,1$.
Jacobi identities imply that
\begin{align}
  \nabla_{p+} W_{q-} + \nabla_{q+} W_{p-}
&=  2i\delta_{pq}F_{\plpl\mimi}
\end{align}
The components of the gauge multiplet are
\begin{align}
  \chi_{0-} &= W_{0-}|
&\chi_{1-} &= W_{1-}|  \nonumber\\
i F_{\plpl\mimi} &= \nabla_{0+}W_{0-}|
& f &= \nabla_{0+}W_{1-}| ~.
\end{align}
The components, $(\chi_{0-}, \chi_{1-})$, are the gaugini which are real
chiral fermions in two dimensions, $F_{\plpl\mimi}$ is the field
strength and $f$ is an auxiliary field. (We have suppressed the gauge indices.)

\subsection{The sigma model multiplet}

We recall that the target manifold $M$ of a (2,0)-supersymmetric
ungauged sigma model is a K\"ahler manifold with torsion (KT).
Therefore $M$ is a hermitian manifold with metric $g$ and equipped
with a complex structure $J$ which is parallel with respect to the
$\nabla^{(+)}$ connection. (For the definition of these geometries see
\cite{Howe:1996kj, Gibbons:1997iy}).  To gauge the model, we assume as
in section \ref{sub:data} that the gauge group $G$ acts on $M$ and
leaving invariant the metric $g$ and the Wess-Zumino term $H$. In
addition we require that the action of the group $G$ is holomorphic.
This means that
\begin{align}
  \mathcal{L}_a J=0 \ ,
\end{align}
where the Lie derivative is along vector fields $\xi_a$ generating
the group action of $G$.  The sigma model fields are
maps $\phi:\Xi^{2,0}\rightarrow M$ into a complex manifold $M$ which
in addition satisfy
\begin{align}
\nabla_{1+}\phi^i= J^i{}_j \nabla_{0+}\phi^j \ ,
\end{align}
where $\nabla_{p+}\phi=D_{p+}\phi^i+ A_{p+}^a \xi_a^i$.  Note that the
requirement for $M$ to be a complex manifold can be derived from the
above condition.

The components of the sigma model multiplet $\phi$ are as follows,
\begin{align}
\phi^i&=\phi^i| & \lambda^i_{+}=&\nabla_{0+}\phi^i|~.
\end{align}

\subsection{The fermionic multiplet}

Let $E$ be a vector bundle over $M$ equipped with a connection $B$ and a fibre (almost)
 complex structure $I$.
The fermionic multiplet $\psi_-$ is a section of $\phi^*E\otimes S_-$
over the $\Xi^{2,0}$, where $S_-$ is a spin bundle over $\Xi^{2,0}$.
In addition we require that the fermionic multiplet $\psi_-$ satisfies
\begin{align}
  \Nabla_{1+}\psi_-^A= I^A{}_B \Nabla_{0+}\psi_-^B+ \frac{1}{2}m L^A
  \label{con20con}
\end{align}
where $L$ is a section of $E$
and
\begin{align}
  \Nabla_{p+}\psi_-^A=
  D_{p+}\psi^A+\nabla_{p+}\phi^i B_i{}^A{}_B \psi_-^B+ A^a_{p+}
  U_a{}^A{}_B  ~,
\end{align}
for $p=0,1$.

Compatibility of the  condition \eqref{con20con} with gauge transformations
requires that
\begin{align}\begin{split}
{\cal L}_a I^A{}_B&= U_a{}^A{}_C I^C{}_B-I^A{}_C U_a{}^C{}_B
\\
{\cal L}_a L^A&=  U_a{}^A{}_B L^B\ .
\end{split}\end{align}
These are the conditions for the gauge transformations and the (2,0)-supersymmetry
transformations to commute.

The compatibility of this constraint with the algebra of covariant
derivatives $\nabla$ implies the following conditions:
\begin{align}\begin{split}
G_{kl\phantom{A}B}^{\phantom{kl}A} \ud Jku\ud Jlj&=G_{ij\phantom{A}B}^{\phantom{ij}A}
\\
J^k{}_i\nabla_k L^A- I^A{}_B \nabla_iL^B&=0
\\
J^k{}_i\nabla_k I^A{}_B- I^A{}_C\nabla_iI^C{}_B&=0~.
\label{susyalcon}
\end{split}\end{align}
These are precisely the conditions required for the off-shell closure of
(2,0) supersymmetry algebra.

It is always possible to find a connection $B$ on the bundle $E$ such
that $\nabla I=0$. In such case the last condition in
\eqref{susyalcon} is satisfied.  Decomposing $E\otimes \bC$ as
$E\otimes \bC={\cal E}\oplus \bar {\cal E}$ using $I$, the first
condition implies that ${\cal E}$ is a holomorphic vector bundle.
Then the second condition in \eqref{susyalcon} implies that the
section $L$ is the real part of a holomorphic section of ${\cal E}$.

The components of the fermionic multiplet are as follows:
\begin{align}
\psi_-^A&=\psi_-^A| & \ell^A=&\Nabla_{0+}\psi_-^A| \ ,
\end{align}
where $\psi_-$ is a two-dimensional real chiral fermion and $\ell$ is
an auxiliary field.

\subsection{Action}

The action of the (2,0)-supersymmetric gauged sigma model can be written
as
\begin{align}
  S=S_g+S_\sigma+S_f  \ ,
\end{align}
where $S_g$ is the action of the gauge multiplet, $S_\sigma$ is the
action of the sigma-model multiplet and $S_f$ is the action of the
fermionic multiplet. We shall describe each term separately.

\subsection{The gauge multiplet action}

The most general action for the (2,0)-supersymmetric gauge multiplet
up to terms quadratic in the
field strength  is
\begin{align}
  \label{eq:cpt20action}
  S_g&=\intd{^2x\rd\theta_{0}^+} \bigl(
  -u^0_{ab}\delta^{pq}W_{p-}^a\nabla_{0+}W_{q-}^b
  +u^1_{ab}\delta^{pq}W_{p-}^a\nabla_{1+}W_{q-}^b+i z^p_a W^a_{p-} \bigr)\ ,
\end{align}
where $u^0$ and $u^1$ are the gauge coupling constants which in
general depend on the superfield $\phi$ and similarly for the
theta terms $z^p$. Both $u^0$ and $u^1$
are not necessarily symmetric in the gauge indices. The above action can be
written in different ways. However there are always field and coupling
constant redefinitions which can bring the action to the above form.

Observe that this action is not an integral over the full $\Xi^{2,0}$
superspace.  Therefore it is not manifestly (2,0)-supersymmetric.
The requirement of invariance under (2,0) supersymmetry imposes the
conditions
\begin{align}
\begin{split}
 J^j{}_i\partial_ju^0 &= -\partial_iu^1 \\
 J^j{}_i\partial_jz^1 &= -\partial_iz^0
 ~.
\end{split}
 \label{eq:cr}
\end{align}
This is most easily seen by verifying that the Lagrangian density is
independent of $\theta_{1}^+$ up to $\theta_{0}^+, x^\plpl,
x^\mimi$-surface terms.  The conditions~\eqref{eq:cr} are the
Cauchy-Riemann equations which imply that $u^0+iu^1$ and $z^1+iz^0$
are {\it holomorphic}.

Indeed provided that the holomorphicity conditions~\eqref{eq:cr} hold,
the action~\eqref{eq:cpt20action} apart from the theta terms can be
written as an integral over the $\Xi^{2,0}$ superspace as
\begin{align}
\label{eq:manifest20action}
\begin{split}
  S_g=\intd{^2x\rd\theta_{0}^+\rd\theta_{1}^+} \bigl(&
  \alpha u^0_{ab} W_{0-}^a W_{1-}^b
  + (\alpha-1) u^1_{ab} W_{0-}^a W_{0-}^b \\
&  - \alpha u^1_{ab} W_{1-}^a W_{1-}^b
  + (\alpha-1) u^0_{ab} W_{1-}^a W_{0-}^b
  \bigr)
\end{split}
\end{align}
for any constant $\alpha$. After integrating over the odd coordinate
$\theta_{1}^+$ we recover the action~\eqref{eq:cpt20action}. Observe
that the action~\eqref{eq:manifest20action} simplifies if one takes $u^0, u^1$
to be  symmetric matrices.  In particular one finds that
\begin{align}
  S_g=\intd{^2x\rd\theta_{0}^+\rd\theta_{1}^+} u^0_{ab}W_{0-}^aW_{1-}^b~.
\end{align}

Invariance of the action  \eqref{eq:cpt20action} under gauge transformations
requires that the couplings $u^0$, $u^1$ and $z^p$ satisfy
\begin{align}
\begin{split}
{\cal L}_a u^0_{bc}&=-f^e{}_{ab} u^0_{ec}-f^e{}_{ac} u^0_{be}
\\
{\cal L}_a u^1_{bc}&=-f^e{}_{ab} u^1_{ec}-f^e{}_{ac} u^1_{be}
\\
{\cal L}_a z^p_{b}&=-f^c{}_{ab} z^p_c
\ .
\end{split}
\end{align}
The gauge transformations of the gauge multiplet and the sigma model
multiplet that are required to derive the above result are as in
the (1,0)-supersymmetric models studied in the previous sections.

\subsection{The sigma model action}

The part of the action which describes the coupling of the sigma model
(2,0)-multiplet to the gauge multiplet has already been given in
\cite{Hull:1991uw}.  This action is
\begin{align}
\begin{split}
  S_\sigma =&-i \intd{^2x\rd\theta_{0}^+}\bigl(
  g_{ij}\nabla_{0+}\phi^i\nabla_\mimi\phi^j
  + \nu_{a}W^a_{1-}\bigr)\\
  &-i \intd{^2x\rd t\rd\theta_{0}^+}\bigl(
  H_{ijk}\partial_t\phi^i\nabla_{0+}\phi^j\nabla_{\mimi}\phi^k -
  w_{ia}\partial_t\phi^iW_{0-}^a \bigr)
\end{split}
\end{align}
where $\nu_{a}$ is a function on $M$, possibly locally defined,
given by
\begin{align}
\label{eq:moment20def}
  I^j_{\phantom{j}i} (\xi_{aj} + w_{aj}) &=  -\partial_i \nu_{a}
\end{align}
Under certain conditions~\cite{Grantcharov:2002fk} the maps $\nu$ can
be thought of as the moment maps of KT geometry.

Gauge invariance of the above part of action requires in addition to the
conditions on  $w$, which we have already mentioned in section  \ref{sub:gaugecon},
that $\nu_a$ is
globally defined on $M$ and that
\begin{align}
\mathcal{L}_a \nu_b=-f_{ab}{}^c \nu_c~.
\end{align}

\subsection{The action of the fermionic multiplet}

This part of the action is
\begin{align}
  S_f=\intd{^2x\rd\theta_{0}^+}\bigl(
    h_{AB} \psi_-^A \Nabla_{0+} \psi_-^B +m s_A \psi_-^A
  \big)
\end{align}

Gauge invariance of this part of the action requires the same
conditions as those appearing for the couplings of (1,0)-multiplet
in~\eqref{eq:psigaugecnd}.
 
The conditions required by (2,0)-supersymmetry on the couplings of the
above action are the same as those of the ungauged model and have been
given in~\cite{Hull:1985jv}. These can be easily derived by requiring that the
Lagrangian density is independent from $\theta_{1}^+$ up to $x^\plpl,
x^\mimi, \theta_{0}^+$ surface terms. In particular, we find that
\begin{align}
 \label{eq:susyfcon}
 \begin{split}
 h_{CB} I^C{}_A+h_{CA} I^C{}_B&=0
 \\
 J^j{}_i \nabla_j h_{AB}+\nabla_i h_{AC} I^C{}_B&=0
 \\
J^j{}_i \nabla_j s_A-\nabla_i (s_B I^B{}_A)-{1\over2} \nabla_i
h_{AB} L^B&=0
\\
s_A L^A&={\rm const}\ .
\end{split}
\end{align}
The first condition implies that the fibre metric is hermitian with
respect to the fibre complex structure. It is always possible to
choose such a fibre metric given a fibre complex structure on a bundle
vector bundle $E$.  In the context of sigma models this has been
explained in \cite{Howe:1988cj}.  The rest of the conditions can be
considerably simplified if the connection $B$ is chosen such that
$\nabla I=\nabla h=0$. Such connection always exists on a hermitian
vector bundle $E$. In such case, the third equation in
\eqref{eq:susyfcon} implies that $s$ is the real part of a holomorphic
section of ${\cal E}^*$.

\subsection{The action in components}

It is straightforward to write the action $S$ of the
(2,0)-supersymmetric gauge sigma model in components.
In particular we find that the component action of the gauge
multiplet~\eqref{eq:cpt20action} is
\begin{align}
  \begin{split}
    S_g = \intd{^2x}\bigl(& u^0_{ab} F_{\plpl\mimi}^a F_{\plpl\mimi}^b+ z^0_a F_{\plpl\mimi}^a
-u^0_{ab} f^a f^b + i z_a^1 f^a
    \\
    &+
    i u^0_{ab} \chi_{0-}^a \nabla_\plpl\chi_{0-}^b +
    i u^0_{ab} \chi_{1-}^a \nabla_\plpl\chi_{1-}^b
     \\
    & -2i u^1_{[ab]} F_{\plpl\mimi}^a f^b
    +2i u^1_{[ab]} \chi_{[0-}^a \nabla_\plpl\chi_{1]-}^b
    \\
    &+i\partial_iz^0_a \lambda_+^i \chi_{0-}^a+ i\partial_i z^1_a \lambda^i_+  \chi_{1-}^a
    \\
    & -\lambda^i_+ \partial_iu^0_{ab}(i\chi_{0-}^aF_{\plpl\mimi}^b
    +\chi_{1-}^af^b)  \\
&   +\lambda^i_+\partial_iu^1_{ab}(-\chi_{0-}^af^b
    +i\chi_{1-}^aF_{\plpl\mimi}^b) \bigr)~.
  \end{split}
\end{align}
The component action of the sigma model part is
\begin{align}
  \begin{split}
    S_\sigma = \intd{^2x} \bigl(&
    g_{ij} \nabla_{\plpl}\phi^i \nabla_\mimi\phi^j
    + b_{ij} \partial_\plpl\phi^i \partial_\mimi\phi^j
      +i \lambda_+^i \Nabla^{(+)}_\mimi \lambda_+^j \\
&      -i  \partial_i \lambda_+^i \nu_a \chi_{1-}^a
      -i \nu_a f^a 
      +i g_{ij} \lambda_+^i \chi_{0-}^a \xi_a^{\phantom{a}j}
   +i w_{ia} \lambda_+^i \chi_{0-}^a \\
&    -  A_\plpl^a w_{ia} \partial_\mimi\phi^i
    +  A_\mimi^a w_{ia} \partial_\plpl\phi^i
    -  A_\plpl^a A_\mimi^b \xi_{[b}^{\phantom{[b}i} w_{a]i}
    \bigr)
  \end{split}
\end{align}
and the component action of the fermionic multiplet is
\begin{align}
  \begin{split}
    S_f = \intd{^2x} \bigl(&
    -i h_{AB} \psi_-^A \Nabla_\plpl \psi_-^B
    + h_{AB} \ell^A \ell^B \\
&    - \frac{1}{2} h_{AB} \psi_-^A \psi_-^B \lambda_+^i \lambda_+^j
    G_{ijAB} 
    + m \nabla_i s_A \lambda_+^i \psi_-^A
    + m s_A \ell^A
    \bigr)\ .
  \end{split}
\end{align}
Eliminating the auxiliary fields of the gauge and fermionic
multiplets, we find that
\begin{align}
  \begin{split}
    S = \intd{^2x}\bigl(& 
    g_{ij} \nabla_{\plpl}\phi^i \nabla_\mimi\phi^j
    + b_{ij} \partial_\plpl\phi^i \partial_\mimi\phi^j
    + u^0_{ab} F_{\plpl\mimi}^a F_{\plpl\mimi}^b \\
&   +i u^0_{ab} \chi_{0-}^a \nabla_\plpl\chi_{0-}^b +
    i u^0_{ab} \chi_{1-}^a \nabla_\plpl\chi_{1-}^b
     \\
&     +i g_{ij} \lambda_+^i \tilde\nabla^{(+)}_\mimi \lambda_+^j 
   -i h_{AB} \psi_-^A \Nabla_\plpl \psi_-^B 
   + z^0_a F_{\plpl\mimi}^a \\
&    -\frac{1}{4} m^2 h^{AB} s_A s_B 
     -\frac{1}{4} u_0^{ab} (\nu_a - z^1_a)(\nu_b - z^1_b) \\
&    - \frac{1}{2} h_{AB} \psi_-^A \psi_-^B \lambda_+^i \lambda_+^j
    G_{ijAB} 
    + m \nabla_i s_A \lambda_+^i \psi_-^A \\
&    -  A_\plpl^a w_{ia} \partial_\mimi\phi^i
    +  A_\mimi^a w_{ia} \partial_\plpl\phi^i
    -  A_\plpl^a A_\mimi^b \xi_{[b}^{\phantom{[b}i} w_{a]i} \\
&    -  u_0^{ab} (\nu_a - z^1_a)  u^1_{[cb]} F_{\plpl\mimi}^c 
   - u_0^{ab} u^1_{[ac]} u^1_{[bd]} F_{\plpl\mimi}^c F_{\plpl\mimi}^d \\
&+i\partial_i z^0_a \lambda_+^i \chi_{0-}^a+ i\partial_i z^1_a \lambda^i_+  \chi_{1-}
    \\
    & -i \partial_i  u^0_{ab} \lambda^i_+ \chi_{0-}^aF_{\plpl\mimi}^b  
   +i \partial_i u^1_{ab} \lambda^i_+\chi_{1-}^aF_{\plpl\mimi}^b \\
&      -i \partial_i \nu_a \lambda_+^i \chi_{1-}^a
      +i g_{ij} \lambda_+^i \chi_{0-}^a \xi_a^{\phantom{a}j}
   +i w_{ia} \lambda_+^i \chi_{0-}^a \\
&  +\frac{1}{2}i u_0^{ab} (\nu_a - z^1_a)(\partial_i u^0_{cb}\chi_{1-}^c
    + \partial_i u^1_{cb}\chi_{0-}^c)\lambda_+^i \\
&   -\frac{1}{4} u_0^{ab}(\partial_i u^0_{ca}\chi_{1-}^c
      + \partial_i u^1_{cb}\chi_{0-}^c)
    (\partial_i u^0_{db}\chi_{1-}^d + \partial_i u^1_{db}\chi_{0-}^d)\lambda_+^i\lambda_+^j\\
&   -i u_0^{ab} u^1_{[ac]} F_{\plpl\mimi}^c 
    (\partial_i u^0_{db}\chi_{1-}^d + \partial_i u^1_{db}\chi_{0-}^d)\lambda_+^i\\
&     +2i u^1_{[ab]} \chi_{[0-}^a \nabla_\plpl\chi_{1]-}^b
    \bigr)
  \end{split}
\end{align}
where $u_0^{ab}$ is the matrix inverse of $u^0_{(ab)}$,
\begin{align}
  u_0^{ab} u^0_{(bc)} &= \delta^a_{\phantom{a}c}~.
\end{align}
Note that we have assumed that $u^0$ is invertible.

\subsection{Scalar potential and classical vacua}

The scalar potential of the
(2,0)-supersymmetric  gauge theories coupled to sigma model matter is
is
\begin{align}
V=\frac{1}{4}u_0^{ab}(\nu_a-z^1_a) (\nu_b-z^1_b)+\frac{1}{4}m h^{AB} s_A s_B\ .
\end{align}
The scalar potential in these models is
written as a sum of a `$D$' and an `$F$' term. The classical
supersymmetric vacua of
the theory are those for which
\begin{align}
\nu_a-z^1_a&=0 & s_A&=0\ .
\end{align}
The inequivalent classical vacua are the space of orbits of the gauge group
on the zero set of the section $s$ and  $\nu-z^1$. If the section $s$ and $z^1$
vanish, then the space of inequivalent vacua is the KT reduction $M//G$ of the sigma model
target space $M$. It has been shown in \cite{Grantcharov:2002fk} that the space of vacua inherits the KT
structure of the sigma model manifold $M$ and under certain assumptions
is a smooth manifold. However the three-form of the Wess-Zumino term
 on $M//G$ is not necessarily
closed.

\section{$(4,0)$ supersymmetry}

\subsection{The gauge multiplet}

The $(4,0)$ superspace $\Xi^{4,0}$ has coordinates
$(x^\plpl,x^\mimi,\theta_{p}^+)$,  where
$\theta_{p}^+$, $p=0,1,2,3$, are the odd coordinates. The
(4,0)-supersymmetric Yang-Mills multiplet is described by a
connection $A$ in superspace with components
$(A_\plpl,A_\mimi,A_{p+})$, $p=0,1,2,3$. In addition it is required
that these satisfy the supersymmetry constraints \cite{Hull:1991uw}
\begin{align}\begin{split}
  [ \nabla_{p+},  \nabla_{q+} ] &= 2i\delta_{pq}\nabla_\plpl \\
 [ \nabla_\plpl, \nabla_\mimi ] &= F_{\plpl\mimi} \\
 [ \nabla_{p+},  \nabla_\mimi ] &= W_{p-} \\
\nabla_{p+} W_{q-}&=\frac{1}{2} \epsilon_{pq}{}^{p'q'}\nabla_{p'+} W_{q'-} \ ,
\end{split}\end{align}
where $p,q,p'q'=0,\ldots,3$ and $p\neq q$ in the last condition. (We have suppressed
gauge indices.)
Jacobi identities imply that
\begin{align}
  \nabla_{p+} W_{q-} + \nabla_{q+} W_{p-}
&=  2i\delta_{pq}F_{\plpl\mimi}
\end{align}

The components of the gauge multiplet are
\begin{gather}
  \chi_{p-} = W_{p-}|
\mspace{100mu} i F_{\plpl\mimi} = \nabla_{0+}W_{0-}| 
 \nonumber\\
  f_r = \nabla_{0+}W_{r-}| \quad (r=1,2,3)
\end{gather}
The first four fields, $(\chi_{p-}: p=0,1,2,3)$, are the gaugini which are real
chiral fermions in two dimensions, $F_{\plpl\mimi}$ is the field
strength and $\{f_r: r=1,2,3\}$ are the auxiliary fields.

\subsection{The sigma model multiplet}

Let $M$ be a hyper-K\"ahler manifold with torsion (HKT).  This implies
that $M$ admits a hypercomplex structure $\{J_r: r=1,2,3\}$ the metric
$g$ on $M$ is tri-hermitian and the hypercomplex structure is parallel
with respect to a metric connection with as torsion the three-form $H$,
$\nabla^{(+)}J_r=0$; (see \cite{Howe:1996kj, Gibbons:1997iy} for more
details).  In addition we assume that the gauge group $G$ acts on $M$
preserving the metric, three-form $H$ and the hypercomplex structure.
The latter condition implies that
\begin{align}
  \mathcal{L}_a J_r=0 \ ,
\end{align}
where the Lie derivative is along vector fields $\xi_a$ generated by the
action of $G$ on $M$.  The sigma model fields are
maps $\phi:\Xi^{4,0}\rightarrow M$ into the HKT manifold $M$ which
in addition satisfy
\begin{align}
\nabla_{r+}\phi^i= J_r{}^i{}_j \nabla_{0+}\phi^j \ ,
\end{align}
where $\nabla_{p+}\phi=D_{p+}\phi^i+ A_{p+}^a \xi_a^i$ and $r=1,2,3$.
We remark that the algebra of (4,0) supersymmetry transformations
closes as a consequence of the HKT condition we imposed on $M$.

The components of the sigma model multiplet $\phi$ are as follows,
\begin{align}
\phi^i&=\phi^i| & \lambda^i_{+}=&\nabla_{0+}\phi^i|~.
\end{align}

\subsection{The fermionic multiplet}

Let $E$ be a vector bundle over $M$ equipped with a connection $B$ and a
fibre almost hypercomplex structure $\{I_r: r=1,2,3\}$.
The fermionic multiplet $\psi_-$ is a section of $\phi^*E\otimes S_-$
over the $\Xi^{4,0}$, where $S_-$ is a spin bundle over $\Xi^{4,0}$.
In addition the fermionic multiplet $\psi_-$ satisfies
\begin{align}
  \Nabla_{r+}\psi_-^A= I_r{}^A{}_B \Nabla_{0+}\psi_-^B+ \frac{1}{2}m L_r^A
  \label{con40con}
\end{align}
where $\{L_r: r=1,2,3\}$ are sections of $E$
and
\begin{align}
  \Nabla_{p+}\psi_-^A=
  D_{p+}\psi^A+\nabla_{p+}\phi^i B_i{}^A{}_B \psi_-^B+ A^a_{p+}
  U_a{}^A{}_B  ~,
\end{align}
$p=0,1,2,3$.

Compatibility of the condition \eqref{con40con} with gauge
transformations requires that
\begin{align}
\begin{split}
{\cal L}_a I_r^A{}_B&= U_a{}^A{}_C I_r^C{}_B-I_r^A{}_C U_a{}^C{}_B
\\
{\cal L}_a L_r^A&=  U_a{}^A{}_B L_r^B\ .
\end{split}
\end{align}
These are the conditions for the gauge transformations and the
(2,0)-supersymmetry transformations to commute.

The conditions required for the closure of the (4,0) supersymmetry
algebra are similar to those found for the ungauged (4,0) model in
\cite{Hull:1993ct,Howe:1987qv}. Here to simplify the analysis, we
shall in addition assume that the fibre hypercomplex structure is
parallel with respect to $\nabla$, ie $\nabla I_r=0$.  (See
\cite{Howe:1988cj} for a discussion on the conditions required for the
existence of such a connection $\nabla$ on the vector bundle $E$.) The
more general case can be easily derived but we shall not use these
results later.  In the special case, we find that
\begin{align}\begin{split}
G_{kl}{}^A{}_B J_{r}{}^k{}_i J_{s}{}^l{}_j+G_{kl}{}^A{}_B J_{s}{}^k{}_i J_{r}{}^l{}_j
=2\delta_{rs}G_{ij}{}^A{}_B
\\
J_r{}^j{}_i \nabla_j L_s^A+ J_s{}^j{}_i \nabla_j L_r^A- I_r{}^A{}_B
\nabla_i L_s^B-I_s{}^A{}_B \nabla_i L_r^B =0\ .
\end{split}\end{align}
The first condition implies that the curvature $G$ of the vector
bundle $E$ is a (1,1)-form with respect to all three complex
structures $J_r$. Observe that the diagonal relation $r=s$ implies all
the rest.  The diagonal part of the second condition implies that each
section $L_r$ is a holomorphic section with respect to the pair $(J_r,
I_r)$.

The components of the fermionic multiplet are as follows:
\begin{align}
\psi_-^A&=\psi_-^A| & \ell^A=&\Nabla_{0+}\psi_-^A| \ ,
\end{align}
where $\psi_-$ is a two-dimensional real chiral fermion and $\ell$ is
an auxiliary field.

\subsection{Action}

The action of the (4,0)-supersymmetric gauged sigma model can be written
as
\begin{align}
  S=S_g+S_\sigma+S_f  \ ,
\end{align}
where $S_g$ is the action of the gauge multiplet, $S_\sigma$ is the
action of the sigma-model multiplet and $S_f$ is the action of the
fermionic multiplet. We shall describe each term separately.

\subsection{The gauge multiplet action}

The most general action for the (4,0)-supersymmetric gauge multiplet
up to terms quadratic in the
field strength  is
\begin{align}
  \label{eq:cpt40action}
  S_g&=\intd{^2x\rd\theta_{0}^+} \bigl(
  -u^0_{ab}\delta^{pq}W_{p-}^a\nabla_{0+}W_{q-}^b
  +\sum^3_{r=1}u^r_{ab}\delta^{pq}W_{p-}^a\nabla_{r+}W_{q-}^b+i z^p_a W^a_{-p} \bigr)
\end{align}
where $\{u^p: p=0,1,2,3\}$ and $\{z^p: p=0,1,2,3\}$ are the gauge
coupling constants which in general depend on the superfield $\phi$
and $u^p$ are not necessarily symmetric in the gauge indices. The
above action can be written in different ways. However there are
always field and coupling constant redefinitions which bring the
action to the above form.

Observe that this action is not an integral over the full $\Xi^{4,0}$
superspace.  Therefore it is not manifestly (4,0)-supersymmetry.
Define $J_{0\phantom{i}j}^{\phantom{0}i}=\delta^i_{\phantom{i}j}$. The
requirement of invariance the action \eqref{eq:cpt40action} under
(4,0) supersymmetry imposes the conditions
\begin{align}\begin{split}
  \label{eq:cr40}
      J_{p\phantom{j}i}^{\phantom{p}j} \partial_j u_q
  &= \frac{1}{2} \epsilon_{pq}{}^{p'q'}
     J_{p'}{}^j{}_i \partial_ju_{q'}
  \quad (p\neq q)
   \\
  \partial_i u_0
    = J_{1\phantom{j}i}^{\phantom{1}j} \partial_j u_1
    &= J_{2\phantom{j}i}^{\phantom{2}j} \partial_j u_2
    = J_{3\phantom{j}i}^{\phantom{3}j} \partial_j u_3
  ~,
\end{split}\end{align}
and
$\{u^r: r=1,2,3\}$ are {\it symmetric} in the gauge indices. In addition
\begin{align}
\label{eq:sus40}
\begin{split}
J_r{}^j{}_i \partial_jz^r&=-\partial_i z^0\\
J_p{}^j{}_i \partial_j z_q&=-{1\over2} \epsilon_{pq}{}^{p'q'} J_{p'}{}^k{}_i \partial_k z_{q'}\ .
\end{split}
\end{align}
The conditions \eqref{eq:cr40} and \eqref{eq:sus40} imply that in fact
$\{u^p: p=0,\dots, 3\}$ and $\{z^p: p=0,\dots, 3\}$ are constant, i.e.\ 
independent from the sigma model superfield $\phi$.  The above
conditions are most easily derived by verifying that the Lagrangian
density is independent of $\theta_{r}^+$ up to surface terms in
$x^\plpl$, $x^\mimi$ and $\theta_{0}^+$.  In addition gauge invariance
requires that the coupling constants $u^p$ and $z^p$ satisfy the
condition
\begin{align}\begin{split}
f^d{}_{ab} u^p_{dc}+ f^d{}_{ac} u^p_{bd}&=0
\\
f^c{}_{ab} z^p_{c}&=0\ .
\end{split}\end{align}
In particular $u^r$ should be proportional to an invariant quadratic
form on the Lie algebra of the gauge group $G$. If $G$ semi-simple,
the condition on $z^p$ implies that $z^p=0$. If $G$ is abelian, then
the above conditions are satisfied for any constants $u^p$ and $z^p$.

\subsection{The sigma model multiplet action}

This part of the action has already been described in \cite{Hull:1991uw}.
Here we shall summarise the some of results relevant to this chapter.
The action of this multiplet is
\begin{align}
\begin{split}
  S_\sigma =&-i \intd{^2x\rd\theta_{0}^+}\bigl(
    g_{ij} \nabla_{0+}\phi^i \nabla_\mimi\phi^j
    +\sum_{r=1}^3 \nu_{ra} W_{r-}^a
  \bigr)\\
  &-i \intd{^2x\rd t\rd\theta_{0}^+}\bigl(
  H_{ijk}\partial_t\phi^i\nabla_{0+}\phi^j\nabla_{\mimi}\phi^k -
  w_{ia}\partial_t\phi^iW_{0-}^a \bigr)
\end{split}
\end{align}
where $\nu_{a}$ is a function on $M$, possibly locally defined,
given by
\begin{align}
\label{eq:momentp0def}
  I_{r\phantom{j}i}^{\phantom{r}j} (\xi_{aj} + w_{aj}) &=  -\partial_i\nu_{ra}
\end{align}
It has been shown in~\cite{Grantcharov:2002fk} that under certain
conditions $\nu$ is the moment map of HKT geometry.

The gauge transformations of the superfield $\phi$ are
\begin{align}
\delta\phi^i &= \epsilon^a\xi_a^{\phantom{a}i}(\phi)
\end{align}
Gauge invariance of the above action requires that the one-form $w$
should satisfy the conditions  mentioned in section \ref{sub:gaugecon},
the moment maps should be globally defined on the sigma model target space $M$ and
\begin{gather}
  \mathcal{L}_a \nu_{ra} = -f_{ab}^{\phantom{ab}c} \nu_{rc} ~.
\end{gather}

\subsection{The action of the fermionic multiplet}

This part of the action is
\begin{align}
  S_f=\intd{^2x\rd\theta_{0}^+}\bigl(
    h_{AB} \psi_-^A \Nabla_{0+} \psi_-^B +m s_A \psi_-^A
  \big)
\end{align}

Gauge invariance of this part of the action requires the same
conditions as those appearing for the couplings of (1,0)-multiplet
 in~\eqref{eq:psigaugecnd}.

The conditions required by (4,0)-supersymmetry on the couplings of the
above action are the same as those of the ungauged model and have been
given in~\cite{Howe:1987qv}. These can be easily derived by requiring that
the Lagrangian density is independent from $\theta_{r}^+$ up to
 $x^\plpl, x^\mimi, \theta_{0}^+$ surface terms. In particular, we find that
 \begin{align}
 \label{eq:susyf40con}
 \begin{split}
 h_{CB} I_r^C{}_A+h_{CA} I_r^C{}_B&=0
 \\
 J_r^j{}_i \nabla_j h_{AB}+\nabla_i h_{AC} I_r^C{}_B&=0
 \\
J_r^j{}_i \nabla_j s_A-\nabla_i (s_B I_r^B{}_A)-{1\over2} \nabla_i
h_{AB} L_r^B&=0
\\
s_A L_r^A={\rm const}\ .
\end{split}
\end{align}
To derive the above conditions we have used that $\nabla I_r=0$ as we
have assumed in the construction of the fermionic multiplet. The first
condition implies that the fibre metric is tri-hermitian.  These
conditions can be further simplified if the connection $B$ is chosen
such that $\nabla h=0$.  In such case, the third equation in
(\ref{eq:susyf40con}) implies that $s$ is the real part of three
holomorphic sections of ${\cal E}^*$ each with respect to the three
doublets $(J_r, I_r)$ of complex structures, ie $s$ is triholomorphic.

\subsection{The action in components and scalar potential}

The part of the action of the theory involving the
kinetic term of the gauge multiplets \eqref{eq:cpt40action} can be easily expanded in components
as follows,
\begin{align}
  \begin{split}
    S_g = \intd{^2x} \bigl( & u^0_{ab} F_{\plpl\mimi}^a F_{\plpl\mimi}^b
      +i u^0_{ab} \delta^{pq} \chi_{p-}^a \nabla_{\plpl}\chi_{q-}^b\\
&   + z^0_a F^a_{\plpl\mimi}
    - \sum_r (u^0_{ab} f_r^a f_r^b+i z^r_a f_r^a )\\
&   + 2i u^1_{ab} \chi_{[0-}^a \nabla_\plpl\chi_{1]-}^b
    + 2i u^1_{ab} \chi_{[2-}^a \nabla_\plpl\chi_{3]-}^b \\
&   + 2i u^2_{ab} \chi_{[0-}^a \nabla_\plpl\chi_{2]-}^b
    + 2i u^2_{ab} \chi_{[1-}^a \nabla_\plpl\chi_{3]-}^b \\
&   + 2i u^3_{ab} \chi_{[0-}^a \nabla_\plpl\chi_{3]-}^b
    + 2i u^3_{ab} \chi_{[1-}^a \nabla_\plpl\chi_{2]-}^b     
    \bigr) \ .
  \end{split}
\end{align}
To derive this we have used that $\{u^p:0,\dots, 3\}$ are constant and $\{u^r:1,2,3\}$ symmetric
in the gauge indices.

The part of the action that contains the kinetic term and the
Wess-Zumino term of the sigma model fields in components is as
follows,
\begin{align}
  \begin{split}
    S_\sigma = \intd{^2x} \bigl(&
    g_{ij} \nabla_{\plpl}\phi^i \nabla_\mimi\phi^j
    + b_{ij} \partial_\plpl\phi^i \partial_\mimi\phi^j
      +i g_{ij} \lambda_+^i \tilde\nabla^{(+)}_\mimi \lambda_+^j \\
&      -i \sum_r ( \nu_{ra} f_r^a + \partial_i \nu_{ra} \lambda_+^i \chi_{r-}^a)\\
&      +i g_{ij} \lambda_+^i \chi_{0-}^a \xi_a^{\phantom{a}j}
   +i w_{ia} \lambda_+^i \chi_{0-}^a \\
&   -  A_\plpl^a w_{ia} \partial_\mimi\phi^i
    +  A_\mimi^a w_{ia} \partial_\plpl\phi^i
    -  A_\plpl^a A_\mimi^b \xi_{[b}^{\phantom{[b}i} w_{a]i}
    \bigr)
  \end{split}
\end{align}
The part of the action that contains the kinetic term the fermionic
multiplet in components is as follows:
\begin{align}
  \begin{split}
    S_f = \intd{^2x} \bigl(&
    -i h_{AB} \psi_-^A \Nabla_\plpl \psi_-^B
    + h_{AB} \ell^A \ell^B 
    - \frac{1}{2} h_{AB} \psi_-^A \psi_-^B \lambda_+^i \lambda_+^j
    G_{ijAB} \\
&    + m \nabla_i s_A \lambda_+^i \psi_-^A
    + m s_A \ell^A
    \bigr)
  \end{split}
\end{align}
After eliminating the auxiliary fields of both the gauge multiplet and
the fermionic multiplet, we find that the action of
(4,0)-supersymmetric gauge theories coupled to sigma models is
\begin{align}
  \begin{split}
    S_g+S_f = \intd{^2x}\bigl(& 
    u^0_{ab} F_{\plpl\mimi}^a F_{\plpl\mimi}^b 
   -i h_{AB} \psi_-^A \Nabla_\plpl \psi_-^B \\
&  +i \delta^{pq} u^0_{ab} \chi_{p-}^a \nabla_\plpl\chi_{q-}^b 
     \\
&   + 2i u^1_{ab} \chi_{[0-}^a \nabla_\plpl\chi_{1]-}^b
    + 2i u^1_{ab} \chi_{[2-}^a \nabla_\plpl\chi_{3]-}^b \\
&   + 2i u^2_{ab} \chi_{[0-}^a \nabla_\plpl\chi_{2]-}^b
    + 2i u^2_{ab} \chi_{[1-}^a \nabla_\plpl\chi_{3]-}^b \\
&   + 2i u^3_{ab} \chi_{[0-}^a \nabla_\plpl\chi_{3]-}^b
    + 2i u^3_{ab} \chi_{[1-}^a \nabla_\plpl\chi_{2]-}^b  \\
&   - \frac{1}{2} h_{AB} \psi_-^A \psi_-^B \lambda_+^i \lambda_+^j
    G_{ijAB} 
    + m \nabla_i s_A \lambda_+^i \psi_-^A \\
&     - \frac{1}{4} u_0^{ab} \sum_r (\nu_{ra} - z^r_a)(\nu_{rb} - z^r_b) 
    -\frac{1}{4} m^2 h^{AB} s_A s_B \\
&   + z^0_a F_{\plpl\mimi}^a 
      + \sum_p \partial_i z^p_a \lambda_+^i \chi_{r-}^a
    \bigr)
  \end{split}
\end{align}
It is straightforward to write the action $S$ of the
(2,0)-supersymmetric gauge sigma model in components. In particular
 we find the following:
\begin{align}
  \begin{split}
    S = \intd{^2x}\bigl(& 
    g_{ij} \nabla_{\plpl}\phi^i \nabla_\mimi\phi^j
    + b_{ij} \partial_\plpl\phi^i \partial_\mimi\phi^j
    + u^0_{ab} F_{\plpl\mimi}^a F_{\plpl\mimi}^b \\
&   +i u^0_{ab} \delta^{pq} \chi_{p-}^a \nabla_\plpl\chi_{q-}^b 
     \\
&     +i \lambda_+^i \tilde\nabla^{(+)}_\mimi \lambda_+^j 
   -i h_{AB} \psi_-^A \Nabla_\plpl \psi_-^B 
   + z^0_a F_{\plpl\mimi}^a \\
&    -\frac{1}{4} m^2 h^{AB} s_A s_B 
     - \frac{1}{4} u_0^{ab} \sum_r (\nu_{ra} - z^r_a)(\nu_{rb} - z^r_b) \\
&    - \frac{1}{2} h_{AB} \psi_-^A \psi_-^B \lambda_+^i \lambda_+^j
    G_{ijAB} 
    + m \nabla_i s_A \lambda_+^i \psi_-^A \\
&    -  A_\plpl^a w_{ia} \partial_\mimi\phi^i
    +  A_\mimi^a w_{ia} \partial_\plpl\phi^i
    -  A_\plpl^a A_\mimi^b \xi_{[b}^{\phantom{[b}i} w_{a]i} \\
&+i\partial_i z^0_a \lambda_+^i \chi_{0-}^a
    \\
&          +i g_{ij} \lambda_+^i \chi_{0-}^a \xi_a^{\phantom{a}j}
   +i w_{ia} \lambda_+^i \chi_{0-}^a 
  + i\sum_r \partial_i (z^r_a-\nu_{ra}) \lambda_+^i \chi_{r-}^a
    \bigr)
  \end{split}
\end{align}

\subsection{Scalar Potential and Classical Vacua}

The scalar potential of the (4,0)-supersymmetric gauged sigma models
is
\begin{align}
V=\frac{1}{4}\sum^3_{r=1}u_0^{ab}\nu_{ra} \nu_{rb}+\frac{1}{4}m h^{AB} s_A s_B\ ,
\end{align}
where we have absorbed the constants $z^r_a$ into the definition
of the moment maps $\nu_{ra}$.
The scalar potential in these models is
written as a sum of a `$D$' and an `$F$' term. The classical
supersymmetric vacua of
the theory are those for which
\begin{align}
\nu_{ra}&=0 & s_A&=0\ .
\end{align}
The inequivalent classical vacua are the space of orbits of the gauge group
on the zero set of the section $s$ and the HKT moment maps $\nu_r$. If the section $s$
vanishes, then the space of inequivalent vacua is  the theory is  the HKT reduction $M//G$ of the sigma model
target space $M$. It has been shown in \cite{Grantcharov:2002fk} that under certain assumptions the space of vacua inherits the HKT
structure of the sigma model manifold $M$ and it is a smooth space.
However the three-form of the Wess-Zumino term on $M//G$ is not necessarily
closed.

\section{$(1,1)$ supersymmetry}

\subsection{The gauge multiplet}

The $(1,1)$ superspace $\Xi^{1,1}$ has coordinates
$(x^\plpl,x^\mimi,\theta^+,\theta^-)$, where $\theta^{\pm}$ are
Grassman valued odd coordinates. The (1,1)-supersymmetric Yang-Mills
multiplet is described by a connection $A$ in superspace with
components $(A_\plpl,A_\mimi,A_+,A_-)$. In addition it is required
that these satisfy the supersymmetry constraints \cite{Hull:1991uw}
\begin{align}
  [ \nabla_+, \nabla_- ] &= W & [ \nabla_\plpl, \nabla_\mimi ] &=
  F_{\plpl\mimi}  \nonumber\\
  [ \nabla_+, \nabla_+ ] &= 2i\nabla_\plpl & [ \nabla_-, \nabla_- ] &=
  2i\nabla_\mimi
\end{align}
(We have suppressed the gauge indices.)
The Jacobi identities imply that
\begin{gather}
  [\nabla_+, \nabla_\mimi ] = i\nabla_-W
\mspace{100mu} [\nabla_-, \nabla_\plpl ] = i\nabla_+W
 \nonumber\\
  F_{\plpl\mimi} = \nabla_+ \nabla_-W
\end{gather}
We mention that the (1,1)-supersymmetric gauge multiplet can be
constructed from a scalar superfield. This allows for the possibility
of non-linear couplings between the sigma model multiplet and the
gauge multiplet.  The components of the gauge multiplet are
\begin{align}
  W &= W|
& F_{\plpl\mimi} &= \nabla_+\nabla_-W|\nonumber \\
  \chi_+ &= \nabla_+ W|
& \chi_- &= \nabla_- W|
\end{align}
The field, $W$, is a scalar, $\chi_+$,$\chi_-$ are
gaugini which are real chiral fermions in two dimensions, and
$F_{\plpl\mimi}$ is the field strength.

\subsection{The sigma model multiplet}

Let $M$ be a Riemannian manifold with metric $g$ and locally defined
two-form $b$.  We take the gauge group $G$ to act on $M$
with isometries preserving the Wess-Zumino three-form $H=\rd b$
and generating the vector fields $\xi_a$ as in section \ref{sub:data}. In addition
we assume that the one-form $w$ satisfies the conditions of section \ref{sub:gaugecon}.

The  sigma model (1,1) multiplet $\phi$ is a map from the (1,1) superspace $\Xi^{1,1}$ into the
Riemannian manifold $M$.
The components of $\phi$ are as follows,
\begin{align}
  \phi^i &= \phi^i|
& \ell^i &= \nabla_+\nabla_-\phi^i| \nonumber\\
  \lambda^i_{+}&=\nabla_{+}\phi^i|
& \lambda^i_{-}&=\nabla_{-}\phi^i|
 ~,
\end{align}
where $\nabla_+\phi^i=D_+\phi^i+ A_+^a\xi_a^i$ and similarly for $\nabla_-$.
Observe that the first two components of $(1,1)$ superfield $\phi$
can be identified with the two components of  a $(1,0)$ superfield $\phi$
while the latter two components can be identified with those of  a $(1,0)$
fermionic superfield $\psi_-$. The vector bundle associated with this
fermionic multiplet is the tangent bundle of $M$.

\subsection{Action}

The action of the (1,1)-supersymmetric gauged sigma model can be written
as sum of three terms,
\begin{align}
  S &= S_g + S_\sigma +S_p  \ ,
\end{align}
where $S_g$ is the action of the gauge multiplet, $S_\sigma$ is the
action of the sigma-model multiplet and $S_p$ is a potential term.
We shall describe each term separately.

\subsection{The gauge multiplet action}

An action of the (1,1)-supersymmetric gauge multiplet is most easily
written in (1,1) superspace.
In particular we have
\begin{align}
  S_g = \intd{^2x\rd\theta^+\rd\theta^-}\bigl(
    - u_{ab} \nabla_+W^a \nabla_- W^b
    + \frac{1}{2} v_{ab} W^a W^b+ z_a W^a
    \bigr)\ ,
    \label{f11act}
\end{align}
where $u_{ab}=u_{ab}(\phi)$, $u$ is not necessarily symmetric in the gauge indices,
$v_{ab}=v_{ab}(\phi)$ and the theta term $z_a=z_a(\phi)$. Of course this action is manifestly (1,1)-supersymmetric
because it is an integral over full superspace. Gauge invariance imposes the additional
conditions
\begin{align}
\begin{split}
{\cal L}_a u_{bc}&=-f^d{}_{ab} u_{dc}-f^d{}_{ac} u_{bd}
\\
{\cal L}_a v_{bc}&=-f^d{}_{ab} v_{dc}-f^d{}_{ac} v_{bd}
\\
{\cal L}_a z_b&=-f^d{}_{ab} z_d
\end{split}
\end{align}
on the couplings $u$,$v$ and $z$.

The action~\eqref{f11act} can be easily expanded in components
to find
\begin{align}
  \begin{split}
    S_g = \intd{^2x} \bigl(& 
    + u_{ab} F_{\plpl\mimi}^a F_{\plpl\mimi}^b
     - u_{ab} \nabla_\plpl W^a \nabla_\mimi W^b \\
&   + i u_{ab} \nabla_\plpl\chi_-^a \chi_-^b
      - i u_{ab} \chi_+^a \nabla_\mimi\chi_+^b \\
&      + \partial_i u_{ab} F_{\plpl\mimi}^a \lambda_+^i \chi_-^b
       - \partial_i u_{ab} \lambda_-^i \chi_+^a F_{\plpl\mimi}^b \\
&       + i \partial_i u_{ab} \lambda_+^i \chi_+^a \nabla_\mimi W^b 
      + i \partial_i u_{ab} \lambda_-^i \nabla_\plpl W^a \chi_-^b\\
&        + u_{ab} f^b_{\phantom{b}cd} \chi_+^a \chi_-^c W^d 
      - \nabla_i\partial_j u_{ab} \lambda_+^i \lambda_-^j \chi_+^a \chi_-^b
       - \partial_i u_{ab} \ell^i \chi_+^a \chi_-^b \\
&      + v_{ab} \chi_+^a \chi_-^b  
      + v_{ab} W^a F_{\plpl\mimi}^b \\
&     + \partial_i v_{ab} \lambda_+^i W^a \chi_-^b
      - \partial_i v_{ab} \lambda_-^i \chi_+^a W^b \\
&      + \frac{1}{2} \nabla_i\partial_j v_{ab} \lambda_+^i
         \lambda_-^j W^a W^b
       + \frac{1}{2} \partial_i v_{ab} \ell^i W^a W^b \\
&      + z_a F^a_{\plpl\mimi}
       +\partial_i z_a \lambda_+^i \chi_-^a
       -\partial_i z_a \lambda_-^i \chi_+^a\\
&     +  \ell^i\partial_i z_a W^a
      +\lambda_+^i \lambda_-^j \nabla_i\partial_j z_a W^a
    \bigr)\ .
  \end{split}
\end{align}

\subsection{The sigma model multiplet action and potential term}

A (1,1)-supersymmetric gauged sigma model action has been given in  \cite{Hull:1991uw}.
This action can be written as
\begin{align}
\begin{split}
  S_\sigma =& \intd{^2x\rd\theta^+\rd\theta^-}
    g_{ij} \nabla_+\phi^i \nabla_-\phi^j \\
  &+ \intd{^2x\rd t\rd\theta^+\rd\theta^-}\bigl(
  H_{ijk} \partial_t\phi^i \nabla_+\phi^j \nabla_-\phi^k -
  w_{ia} \partial_t\phi^i W_-^a
  \bigr)
\end{split}
\end{align}
This can be written without the $t$ integration as
\begin{align}
\begin{split}
  S_\sigma = \intd{^2x\rd\theta^+\rd\theta^-}\bigl(&
    g_{ij} \nabla_+\phi^i \nabla_-\phi^j
    + b_{ij} D_+\phi^i D_-\phi^j \\
&   - A_+^a w_{ia} D_-\phi^i
    - A_-^a w_{ia} D_+\phi^j
    + A_-^a A_+^b \xi_{[b}^i w_{a]i}
  \bigr)\ .
\end{split}
\end{align}

It is straightforward to add a potential term to the above actions as
\begin{align}
S_p=\intd{^2x\rd\theta^+\rd\theta^-} h
\end{align}
where $h=h(\phi)$ is a function of the superfield $\phi$.
Gauge invariance of the above action requires that $w$ should satisfy the
conditions stated in section \ref{sub:gaugecon}.

\subsection{A generalisation of the action}

The action of (1,1)-supersymmetric gauge theory presented above can be
generalized by allowing the various couplings of the theory to depend
on the scalar component of the gauge multiplet superfield. Supersymmetry
then requires  additional fermionic couplings. The
new theory can be organised as a (1,1)-supersymmetric sigma model
which has target space $L=M\times \LieG$, where $\LieG$
is the Lie algebra of the gauge group $G$. The various allowed
couplings are restricted by two-dimensional Lorentz invariance,
supersymmetry and gauge invariance as usual. The sigma model multiplet
is maps $Z=(\phi, W)$ from the (1,1) superspace $\Xi^{1,1}$ into
$L$, where $\phi$ is the usual (1,1) sigma model superfield and $W$
is the (1,1) gauge theory multiplet.
We again allow the gauge group  $G$ to act on $L$ with a group action
on $M$ and the adjoint action on $\LieG$.
The vector fields generated by such a group action are
\begin{align}
\xi_a=\xi_a^A\partial_A= \xi_a^i\partial_i+ W^b f^c{}_{ab} \partial_c
\label{nkill}
\end{align}
where $A=(i,a)$, the component $\xi^i$ is allowed to depend on both $\phi$ and $W$, and
the partial derivative with the gauge index denotes differential with
respect to $W$.

Next we introduce a metric $g$ and a Wess-Zumino term $H$ on $L$ and assume
that the gauge group $G$ acts on $L$ with isometries leaving the Wess-Zumino
term $H$ invariant. We also define $w$ as $i_{\xi_{a}}H=\rd w_a$, where $\xi_a$
is the new Killing vector field \eqref{nkill}.
Then an action can be written for this new sigma model as
\begin{align}
\begin{split}
  S_\sigma =& \intd{^2x\rd\theta^+\rd\theta^-}
   \big( g_{AB} \nabla_+Z^A \nabla_-Z^B+ h \big)\\
  &+ \intd{^2x\rd t\rd\theta^+\rd\theta^-}\bigl(
  H_{ABC} \partial_tZ^A \nabla_+Z^B \nabla_-Z^C -
  w_{Ba} \partial_tZ^B W_-^a
  \bigr)
\end{split}
\end{align}
where $h$ is a function which depends on $Z$.  This action is clearly
supersymmetric because it is a full (1,1) superspace integral.  Gauge
invariance requires that $w$ above satisfies all the conditions
stated in section \ref{sub:gaugecon} but for the group action
\eqref{nkill} and the Wess-Zumino term in $L$.  In addition, it
requires that
\begin{align}
{\cal L}_a h=0
\end{align}
where the Lie derivative is with respect to the vector field \eqref{nkill}.

\subsection{Scalar Potential}

To compute the scalar potential we express the action in components and
eliminate the auxiliary field of the sigma model superfield $\phi$
from the action using the field equations. The scalar potential is
\begin{align}
V(W,\phi)=\frac{1}{4} g^{ij} \partial_i h \partial_j h~,
\end{align}
where $g^{ij}$ is the inverse of the restriction of the metric of $L$ on $M$.
Observe that $V$ depends on both the sigma model scalar $\phi$ and
the gauge multiplet scalar $W$.
 The classical supersymmetric vacua
of the theory are those values of $(\phi,W)$ for which
$\partial_i h=0$. For example, for the special (1,1)-supersymmetric
model investigated in the beginning of the section,
$V=\frac{1}{4} g^{ij} (\partial_i h+\partial_i z_a W^a) (\partial_j h+\partial_j z_b W^b)$,
where in this case $h=h(\phi)$.

\section{$(2,1)$ supersymmetry}

\subsection{The gauge multiplet}

The $(2,1)$ superspace $\Xi^{2,1}$ has coordinates
$(x^\plpl,x^\mimi,\theta_{p}^+,\theta^-)$, where $(x^\plpl,x^\mimi)$ are the even and
$(\theta_{0}^+$,
$\theta_{1}^+$, $\theta^{-})$ are odd coordinates. The
(2,1)-supersymmetric Yang-Mills multiplet is described by a connection
$A$ in superspace with components $(A_\plpl,A_\mimi,A_{p+},A_-)$,
$p=0,1$. In addition it is required that these satisfy the
supersymmetry constraints \cite{Hull:1991uw}
\begin{align}
  [ \nabla_{p+}, \nabla_- ] &= W_p
& [ \nabla_\plpl, \nabla_\mimi ] &=  F_{\plpl\mimi}
\nonumber\\
  [ \nabla_{p+}, \nabla_{q+} ] &= 2i \delta_{pq} \nabla_\plpl
& [ \nabla_-, \nabla_- ] &=  2i\nabla_\mimi \ .
\end{align}
We have suppressed the gauge indices. We remark that $W_p$ are scalar
superfields.
The Jacobi identities imply that
\begin{gather*}
\begin{align}
[\nabla_{p+}, \nabla_\mimi ] &= i\nabla_-W_p
& [\nabla_-, \nabla_\plpl ] &= i\nabla_{0+} W_0
 \nonumber\\
  F_{\plpl\mimi}^a &= \nabla_{0+} \nabla_- W_0^a 
& \nabla_{1+}W_1&=\nabla_{0+}W_0
\end{align}\label{jac21c}  
\\
  \nabla_{1+}W_0+\nabla_{0+}W_1=0\ . 
\end{gather*}
The two scalar superfields $(W_0, W_1)$ can be viewed as a map $W$
from the (2,1) superspace $\Xi^{2,1}$ into $\LieG\otimes \bR^2$,
where $\LieG$ is the Lie algebra of the group $G$. Next introduce
a complex structure $I={\rm Id}\otimes \epsilon$ in $\LieG\otimes \bR^2$ 
where $\epsilon$ is the constant complex structure in $\bR^2$
with $\epsilon^0{}_1=-1$.  The last two conditions in \eqref{jac21c}
can be expressed as
\begin{align}
\nabla_{1+} W^{ap}=\epsilon^p{}_q \nabla_{0+}W^{aq}\ .
\label{g21con}
\end{align}
In fact this implies that $W$ is a covariantly chiral superfield, $(\nabla_{1+}+i\nabla_{0+}) (W_1+iW_0)=0$.

The components of the gauge superfields $W_p$ are
\begin{align}
  W_p &= W_p| 
& F_{\plpl\mimi} &= \nabla_{0+}\nabla_-W_0|
& f &= \nabla_{0+}\nabla_- W_1| \nonumber\\
  \chi_{0+} &= \nabla_{0+} W_0|
& \chi_{1+} &= \nabla_{0+} W_1 |
& \chi_{p-} &= \nabla_- W_p|
\ ,
\end{align}
where $W_p$ are scalars,  $\chi_{p+}$ and $\chi_{p-}$ are the gaugini which are real chiral fermions in two dimensions,
$F_{\plpl\mimi}$ is the field strength and $f$ is a real auxiliary  field.
As in the (1,1)-supersymmetric gauge theory, the gauge multiplet is determined by scalar superfields. This
will lead again to non-linear interactions because the various couplings of the theory can depend
on them.

\subsection{The sigma model multiplet}

Let $M$ be a KT manifold with metric $g$ and complex structure $J$.
We in addition assume that the gauge group $G$ acts on $M$ with
isometries which furthermore preserve the complex structure $J$ and
the Wess-Zumino term $H$. These conditions are the same as in the case
of (2,0)-supersymmetric model. The (2,1) sigma model superfield $\phi$
is a map from the (2,1) superspace $\Xi^{2,1}$ into the sigma model
manifold $M$.  In addition it is required that
\begin{align}
\nabla_{1+}\phi^i= J^i{}_j \nabla_{0+}\phi^j\ ,
\label{s21con}
\end{align}
where $\nabla_{p+}\phi^i=D_{p+}\phi^i+ A^a_{p+} \xi^i_a$ and $\xi_a$
are the vector fields on $M$ generated by the group action. As we have
seen, the (2,1) gauge multiplet satisfies the condition \eqref{g21con}
similar to \eqref{s21con}.  The superfield $\phi$ is also covariantly
chiral, as can be seen by choosing complex coordinates on the sigma
model manifold $M$. These results will be used later for the
construction of actions of (2,1)-supersymmetric gauge theories coupled
to sigma models.

The components of the sigma model multiplet $\phi$ are as follows:
\begin{align}
  \phi^i &= \phi^i|
& \ell^i &= \nabla_{0+}\nabla_-\phi^i|\nonumber \\
  \lambda^i_{+}=&\nabla_{0+}\phi^i|
& \lambda^i_{-}=&\nabla_{-}\phi^i| ~,
\end{align}
where $\phi$ is a scalar, $\lambda_+$ and $\lambda_-$ are real fermions, and $\ell$ is an auxiliary
field.

\subsection{Action}

An action of a (2,1)-supersymmetric gauge theory coupled to sigma model matter can be written
as
\begin{align}
  S &= S_g + S_\sigma+S_p  \ ,
\end{align}
where $S_g$ is the action of the gauge multiplet, $S_\sigma$ is the
action of the sigma-model multiplet and $S_p$ contains the potential term.
 We shall describe each term separately.

\subsection{The gauge multiplet action}

An action for the (2,1)-supersymmetric gauge multiplet
 is
\begin{align}
\label{eq:cpt21action}
  \begin{split}
    S_g=\intd{^2x\rd\theta_{0}^+\rd\theta^-}  \bigl(&
       - u^0_{ab}\delta^{pq} \nabla_{0+}W_p^a \nabla_- W_q^b
        \\
&     + u^1_{ab} \delta^{pq} \nabla_{1+}W_p^a \nabla_- W_q^b+ z^p_a W^a_p
  \bigr)
  \end{split}
\end{align}
where $u^0, u^1$ are the gauge coupling constants  and $z^p$ are theta term type of couplings. All the
couplings are allowed to depend on the superfield $\phi$. We shall assume that
both $u^0, u^1$ are symmetric in the gauge indices but this
restriction can be lifted.

Observe that this action is not an integral over the full $\Xi^{2,1}$
superspace.  Therefore it is not manifestly (2,1)-supersymmetric.  The
requirement of invariance under (2,1) supersymmetry imposes the
conditions
\begin{align}\begin{split}
J^j{}_i\partial_j u^0_{ab}&= -\partial_i u^1_{ab}
\\
J^j{}_i\partial_j z^1_{a}&= -\partial_i z^0_{a}
 \ .
\end{split}\end{align}
Therefore the couplings $u^0+i u^1$ and $z^1+i z^0$ are holomorphic.

In addition gauge invariance of the action \eqref{eq:cpt21action} implies that
\begin{align}\begin{split}
{\cal L}_a u^p_{bc}&= -f^d{}_{ab} u^p_{dc}- f^d{}_{ac} u^p_{bd}
\\
{\cal L}_a z^p_{b}&= -f^d{}_{ab} z^p_{d}
\ .
\end{split}\end{align}

\subsection{The sigma model multiplet action and potential}

The action of the gauged (2,1)-supersymmetric sigma model with Wess-Zumino
term has been given in
\cite{Hull:1991uw}.  Here we shall summarise some of results
relevant to this chapter. The action of this multiplet is
\begin{align}
\begin{split}
  S_\sigma =& \intd{^2x\rd\theta_{0}^+\rd\theta^-}
    \big(g_{ij} \nabla_{0+}\phi^i \nabla_-\phi^j+\nu_a W^a_1 \bigr) \\
  &+ \intd{^2x\rd t\rd\theta^+\rd\theta^-}\bigl(
  H_{ijk} \partial_t\phi^i \nabla_{0+}\phi^j \nabla_-\phi^k -
  w_{ia} \partial_t\phi^i W_-^a
  \bigr)
\end{split}
\end{align}
This can be written without the $t$ integration as
\begin{align}
\begin{split}
  S_\sigma = \intd{^2x\rd\theta^+\rd\theta^-}\bigl(&
    g_{ij} \nabla_+\phi^i \nabla_-\phi^j
    + b_{ij} D_+\phi^i D_-\phi^j \\
&   - A_+^a w_{ia} D_-\phi^i
    - A_-^a w_{ia} D_+\phi^j
    + A_-^a A_+^b \xi_{[b}^i w_{a]i}
  \bigr)
\end{split}
\end{align}

The gauge transformations of $\phi$ are $\delta\phi^i =
\epsilon^a\xi_a^{\phantom{a}i}(\phi)$.  Gauge invariance of the above
action requires that $w$ should satisfy the conditions described in
section \ref{sub:gaugecon}. As in the case of (2,0)-supersymmetric
gauged sigma model, $\nu$ should satisfy $\mathcal{L}_a \nu_b =-
f_{ab}^{\phantom{ab}c} \nu_c $. In fact $\nu$ is a moment map
associated with the action of the gauge group on the KT manifold $M$.

The part of the action involving the potential is
\begin{align}
S_p=\intd{^2x \rd\theta_{0}^+ \rd\theta^{-}} h\ ,
\end{align}
where $h=h(\phi)$
Invariance under (2,1) supersymmetry requires that
\begin{align}
\partial_i h= J^k{}_i \partial_k h^1
\end{align}
where $h^1 =h^1(\phi)$. This implies that $h$ is the real part of a holomorphic function on $M$.

The scalar potential of (2,1)-supersymmetric gauge theories  coupled to sigma models described above is
\begin{align}
V={1\over4}  u_0^{ab} (\nu_a+ z_a^1) (\nu_b+z_b^1)+ {1\over4} g^{ij} (\partial_i h+\partial_i z^p_a W^a_p)
 (\partial_j h++\partial_j z^p_a W^a_p)\ .
\label{morepot}
\end{align}

\subsection{A generalisation}

As we have shown both the (2,1) gauge multiplet and the (2,1) sigma model multiplet
are constructed from covariantly chiral scalar superfields, ie both satisfy the  conditions
\eqref{s21con} and \eqref{g21con}. Because of this, these two superfields can be
combined to a single superfield $Z=( W_0, W_1, \phi)$ which
is a map from the $(2,1)$ superspace $\Xi^{2,1}$
into $(\LieG\otimes \bR^2)\times M$. In addition we can take $Z$ to satisfy
a chirality condition which is the combination of \eqref{g21con} and \eqref{s21con}.
Next we can take the gauge group $G$ to act on $(\LieG\otimes \bR^2)\times M$
with the adjoint action in the first factor and a group action on $M$. The
vector fields associated by such a group action are
\begin{align}
\xi_a= \sum_p f^c{}_{ab} W_p^b {\partial\over\partial W^c_p} +\xi^i \partial_i\ .
\end{align}

Treating the (2,1)-supersymmetric gauge theory coupled to sigma model matter as
a sigma model with superfield $Z$, which satisfies \eqref{g21con} and \eqref{s21con},
we can write the action
\begin{align}
\begin{split}
  S_\sigma =& \intd{^2x\rd\theta_{0}^+\rd\theta^-}
   \big( g_{AB} \nabla_{0+}Z^A \nabla_-Z^B+ \nu_a W^a_1+ h \big)\\
  &+ \intd{^2x\rd t\rd\theta^+\rd\theta^-}\bigl(
  H_{ABC} \partial_tZ^A \nabla_{0+}Z^B \nabla_-Z^C -
  w_{Ba} \partial_tZ^B W_-^a
  \bigr)
\end{split}
\label{g21act}
\end{align}
where now all the couplings are defined using the geometry of $(\LieG\otimes \bR^2)\times M$.
Of course (2,1)-supersymmetric requires that $(\LieG\otimes \bR^2)\times M$ is a
KT manifold with respect to the complex structure $(J, {\rm id}\otimes \epsilon)$.
In particular the metric and the rest of the couplings depend on the coordinates of
$(\LieG\otimes \bR^2)\times M$. The conditions for gauge invariance are easily
determined from those of the (2,1)-supersymmetric gauged sigma model.
We remark that the couplings of \eqref{g21act} can be arranged such that the $SO(2)$ R-symmetry of the
$(2,1)$-supersymmetry algebra is broken. In particular the $SO(2)$ rotation that rotate
the $W_p$ scalar components is not a symmetry of the action. However if one insists in preserving
the R-symmetry, then the KT manifold $(\LieG\otimes \bR^2)\times M$ should admit
a $SO(2)$ action preserving all the geometric data.

\section{$(4,1)$ supersymmetry}

\subsection{The gauge multiplet}

The $(4,1)$ superspace $\Xi^{4,1}$ has coordinates
$(x^\plpl,x^\mimi,\theta_{p}^+,\theta^-)$, where $(x^\plpl,x^\mimi)$ are the even and $\{\theta^-,\theta_{p}^+,
p=0,\ldots,3\}$ are the odd coordinates.
The (4,1)-supersymmetric Yang-Mills multiplet is described by a
connection $A$ in $\Xi^{4,1}$ superspace with components
$(A_\plpl,A_\mimi,A_{p+},A_-)$ with $p=0,\ldots 3$. In addition it is
required that these satisfy the supersymmetry constraints \cite{Hull:1991uw}
\begin{gather*}
  \begin{align}
  [ \nabla_{p+}, \nabla_- ] &= W_p
& [ \nabla_\plpl, \nabla_\mimi ] &=  F_{\plpl\mimi} \nonumber\\
  [ \nabla_{p+}, \nabla_{q+} ] &= 2i \delta_{pq} \nabla_\plpl
& [ \nabla_-, \nabla_- ] &=  2i\nabla_\mimi
\end{align}  \label{con41v}  
\\
  \nabla_{p+} W_q = \epsilon_{pq}^{\phantom{pq}p'q'} \nabla_{p'+} W_{q'}\ .
\end{gather*}
(We have suppressed all the gauge indices.)
The Jacobi identities imply that
  \begin{gather*}
  \begin{align}
    [\nabla_{p+}, \nabla_\mimi ] &= i\nabla_-W_p \nonumber\\
    [\nabla_-, \nabla_\plpl ] &= i\nabla_{0+} W_0 \nonumber\\
    F_{\plpl\mimi}^a &= \nabla_{0+} \nabla_- W_0^a   
  \end{align}  \label{jac41con}
\\
\nabla_{p+} W_{q}+ \nabla_{q+} W_p=0 \quad (p\neq q) \nonumber
\\
\nabla_{0+} W_{0}=\nabla_{1+} W_{1}= \nabla_{2+} W_{2}=\nabla_{3+}
W_{3}
  \end{gather*}

The (4,1) gauge multiplet is determined by four scalar superfields. Some of the conditions
on these superfields given in \eqref{jac41con}, like in the (2,1) model previously, can be
expressed as conditions of a (4,1) sigma model multiplet.  For this view the four-scalar
superfields $W_p$ as maps from the superspace $\Xi^{4,1}$ into $\LieG\otimes \bR^4$, where $\LieG$
is the Lie algebra of the gauge group $G$. Then introduce three constant complex structures $\{I_r \}$
in $\bR^4$ such that $(I_r)^0{}_s=\delta_{rs}$ and $(I_r)^s{}_t=- \epsilon_{rst}$ where
$r,s,t=1,2,3$. The conditions on $W_p$ in \eqref{con41v} and \eqref{jac41con} can be expressed
as
\begin{align}
\nabla_{r+} W^a_p= I_r{}^q{}_p \nabla_{0+}W^a_q\ .
\label{g41con}
\end{align}

The components of the gauge multiplet are
\begin{gather}
\begin{align}
  W_p &= W_p| 
& F_{\plpl\mimi} &= \nabla_{0+}\nabla_-W_0| \nonumber\\
  \chi_{p-} &= \nabla_- W_p| 
& \chi_{p+} &= \nabla_{0+} W_p |
 \end{align}
 \\
 f_r = \nabla_{0+}\nabla_- W_r| \quad r=1,2,3\ , \nonumber
\end{gather}
where $W_p$ are scalars, $\chi_{p+}$,$\chi_{p-}$ are the
gaugini which are real chiral fermions in two dimensions,
$F_{\plpl\mimi}$ is the field strength and $\{f_{r}: r=1,2,3\}$ are auxiliary
fields. The $SO(4)$ R-symmetry of the (4,1)-supersymmetric gauge theory rotates
both the scalars and the fermions of the gauge multiplet.

\subsection{The sigma model multiplet}

Let $M$ be a HKT manifold with metric $g$ and hypercomplex structure $\{J_r ; r=1,2,3\}$.
We in addition assume that the gauge group $G$ acts on $M$ with isometries which in
addition preserve the hypercomplex structure $J_r$ and the Wess-Zumino term $H$. These
conditions are the same as in the case of (4,0)-supersymmetric model. The (4,1) sigma model
superfield $\phi$ is a map  from the (4,1) superspace $\Xi^{4,1}$ into the sigma model manifold $M$.
In addition it is required that
\begin{align}
\nabla_{r+}\phi^i= J_r{}^i{}_j \nabla_{0+}\phi^j\ ,
\label{s41con}
\end{align}
where $\nabla_{p+}\phi^i=D_{p+}\phi^i+ A^a_{p+} \xi^i_a$ and $\xi$ are the vector
fields on $M$ generated by the group action. As we have seen the
(4,1) gauge multiplet satisfies the condition \eqref{g41con} similar to \eqref{s41con}.
 These results
will be used later for the construction of actions of (4,1)-supersymmetric
gauge theories coupled to sigma models.

The components of the sigma model (4,1) multiplet $\phi$ are as follows,
\begin{align}
  \phi^i &= \phi^i|
& \ell^i &= \nabla_{0+}\nabla_-\phi^i| \nonumber\\
  \lambda^i_{+}&=\nabla_{0+}\phi^i|
& \lambda^i_{-}&=\nabla_{-}\phi^i| ~,
\end{align}
where $\phi$ is a scalar, $\lambda_+$ and $\lambda_-$ are real fermions, and $\ell$ is an auxiliary
field.

\subsection{Action}

The action of a (4,1)-supersymmetric gauged theory coupled to sigma model
matter can be written
as
\begin{align}
  S &= S_g + S_\sigma+ S_p  \ ,
\end{align}
where $S_g$ is the action of the gauge multiplet, $S_\sigma$ is the
action of the sigma-model multiplet and $S_p$ is the potential. We
shall describe each term separately.

\subsection{The gauge multiplet action}

An action for the (4,1)-supersymmetric gauge multiplet
 is
\begin{align}
\label{eq:cpt41action}
  \begin{split}
    S_g=\intd{^2x\rd\theta_{0}^+} \bigl(&
       - u^0_{ab}\delta^{pq} \nabla_{0+}W_p^a \nabla_- W_q^b
        \\
&     + u^r_{ab} \delta^{pq} \nabla_{r+}W_p^a \nabla_- W_q^b+ z_a^p W^a_p
  \bigr)
  \end{split}
\end{align}
where $\{u^p\}=\{u^0, u^r\}$ and $z^p$ are the gauge coupling
constants and theta type of terms, respectively, which in general
depend on the superfield $\phi$. We shall assume that both $u^p$ are
symmetric in the gauge indices but this restriction can be lifted.

Observe that this action is not an integral over the full $\Xi^{4,1}$
superspace.  Therefore it is not manifestly (4,1)-supersymmetric.  The
requirement of invariance under (4,1) supersymmetry imposes the
condition
that $u^p$ and $z^p$ are constant. This is similar to the condition that arises in
(4,0) supersymmetric gauge theories. Gauge invariance of the action
\eqref{eq:cpt41action} in addition requires
\begin{align}\begin{split}
 f^d{}_{ab} u^p_{dc}+ f^d{}_{ac} u^p_{bd}&=0
 \\
  f^d{}_{ab} z^p_d&=0
 \ .
\end{split}\end{align}
Thus $u^p$ must be invariant quadratic forms on the Lie algebra of
the group $G$ and $z^p$ must be  invariant elements of the Lie algebra.
Of course $z^p=0$, if $G$ is semi-simple.

\subsection{The sigma model multiplet action and the potential}

An action for the (4,1) sigma model multiplet coupled to gauge fields
has been given in
\cite{Hull:1991uw}.  Here we shall summarise the some of results
relevant to this chapter.  The action of this multiplet is
\begin{align}
\begin{split}
  S_\sigma =& \intd{^2x\rd\theta^+\rd\theta^-}
   \big( g_{ij} \nabla_+\phi^i \nabla_-\phi^j+\sum_r \nu^r_a W^a_r\big) \\
  &+ \intd{^2x\rd t\rd\theta^+\rd\theta^-}\bigl(
  H_{ijk} \partial_t\phi^i \nabla_+\phi^j \nabla_-\phi^k -
  w_{ia} \partial_t\phi^i W_-^a
  \bigr)
\end{split}
\end{align}

The gauge transformations of $\phi$ are $\delta\phi^i =
\epsilon^a\xi_a^{\phantom{a}i}(\phi)$.  Gauge invariance of the above
action requires that $w$ should satisfy the conditions described in
section \ref{sub:gaugecon}. As in the case of (4,0)-supersymmetric
gauged sigma model, $\nu^r$ should satisfy $\mathcal{L}_a \nu^r_b =-
f_{ab}^{\phantom{ab}c} \nu^r_c $. In fact $\nu^r$ is a moment map
associated with the action of the gauge group $G$ on the HKT manifold
$M$.

The part of the action involving the potential is
\begin{align}
S_p=\intd{^2x \rd\theta_{0}^+ \rd\theta^{-}} h\ ,
\end{align}
where $h=h(\phi)$.
Invariance under (4,1) supersymmetry requires that
\begin{align}
\partial_i h= J_r{}^k{}_i \partial_k h^r
\end{align}
where $h^r =h^r(\phi)$. This implies that $h$ is the real
part of three holomorphic functions on $M$, ie $h$
is tri-holomorphic.

The scalar potential of (4,1)-supersymmetric gauge theories coupled to sigma models is
\begin{align}
V={1\over4}  u_0^{ab}\sum_{r=1}^3 \nu^r_a \nu^r_b+ {1\over4} g^{ij} \partial_i h \partial_j h\ ,
\label{more41pot}
\end{align}
where we have shifted the moment maps $\nu_r$ by a constant $z_r$.

\subsection{A generalisation}

As we have shown both the (4,1) gauge multiplet and the (4,1) sigma
model multiplet are constructed from scalar superfields which satisfy
the similar constraints \eqref{g41con} and \eqref{s41con}. Because of
this, these two superfields can be combined to a single superfield
$Z=( W, \phi)$ which is a map from the $(4,1)$ superspace $\Xi^{4,1}$
into $(\LieG\otimes \bR^4)\times M$. In addition we can take $Z$
to satisfy a condition which is the combination of \eqref{s41con} and
\eqref{g41con}.  Next we can take the gauge group $G$ to act on
$(\LieG\otimes \bR^4)\times M$ with the adjoint action in the
first factor and a group action on $M$. The vector fields associated
by such a group action are
\begin{align}
\xi_a= \sum_p f^c{}_{ab} W_p^b {\partial\over\partial W^c_p} +\xi^i \partial_i\ .
\end{align}

Treating the (4,1)-supersymmetric gauge theory coupled to sigma model
matter as a sigma model with superfield $Z$, which satisfies
\eqref{g41con} and \eqref{s41con}, we can write the action
\begin{align}
\begin{split}
  S_\sigma =& \intd{^2x\rd\theta_{0}^+\rd\theta^-}
   \big( g_{AB} \nabla_{0+}Z^A \nabla_-Z^B+\sum_r \nu^r_a W^a_r+ h \big)\\
  &+ \intd{^2x\rd t\rd\theta^+\rd\theta^-}\bigl(
  H_{ABC} \partial_tZ^A \nabla_{0+}Z^B \nabla_-Z^C -
  w_{Ba} \partial_tZ^B W_-^a
  \bigr)
\end{split}
\label{g41act}
\end{align}
where now all the couplings are defined using the geometry of
$(\LieG\otimes \bR^4)\times M$.  Of course (4,1)-supersymmetry
requires that $(\LieG\otimes \bR^4)\times M$ is a HKT manifold with
respect to the hypercomplex structure $(J_r, I_r)$.  In particular the
metric and the rest of the couplings depend on the coordinates of
$(\LieG\otimes \bR^2)\times M$. The conditions for gauge
invariance are easily determined from those of the
(4,1)-supersymmetric gauged sigma model.  We remark that the couplings
of \eqref{g41act} can be arranged such that the $SO(4)$ R-symmetry of
the $(4,1)$-supersymmetry algebra is broken. In particular the $SO(4)$
rotation that rotates the $W_p$ scalar components is not a symmetry of
the action. However if one insists in preserving the R-symmetry, then
the HKT manifold $(\LieG\otimes \bR^4)\times M$ should admit a
$SO(4)$ action preserving all the geometric data.

\section{Summary}

We now summarise the conditions given in this chapter. The conditions
for the general case are
\begin{align}
  \mathcal{L}_a g&=0 \\
 \mathcal{L}_a H &=0  \\
 \mathcal{L}_a B_{i\phantom{A}B}^{\phantom{i}A}&=
 -\nabla_i U_{a\phantom{A}B}^{\phantom{a}A}\\
  \mathcal{L}_c h_{AB}&=-U_a{}^C{}_A h_{CB}-U_a{}^C{}_B h_{AC}
  \\
{\cal L}_a w_b&=-f_{ab}{}^c w_c \\
\xi^i_a w_{ib}+\xi^i_b w_{ia}&=0\\
  \mathcal{L}_c u_{ab} +
 u_{db}f^d_{\phantom{d}ca}+u_{ad}f^d_{\phantom{d}cb}&=0 \\
 \mathcal{L}_a\mu_b - [U_a,\mu_b] &=
  -f_{ab}^{\phantom{ab}c}\mu_c \\
\mathcal{L}_a V &=0
\end{align}

The conditions in the cases of $(p,0)$ and $(p,q)$ supersymmetry are
given in the following tables.

\begin{table}[h]
\small
  \centering
  \begin{tabular}{ccc}
\hline\hline
Supersymmetry & Multiplets & Extra Constraints  \\
\hline
(1,0) & $(\phi^i,\lambda_+^i)$ &  $[ \nabla_+, \nabla_+ ] = 2i\nabla_\plpl$
 \\
& ($\chi_-^a$,$F_{\plpl\mimi}$) & $[ \nabla_\plpl, \nabla_\mimi ] = F_{\plpl\mimi}$ 
\\
& ($\psi^A_-$,$\ell^A$)  & $[ \nabla_+, \nabla_\mimi ] = W_- \ $ \\
& $F_{\plpl\mimi}^a$ &  $\mathcal{L}_a s_A = -U^{\phantom{a}B}_{a\phantom{B}A} s_B$ \\
&& ${\cal L}_a z_b=-f_{ab}{}^c z_c$ \\

\\
(2,0) & $(\phi^i,\lambda_+^i)$ & $[ \nabla_{p+},  \nabla_{q+} ] = 2i\delta_{pq}\nabla_\plpl$ \\
& ($\psi^A_-$,$\ell^A$) & $[ \nabla_\plpl, \nabla_\mimi ] = F_{\plpl\mimi}$ \\
& $(\chi_{0-}^a,\chi_{1-}^a, F_{\plpl\mimi}^a, f^a)$ & $[ \nabla_{p+},  \nabla_\mimi ] = W_{p-}$ \\
&& $\nabla_{1+}\phi^i= J^i{}_j \nabla_{0+}\phi^j$ \\
&& $\mathcal{L}_a J=0$ \\
&&   $\Nabla_{p+}\psi_-^A=
  D_{p+}\psi^A+\nabla_{p+}\phi^i B_i{}^A{}_B \psi_-^B$ \\
&& $\quad + A^a_{p+}  U_a{}^A{}_B$ \\
&& ${\cal L}_a I^A{}_B= U_a{}^A{}_C I^C{}_B-I^A{}_C U_a{}^C{}_B$ \\
&& ${\cal L}_a L^A=  U_a{}^A{}_B L^B$ \\
&& $G_{kl\phantom{A}B}^{\phantom{kl}A} \ud Jku\ud Jlj=G_{ij\phantom{A}B}^{\phantom{ij}A}$\\
&& $J^k{}_i\nabla_k L^A- I^A{}_B \nabla_iL^B=0$ \\
&&$J^k{}_i\nabla_k I^A{}_B- I^A{}_C\nabla_iI^C{}_B=0$\\

\hline
  \end{tabular}
\end{table}

\begin{table}[h]
\small
  \centering
  \begin{tabular}{ccc}
\hline\hline
Supersymmetry & Multiplets & Extra Constraints  \\
\hline
(4,0) & 
  $(\phi^i,\lambda_+^i)$ &
  $[ \nabla_{p+},  \nabla_{q+} ] = 2i\delta_{pq}\nabla_\plpl$ \\

& ($\psi^A_-$,$\ell^A$) 
& $[ \nabla_\plpl, \nabla_\mimi ] = F_{\plpl\mimi}$ \\

& $(\chi_{0-},\chi_{1-},\chi_{2-},\chi_{3-},$ & $[ \nabla_{p+},  \nabla_\mimi ] = W_{p-}$ \\
&  $F_{\plpl\mimi}^a,f^a_1,f^a_2,f^a_3)$  & $\nabla_{p+} W_{q-}=\frac{1}{2} \epsilon_{pq}{}^{p'q'}\nabla_{p'+}
W_{q'-}$ \\
&& $\nabla_{r+}\phi^i= J_r{}^i{}_j \nabla_{0+}\phi^j$ \\
&& ${\cal L}_a I_r^A{}_B= U_a{}^A{}_C I_r^C{}_B-I_r^A{}_C U_a{}^C{}_B$\\
&& ${\cal L}_a L_r^A=  U_a{}^A{}_B L_r^B$ \\
&& $G_{kl}{}^A{}_B J_{r}{}^k{}_i J_{s}{}^l{}_j+G_{kl}{}^A{}_B
J_{s}{}^k{}_i J_{r}{}^l{}_j$ \\
&& $=2\delta_{rs}G_{ij}{}^A{}_B$\\
&&$J_r{}^j{}_i \nabla_j L_s^A+ J_s{}^j{}_i \nabla_j L_r^A$ \\
&& $- I_r{}^A{}_B
\nabla_i L_s^B-I_s{}^A{}_B \nabla_i L_r^B =0$\\
&& $ J_{p\phantom{j}i}^{\phantom{p}j} \partial_j u_q
  = \frac{1}{2} \epsilon_{pq}{}^{p'q'}
     J_{p'}{}^j{}_i \partial_ju_{q'}
  \quad (p\neq q)$   \\
&& $ \partial_i u_0
    = J_{1\phantom{j}i}^{\phantom{1}j} \partial_j u_1
    = J_{2\phantom{j}i}^{\phantom{2}j} \partial_j u_2
    = J_{3\phantom{j}i}^{\phantom{3}j} \partial_j u_3$ \\
&& $J_r{}^j{}_i \partial_jz^r=-\partial_i z^0$ \\
&& $J_p{}^j{}_i \partial_j z_q=-{1\over2} \epsilon_{pq}{}^{p'q'} J_{p'}{}^k{}_i \partial_k z_{q'}$\\
&& $f^d{}_{ab} u^p_{dc}+ f^d{}_{ac} u^p_{bd}=0$ \\
&& $f^c{}_{ab} z^p_{c}=0$ \\
&& $\mathcal{L}_a \nu_{ra} = -f_{ab}^{\phantom{ab}c} \nu_{rc}$ \\
&& $h_{CB} I_r^C{}_A+h_{CA} I_r^C{}_B=0$ \\
&& $J_r^j{}_i \nabla_j h_{AB}+\nabla_i h_{AC} I_r^C{}_B=0$ \\
&& $J_r^j{}_i \nabla_j s_A-\nabla_i (s_B I_r^B{}_A)-{1\over2}
\nabla_i h_{AB} L_r^B=0$ \\
&& $s_A L_r^A={\rm const}$\\

\hline
  \end{tabular}
\end{table}

\begin{table}[h]
\small
  \centering
  \begin{tabular}{ccc}
\hline\hline
Supersymmetry & Multiplets & Extra Constraints \\
\hline
(1,1) &   
$(\phi^i,\lambda_+^i,\lambda_-^i,\ell^i)$  
& $[ \nabla_+, \nabla_- ] = W$ \\
& $(W^a, \chi_+^a,\chi_-^a,F_{\plpl\mimi}^a)$ 
& $[ \nabla_\plpl, \nabla_\mimi ]=F_{\plpl\mimi}$ \\
&& $[ \nabla_+, \nabla_+ ] = 2i\nabla_\plpl$  \\
&& $[ \nabla_-, \nabla_-]=2i\nabla_\mimi$  \\
&& ${\cal L}_a u_{bc}=-f^d{}_{ab} u_{dc}-f^d{}_{ac} u_{bd}$ \\
&& ${\cal L}_a v_{bc}=-f^d{}_{ab} v_{dc}-f^d{}_{ac} v_{bd}$ \\
&& ${\cal L}_a z_b=-f^d{}_{ab} z_d$ \\
&& ${\cal L}_a h=0$ \\

\\
(2,1)
& $(\phi^i,\lambda_+^i,\lambda_-^i,\ell^i)$ 
& $[ \nabla_{p+}, \nabla_- ] = W_p$  \\
& $(W_0^a,W_1^a, \chi_{0+}^a, \chi_{1+}^a,\chi_-^a,F_{\plpl\mimi}^a,f^a)$ 
& $[ \nabla_\plpl, \nabla_\mimi ] =  F_{\plpl\mimi}$ \\
&& $[ \nabla_{p+}, \nabla_{q+} ] = 2i \delta_{pq} \nabla_\plpl$ \\
&& $[ \nabla_-, \nabla_- ] =  2i\nabla_\mimi$ \\
&& $\nabla_{1+}\phi^i= J^i{}_j \nabla_{0+}\phi^j$ \\
&& $J^j{}_i\partial_j u^0_{ab}= -\partial_i u^1_{ab}$ \\
&& $J^j{}_i\partial_j z^1_{a}= -\partial_i z^0_{a}$ \\
&& ${\cal L}_a u^p_{bc}= -f^d{}_{ab} u^p_{dc}- f^d{}_{ac} u^p_{bd}$\\
&& ${\cal L}_a z^p_{b}= -f^d{}_{ab} z^p_{d}$ \\
&& $\partial_i h= J^k{}_i \partial_k h^1$ \\

\\
(4,1) 
& $(\phi^i,\lambda_{p+},\lambda_-,\ell)$ 
& $[Q_-,Q_-]_+ = 2i P_\mimi$ \\
& $(W_p^a.\chi_{p-}^a,\chi_+^a,F_{\plpl\mimi}^a,f_r^a)$ 
& $[Q_{p+}, Q_{q+} ]_+ = 2i\delta_{pq}P_\plpl$ \\ 
&& $\nabla_{r+}\phi^i= J_r{}^i{}_j \nabla_{0+}\phi^j$ \\
&& $f^d{}_{ab} u^p_{dc}+ f^d{}_{ac} u^p_{bd}=0$ \\
&& $f^d{}_{ab} z^p_d=0$ \\
&& $\partial_i h= J_r{}^k{}_i \partial_k h^r$\\
& ($r=1,2,3$) & ($p,q=0,1,2,3$) \\
\hline
  \end{tabular}
\end{table}

\chapter{Vortices and Equivariant cohomology}

\section{Introduction}

Vortices are the instantons of two-dimensional gauge theories coupled
to sigma models. Bogomol'nyi type of bounds for both abelian
\cite{Bogomolny:1976de} and non-abelian vortices \cite{Bradlow:1989,
  Bradlow:1995vn} have been investigated in the context of linear
sigma models. To describe them, it is instructive to begin with a toy
example.

Let $\Xi$ be the usual two-dimensional spacetime, which after a Wick
rotation we can take to be Euclidean, $\Xi=\bR^2$ with metric
$\delta^{\mu\nu}$. The sigma model fields are maps
$\phi:\Xi\rightarrow M$ to a sigma model manifold $M$, which in this
introduction we take to be $M=\bR^2$. The gauge group $G$ acts on $M$
with isometries, for simplicity we take $G$ to be abelian.

The Yang-Mills-Higgs action is
\begin{align}
\label{eq:yangmillshiggs}  S = \intd{^2x} \Bigl(
    \frac{1}{2} \delta_{ij} \delta^{\mu\nu}\nabla_\mu \phi^i \nabla_\nu \phi^j
    + \frac{1}{4} F^{\mu\nu}F_{\mu\nu}
    + \frac{\lambda}{8}(\phi^2-1)^2
    \Bigr)
\end{align}
where $\nabla_\mu\phi^i=\partial_\mu\phi^i+A_\mu\cdot\phi^i$ is ths
usual covariant derivative, and $\phi^2=\delta_{ij}\phi^i\phi^j$. The
final term is the Higgs self-interaction term, and $\lambda\geq 0$ is
constant.

This action admits finite action, solotonic
configurations~\cite{Jaffe:1980mj}, and these configurations are
precsiely those that saturate the following bound, due to
Bogomol'nyi~\cite{Bogomolny:1976de}. We
rewrite~\eqref{eq:yangmillshiggs} as
\begin{align}
\label{eq:ymhbogomolnyi}
\begin{split}    
    S=\intd{^2x}\Bigl(& 
    \delta_{ij}\delta_{\mu\nu}
    (\epsilon^{\mu\rho}\nabla_\rho\phi^i\pm\ud\epsilon{i}{k}\nabla_\mu\phi^k)
    (\epsilon^{\nu\sigma}\nabla_\sigma\phi^j\pm\ud\epsilon{j}{l}\nabla_\mu\phi^l) \\
    &+ \frac{1}{4}(F_{\mu\nu}\mp\epsilon_{\mu\nu}(1-\phi^2))
  (F^{\mu\nu}\mp\epsilon^{\mu\nu}(1-\phi^2))\\
  &+\Bigl[\frac{1}{2}F_{\mu\nu}\epsilon^{\mu\nu}(1-\phi^2) \pm
  \epsilon_{\mu\nu}\epsilon^{ij}\nabla_\mu\phi^i \nabla_\nu\phi^j\Bigr]
  \Bigr)
  \end{split}
\end{align}
where $\epsilon$ is the alternating tensor.  This expression is
obtained by completing squares and collecting the remaining terms
together in the square brackets above. Finiteness of the
action~\eqref{eq:yangmillshiggs} requires that
\begin{align}
\label{eq:ymhbcnds}
  |\phi|&\rightarrow 1, & \nabla\phi&\rightarrow 0 &&\text{as }|x|\rightarrow\infty~,
\end{align}
and with these boundary conditions, using a partial integration, the
action~\eqref{eq:ymhbogomolnyi} may be rewritten as
\begin{align}
\label{eq:ymhbogomolnyi2}
    S&=\intd{^2x}\Bigl(
    \delta_{ij}\delta_{\mu\nu}
    (\epsilon^{\mu\rho}\nabla_\rho\phi^i\pm\ud\epsilon{i}{k}\nabla_\mu\phi^k)
    (\epsilon^{\nu\sigma}\nabla_\sigma\phi^j\pm\ud\epsilon{j}{l}\nabla_\mu\phi^l) \nonumber\\
    &\mspace{80mu} + \frac{1}{4}(F_{\mu\nu}\mp\epsilon_{\mu\nu}(1-\phi^2))
  (F^{\mu\nu}\mp\epsilon^{\mu\nu}(1-\phi^2))\Bigr)\\
  &\phantom{=}\pm 2\intd{^2x} F_{12}~.\nonumber
\end{align}
The last term is a topological charge proportional to the first Chern
character, which is an integer $N$ (assuming sufficiently uniform
limits in~\eqref{eq:ymhbcnds}). $N$ is also called the \emph{vortex
  number} and solutions with $N>0$ are called $N$-vortex solutions.

Because the other integrands are non-negative, we have that
\begin{align}
  S \geq \pi |N|
\end{align}
because we may always choose the signs in~\eqref{eq:ymhbogomolnyi2} so
that the topological charge is positive. For $N\geq 0$, there is
equality if and only if
\begin{align}
  \begin{split}
    \epsilon^{\mu\rho}\nabla_\rho\phi^i+\ud\epsilon{i}{k}\nabla_\mu\phi^k&=0 \\
  F^{\mu\nu}-\epsilon^{\mu\nu}(1-\phi^2)&=0~.
    \end{split}
\end{align}
The appearance of the alternating tensor $\epsilon^{ij}$ suggests that
we should consider $M$ as a K\"ahler manifold, which implies that we
should consider models with (2,0) supersymmetry. We would also like to
derive a bound for those gauge theories introduced in the previous
chapter. This will be done in the remainder of the chapter.

\section{A bound for vortices in the (2,0) model}

In this section we shall establish bounds for vortices for non-linear
sigma models considered in the previous chapter.  For this we shall
consider the Euclidean action of the (2,0)-supersymmetric gauged sigma
model without Wess-Zumino term. The sigma model target space $M$ is
K\"ahler with metric $g$, complex structure $J$ and associated
K\"ahler form $\Omega_J$. After a Wick rotation the two-dimensional
spacetime is $\bR^2$ with the standard Euclidean metric.  The relevant
part of the bosonic Euclidean action of a (2,0)-supersymmetric gauge
theory coupled to a sigma model is
\begin{align}
  S_E = \ints{\bR^2}{\rd^2x} 
  \bigl(\frac{1}{2} g_{ij}\delta^{\mu\nu} \nabla_\mu \phi^i  \nabla_\nu \phi^j
+ \frac{1}{2} u_{ab} F^a_{\mu\nu} F^b_{\lambda\rho} \delta^{\mu\lambda} \delta^{\nu\rho}
+\frac{1}{4} u^{ab} \nu_a \nu_b\bigr)\ .
\label{eu20act}
\end{align}
Next we introduce $I$ a constant complex structure on $\bR^2$ such
that $\bR^2$ is a K\"ahler manifold. The associated K\"ahler form
$\Omega_I$ is the volume form of $\bR^2$.  In such a case the
Euclidean action \eqref{eu20act} can be rewritten as
\begin{align}
\begin{split}
  S_E=&\intd{^2x} \bigl[
    \frac{1}{4} u_{ab} \bigl((\Omega_I \cdot F^a\mp \nu^a)
    (\Omega_I \cdot F^b\mp \nu^b)
\\
&+ {1\over4} g_{ij} \delta^{\mu\nu}(I^\rho{}_\mu\nabla_\rho\phi^i\mp \nabla_\mu \phi^k J^i{}_k)
(I^\sigma{}_\nu\nabla_\sigma\phi^j\mp \nabla_\nu \phi^\ell J^j{}_\ell)] \\
&\pm \int_{\bR^2} ((\Omega_J)_{ij} \nabla\phi^i\wedge \nabla\phi^j +\nu_a F^a)
\end{split}
\label{bogsquare}
\end{align}
where $\Omega_I\cdot F=(\Omega_I)^{\mu\nu} F_{\mu\nu}$, $\nu^a=u^{ab}
\nu_b$, $u_{ab}=u^0_{ab}$ and $u^{ac}u_{cb}=\delta^a{}_b$
($u_{ab}=u_{(ab)}$). The above expression for the Euclidean action has
been constructed from \eqref{eu20act} by completing squares and
collecting all the remaining terms which organise themselves in the
last term of \eqref{bogsquare}.

The last term in \eqref{bogsquare} is the topological charge,
\begin{align}
 {\cal Q}=\int_{\bR^2}  \omega_J\ ,
 \end{align}
where the form
 \begin{align}
 \omega_J=(\Omega_J)_{ij} \nabla\phi^i\wedge \nabla\phi^j +\nu_a F^a
\end{align}
is the \emph{equivariant extension of the K\"ahler form} $\Omega_J$ of
the sigma model target space $M$, see section~\ref{sec:equivarext}.
That $\omega_J$ is closed follows from dimensional grounds.  In
fact~\eqref{eq:equivarclosed} easily generalises to show that
$\omega_J$ is closed as a two-form on any manifold $N$ for any map
$\phi$ from $N$ into the sigma model manifold $M$ and for any choice
of connection $A$.
 
The Euclidean action of the (2,0)-supersymmetric two-dimensional sigma
model is bounded by the absolute value of the topological charge
${\cal Q}$, $S_E\geq |{\cal Q}|$. This is because it is always
possible to choose the signs in the Bogomol'nyi bound above such that
the topological term is positive. If the topological charge is
positive, then the bound is attained if
\begin{align}\begin{split}
\Omega_I\cdot F^a-\nu^a&=0
\\
J^i{}_j\nabla_\mu\phi^j-\nabla_\nu \phi^i I^\nu{}_\mu&=0\ .
\end{split}\end{align}
In two-dimensions, the curvature $F$ is a (1,1)-form.  Choosing
complex coordinates $(z, \bar z)$ on $\bR^2$ with respect to the
complex structure $I$, it is always possible to arrange using a
(complex) gauge transformation that $A_{\bar z}=0$. Choosing complex
coordinates in the sigma model target space $M$ as well, the second
BPS condition becomes $\nabla_{\Bar{z}}\phi^\alpha=0$ which means that
the map $\phi$ is holomorphic from the spacetime $\bR^2$ into the
sigma model manifold $M$.

A special case of this bound arises for gauge theories couple to
linear sigma models for which the sigma model manifold $M=\bR^{2n}$
with the Euclidean metric and equipped with a constant compatible
complex structure $J$. This case includes the Nielsen-Olesen vortices
\cite{Nielsen:1973cs}. (For these, existence of a solution was shown
in \cite{Taubes:1980tm} and the moduli were studied in
\cite{Weinberg:1979er}, \cite{Bradlow:1989} and more recently in
\cite{Manton:2002wb}, see also \cite{Losev:1999tu}).  The case with a
single complex scalar has been analysed in \cite{Bogomolny:1976de}.
Choosing complex coordinates $\{q^\alpha; \alpha=1,\dots,n\}$ in
$\bR^{2n}$, we write
\begin{align}\begin{split}
\rd s^2&=\sum_{\alpha} \rd q^\alpha \rd q^{\bar\alpha}
\\
\Omega_J&=-i \sum_{\alpha} \rd q^{\alpha}\wedge  \rd q^{\bar\alpha}\ .
\end{split}\end{align}
Next consider the abelian group $U(1)$-action $q^\alpha\rightarrow
e^{i Q_\alpha t} q^\alpha$ which generates the holomorphic Killing
vector fields
\begin{align}
\xi=i \sum_{\alpha} Q_\alpha (q^\alpha {\partial\over \partial q^\alpha}-q^{\bar \alpha}
{\partial\over \partial q^{\bar\alpha}})
 \ .
\end{align}
The moment map is~\eqref{eq:moment20def}
\begin{align}
\nu=- \sum_{\alpha}( Q_\alpha q^\alpha q^{\bar\alpha})-\Lambda\ ,
\end{align}
where $\Lambda$ is a (cosmological) constant. This is an example of a
$(2,0)$-supersymmetric gauged linear sigma model with gauge group
$U(1)$ of the type considered in \cite{Witten:1993yc}.  The
topological charge is
\begin{align}
{\cal Q}=\ints{\bR^2}{\rd^2z} \big(\sum_{\alpha}(\nabla_z q^\alpha \nabla_{\bar z} q^{\bar \alpha}-
\nabla_{\bar z} q^\alpha \nabla_{ z} q^{\bar \alpha})+ \nu F_{z\bar z}\big)
\label{topcharge}
\end{align}
where $\nabla_z q^\alpha=\partial_z q^\alpha+i A_z Q_\alpha q^\alpha$,
$\nabla_z q^{\bar \alpha}=\partial_z q^{\bar\alpha}-i A_z Q_\alpha
q^{\bar\alpha}$, $\nabla_{\bar z} q^{\bar \alpha}=(\nabla_z
q^\alpha)^*$ and $\nabla_{\bar z} q^{\alpha}=(\nabla_z q^{\bar
  \alpha})^*$, and $F_{z\bar z}=\partial_z A_{\bar z}- \partial_{\bar
  z} A_z$.  To compare the bound above \eqref{bogsquare} with that of
vortices in \cite{Witten:1993yc}, we observe that {\it after some
  integration by parts} we have
\begin{align}
{\cal Q}=\int_{\bR^2} d^2z \big(\sum_{\alpha}(\partial_z q^\alpha \partial_{\bar z} q^{\bar \alpha}-
\partial_{\bar z} q^\alpha \partial_{ z} q^{\bar \alpha})-\Lambda F_{z\bar z}\big)+ {\rm surfaces}
\end{align}
The first term in the above expression is the topological charge
expected for the vortices (instantons) of ungauged two-dimensional
sigma models. The same topological charge also appears in the kink
solitons of three-dimensional non-linear sigma
models~\cite{Duff:1976av}. The last part in the above expression
involving the cosmological constant and the Maxwell field is the usual
degree of an abelian vortex.  The relation between the topological
charge ${\cal Q}$ in \eqref{topcharge} and the degree of an abelian
vortex involves integration by parts. Under certain boundary
conditions the two topological charges are the same.  However as we
have shown, the bound that involves the equivariant extension of the
K\"ahler form generalizes in the context of gauge theories coupled to
non-linear sigma models.

\section{A bound for vortices in the (4,0) model}

A bound similar to the one we have described in the previous section
for the Euclidean action of (2,0)-supersymmetric gauged sigma model
can also be found for the Euclidean action of (4,0)-supersymmetric
gauged sigma model.  The Euclidean action of the (4,0)-supersymmetric
gauged sigma model with vanishing Wess-Zumino term is
\begin{align}
S_E= \ints{\bR^2}{\rd^2x} \big({1\over2}
 g_{ij}\delta^{\mu\nu} \nabla_\mu \phi^i  \nabla_\nu \phi^j
+ {1\over2} u_{ab} F^a_{\mu\nu} F^b_
{\lambda\rho} \delta^{\mu\lambda} \delta^{\nu\rho}
+{1\over4} \sum_{r=1}^3 u^{ab} \nu^r_a \nu^r_b\big)
\end{align}
The sigma model target space is hyper-K\"ahler with metric $g$,
hypercomplex structure $\{J_r: r=1,2,3\}$ and associated K\"ahler
forms $\Omega_{J_r}$.  After a Wick rotation the two-dimensional
spacetime is $\bR^2$ with the standard Euclidean metric.  Let $I$ be a
compatible constant complex structure such that $\bR^2$ is a K\"ahler
manifold with associated K\"ahler form $\Omega_I$. In the same ways as
before, Euclidean action can be written as
\begin{align}
\begin{split}
S_E=&\intd{^2x} [{1\over4}\sum_{r=1}^3 u_{ab} (a_r \Omega_{I}\cdot F^a\mp \nu_r^a )
( a_r \Omega_{I}\cdot F^b\mp \nu_r^b)
\\
&+ {1\over4}\sum_{r=1}^3 \delta^{\mu\nu} g_{ij} ( a_r I^\rho{}_\mu\nabla_\rho\phi^i\mp \nabla_\mu \phi^k J_r^i{}_k)
( a_r I^\sigma{}_\nu\nabla_\sigma\phi^i\mp \nabla_\nu \phi^\ell J_r^j{}_\ell)]
\\ & \pm
\int_{\bR^2} \sum_{r=1}^3 a_r \omega_{J_r}\ ,
\end{split}
\label{bou40}
\end{align}
where $\{a_r :r=1,2,3\}$ is a constant vector with length one, $\sum_{r=1}^3 (a_r)^2=1$,
$u_{ab}=u^0_{ab}=u^0_{ba}$, $\nu^a_r=
u^{ac} \nu_{rc}$, $u^{ac} u_{cb}=\delta^a{}_b$ and
\begin{align}
\omega_{J_r}=(\Omega_{J_r})_{ij} \nabla\phi^i\wedge \nabla\phi^j +\nu_a F^a
\end{align}
is the equivariant extension of the K\"ahler form $\Omega_{J_r}$.

The strictest bound is attained whenever the unit vector $\{a_r:
r=1,2,3\}$ is parallel to the vector of the topological charges
$\{{\cal Q}_r: r=1,2,3\}$, where
\begin{align}
{\cal Q}_r=\int_{\bR^2} \omega_{J_r}
\end{align}
and the sign is chosen such that the topological term in the bound is
positive.  If the inner product of $\{a_r: r=1,2,3\}$ and $\{{\cal
  Q}_r: r=1,2,3\}$ in \eqref{bou40} is positive, we have that
\begin{align}
S_E\geq \sqrt {{\cal Q}_1^2+{\cal Q}_2^2+{\cal Q}_3^2}\ .
\end{align}
This bound is attained whenever
\begin{align}
\begin{split}
 a_r \Omega_{I}\cdot F^a- \nu_r^a&=0
\\
{J_r}^i{}_j\nabla_\mu\phi^j-a_r \nabla_\nu \phi^i I^\nu{}_\mu&=0\ .
\end{split}
\label{bog40eqn}
\end{align}

Using a rotation in the space of three complex structures, we can
arrange such that $a_1=1$ and $a_2=a_3=0$.  In such case, the last
equation in \eqref{bog40eqn} implies that
\begin{align}
\nabla_\mu \phi^i=0\ .
\end{align}
This in turn gives
\begin{align}
F_{\mu\nu}^a \xi_a^i=0
\end{align}
Therefore either $\phi$ takes values in the fixed point set $M_f$ of
the group action of $G$ in $M$ or the curvature $F$ of the connection
$A$ vanishes. In the latter case, the first equation in
\eqref{bog40eqn} implies that $\nu_r=0$ and these are the vacua of the
theory. If these are no non-trivial flat connections and $M_f\cap
\nu_1^{-1}(0)\cap \nu_2^{-1}(0)\cap \nu_3^{-1}(0)=0$, then the
space of solutions is the hyper-K\"ahler reduction $M//G$ of $M$ and
it is a hyper-K\"ahler manifold. On the other hand if $\phi$ takes
values in $M_f$, then the second equation in \eqref{bog40eqn} implies
that $\phi$ is constant. Substituting this in the first equation in
\eqref{bog40eqn} implies that $\phi$ is in $M_f\cap \nu_2^{-1}(0)\cap
\nu_3^{-1}(0)$. In addition we have that
\begin{align}
\Omega_{I}\cdot F^a- \nu_1^a=0\ .
\end{align}
This is the Hermitian-Einstein equation in two dimensions.

The first condition in~\eqref{bog40eqn} implies that the curvature $F$
is a (1,1)-form.  Choosing complex coordinates $(z, \bar z)$ on the
two-dimensional spacetime associated with the complex structure $I$,
it is always possible to arrange using a gauge transformation such
that $A_{\bar z}=0$. Choosing complex coordinates in the sigma model
target space $M$ as well, it is easy to see that the second BPS
condition implies that the map $\phi$ is holomorphic from the
spacetime into the sigma model manifold $M$.

\section{K\"ahler manifolds and  non-abelian vortices}

The bounds that we have described in the previous sections can be
generalized as follows. Consider two K\"ahler manifolds $(N,h,I)$ and
$(M,g,J)$ of dimensions $2k$ and $2n$, and K\"ahler forms $\Omega_I$
and $\Omega_J$, respectively. Next allow $M$ to admit a holomorphic
$G$-action with associated killing vector fields $\xi$ and moment map
$\nu$. In our conventions $i_{\xi}\Omega_J=-d\nu$.  Next consider the
functional
\begin{align}
S_E=\int_N d{\rm vol}(N) \big({1\over2}
|\nabla\phi|^2+ {1\over2} |F|^2+{1\over4} |\nu|^2\big)\ ,
\end{align}
where $|\nabla\phi|^2= g_{ij} h^{\mu\nu} \nabla_\mu\phi^i \nabla_\nu
\phi^j$, $\nabla_\mu\phi^i=\partial_\mu \phi^i+ A^a \xi_a^i$,
$|F|^2=u_{ab} F^a_{\mu\nu} F^b_{\rho\sigma} h^{\mu\rho}
h^{\nu\sigma}$, $|\nu|^2= u^{ab} \nu_a \nu_b$ and $u$ is a fibre inner
product on the gauge bundle which we can set $u_{ab}=\delta_{ab}$.

The functional $S_E$ can be rewritten as follows:
\begin{align}
\begin{split}
S_E=&\int_N d{\rm vol}(N) \big[ {1\over 4} |\Omega_I\cdot F\mp \nu|^2+ |F^{2,0}|^2
+ {1\over 4} |I\nabla\phi\mp J\nabla\phi|^2\big]
\\
&\pm {1 \over (k-1)!}\int_N \omega_J\wedge \Omega_I^{k-1}
-{1 \over (k-2)!}\int_N  u_{ab} F^a\wedge F^b\wedge \Omega_I^{k-2}
\end{split}
\label{bogkk}
\end{align}
where we have chosen the normalisation $d{\rm vol}(N)={1\over k!}
\Omega_I^k$, $\Omega_I\cdot F=\Omega_I^{\mu\nu} F_{\mu\nu}$, $F^{2,0}$
is the (2,0) part of the curvature $F$ that corresponds to the
splitting
$\Omega^2(M)=\Omega^{2,0}(M)\oplus\Omega^{1,1}(M)\oplus\Omega^{0,2}(M)$
and
\begin{align}
\omega_J=(\Omega_J)_{ij} \nabla\phi^i\wedge \nabla\phi^j+ \nu_a F^a
\end{align}
is the equivariant extension of the K\"ahler form $\Omega_J$. (The
inner products are taken with respect to the Riemannian metrics $h$
and $g$.) The rest of the notation is self-explanatory.  We remark
that if $\Omega_J$ represents the first Chern class of a line bundle,
i.e.\ the K\"ahler manifold is Hodge, then $\omega_J$ can be thought
of as the equivariant extension of the first Chern class (see
\cite{Atiyah:1984}).

If $u_{ab}$ is a constant invariant quadratic form on the Lie algebra
of the gauge group $G$, it is clear that the functional $S_E$ is
bounded by a topological term ${\cal Q}$ which involves the
equivariant extension of the K\"ahler form and the second Chern
character of the bundle $P\times_G \LieG$, where $P$ is a principal
bundle of the gauge group $G$ and $G$ acts on $\LieG$ with the adjoint
representation.  In particular we can write
\begin{align}
\begin{split}
S_E=&\int_N d{\rm vol}(N) \big[ {1\over 4} |\Omega_I\cdot F\mp \nu|^2+ |F^{2,0}|^2
+ {1\over 4} |I\nabla\phi\mp J\nabla\phi|^2\big]
\\
&\pm {1 \over (k-1)!}\int_N \omega_J\wedge \Omega_I^{k-1}
-{8\pi^2 \over \lambda  (k-2)!}\int_N  ch_2 \wedge \Omega_I^{k-2}\ ,
\end{split}
\label{bogkkk}
\end{align}
where $\lambda$ is an appropriate normalisation factor involving the
ratio between the fibre inner product on $P\times_G \LieG$ and $u$;
where $G$ is simple.  It is worth pointing out that the term involving
the second Chern character is not affected by the choice of sign in
writing \eqref{bogkk}.  Therefore there are three cases to consider
the following: (i) there is no choice of sign such that the
topological charge ${\cal Q}$ is positive. In such a case the bound
cannot be attained.  (ii) There is a critical case in which for one
sign the topological charge is negative while for the other is zero.
This case implies that the Euclidean action vanishes and so every term
should vanish. Solutions exist for $F=\nabla\phi=\nu=0$. (iii) For one
of the choice of signs the topological charge is positive.  Suppose
that ${\cal Q}$ is positive in \eqref{bogkk} for the first choice of
sign.  In such case the bound is attained provided that the equations
\begin{align}
\begin{split}
F^{2,0}&=0
\\
\Omega_I F^a-\nu^a&=0
\\
I^\nu{}_\mu \nabla_\nu\phi^i- J^i{}_j \nabla_\mu \phi^j&=0
\end{split}
\label{bogkkeqn}
\end{align}
hold.  The first equation implies that $F$ is a (1,1)-form. The last
equation in \eqref{bogkkeqn} implies that the maps $\phi$ are
holomorphic. Finally the middle equation is the standard non-abelian
vortex equation.  If the term involving the moment map is constant,
then the resulting equation is the Hermitian-Einstein equation.

\section{Hyper-K\"ahler manifolds and  non-abelian vortices}

Let $(N,h,I)$ be a K\"ahler manifold of dimension $2k$ with associated
K\"ahler form $\Omega_I$ and $(M,g,J_r)$ be a hyper-K\"ahler manifold
of dimension $4n$ with associated K\"ahler forms $\Omega_{J_r}$. Next
allow $M$ to admit a tri-holomorphic $G$-action with associated
killing vector fields $\xi$ moment maps $\nu_r$.  In our conventions
$i_{\xi}\Omega_{J_r}=-\rd\nu_r$.  Next consider the functional
\begin{align}
S_E=\int_N d{\rm vol}(N) \big({1\over2}
|\nabla\phi|^2+ {1\over2} |F|^2+{1\over4}\sum_{r=1}^3 |\nu_r|^2\big)\ ,
\end{align}
where $|\nabla|^2= g_{ij} h^{\mu\nu} \nabla_\mu\phi^i \nabla_\nu
\phi^j$, $\nabla_\mu\phi^i=\partial_\mu \phi^i+ A^a \xi_a^i$,
$|F|^2=u_{ab} F^a_{\mu\nu} F^b_{\rho\sigma} h^{\mu\rho}
h^{\nu\sigma}$, $|\nu_r|^2= u^{ab} \nu_{ra} \nu_{rb}$ and $u$ is a
fibre inner product on the gauge bundle which we can set
$u_{ab}=\delta_{ab}$.

The functional $S_E$ can be rewritten as follows:
\begin{align}
\begin{split}
S_E=&\int_N d{\rm vol}(N) \big[ {1\over 4}\sum_{r=1}^3 |a_r \Omega_I\cdot F\mp \nu_r|^2+ |F^{2,0}|^2
+ {1\over 4}\sum_{r=1}^3 |a_r I\nabla\phi\mp J_r\nabla\phi|^2\big]
\\
&\pm {1 \over (k-1)!}\int_N\sum_{r=1}^3 a_r \omega_{J_r}\wedge \Omega_I^{k-1}
-{1 \over (k-2)!}\int_N  u_{ab} F^a\wedge F^b\wedge \Omega_I^{k-2}
\end{split}
\label{bogkh}
\end{align}
where $d{\rm vol}(N)={1\over k!} \Omega_I^k$, $\{a_r: r=1,2,3\}$ is a
constant vector of length one, $\sum_{r=1}^3 (a_r)^2=1$,
$\Omega_I\cdot F=\Omega_I^{\mu\nu} F_{\mu\nu}$, $F^{2,0}$ is the (2,0)
part of the curvature $F$ and
\begin{align}
\omega_{J_r}=(\Omega_{J_r})_{ij} \nabla\phi^i\wedge \nabla\phi^j+ \nu_{ra} F^a
\end{align}
is the equivariant extension of the K\"ahler form $\Omega_{J_r}$. (The
inner products are taken with respect to the Riemannian metrics $h$
and $g$.) The rest of the notation is self-explanatory.

It is clear that the functional $S_E$ is bounded by a topological
charge ${\cal Q}$ which involves the equivariant extensions of the
K\"ahler forms $\Omega_{J_r}$ and, if $u$ is a constant invariant
quadratic form on $\LieG$, the second Chern character of the gauge
bundle $P\times_G \LieG$.  It is worth pointing out that the term
involving the second Chern character is not affected by the choice of
sign in writing \eqref{bogkh}. Therefore as in the K\"ahler case,
there are several cases to consider but we shall not repeat the
analysis. Suppose that both ${\cal Q}$ and the inner
product of the vector $\{a_r :r=1,2,3\}$ with $\{\tilde{\cal Q}_r :
r=1,2,3\}$ are positive in \eqref{bogkh}, where
\begin{align}
\tilde {\cal Q}_r={1 \over (k-1)!}\int_N  \omega_{J_r}\wedge \Omega_I^{k-1}\ .
\end{align}
 Then the bound is attained provided that the equations
\begin{align}
\begin{split}
F^{2,0}&=0
\\
a_r \Omega_I F^a-\nu_r^a&=0
\\
 a_r I^\nu{}_\mu \nabla_\nu\phi^i- J_r{}^i{}_j \nabla_\mu \phi^j&=0
 \end{split}
\label{bogkheqn}
\end{align}
hold.  The first equation implies that $F$ is a (1,1)-form. It is
always possible with a rotation in the space of complex structures of
the hyper-K\"ahler manifold to set $a_1=1$ and $a_2=a_3=0$.  Then last
equation in \eqref{bogkheqn} implies that
\begin{align}
\nabla_\mu\phi^i=0\ .
\end{align}
This in turn implies that
\begin{align}
F_{\mu\nu}^a \xi_a^i=0\ .
\end{align}
Therefore either the connection $A$ is flat or the maps $\phi$ take
values in the fixed set $M_f$ of the $G$-group action on $M$. In the
former case, in the absence of non-trivial flat connections, the
moduli space of solutions to these equations is the hyper-K\"ahler
reduction $M//G$ of $G$ and it is a smooth manifold provided that the
level set does not intersect $M_f$.  In the latter case, the maps
$\phi$ are constant and the two remaining equations in
\eqref{bogkheqn} are the Hermitian-Einstein equations for the
connection $A$; see \cite{Grantcharov:2002fk} and references therein.

One can also consider the case where $(N,h,I_r)$ is a hyper-K\"ahler
manifold while $(M,g,J)$ is a K\"ahler manifold which admits a
$G$-holomorphic action of isometries.  This case can be treated as
that considered in the previous section involving only K\"ahler
manifolds. A K\"ahler structure on $N$ can chosen with respect to any
complex structure which lies in the two-sphere of complex structures
of $N$.

\chapter{Conclusions}

We have gauged the supersymmetry of the most general $N=1$
supersymmetric one dimensional sigma model; we have investigate the
dynamics of the theory by deriving the Hamiltonian and the first and
second class constraints. We have then quantised the theory using the
Dirac method and checked that the Dirac operator squares to the Klein
Gordon operator.

We have commented that the zero modes of the Dirac operator do not
exist if the target manifold $M$ is compact, without boundary and with
positive scalar curvature. It would be interesting to investigate
further the relationship between the geometry of the target manifold
and existence of zero modes of the Dirac operator.

We have constructed the actions of two-dimensional (p,0)- and
(p,1)-supersymmetric gauge theories coupled to sigma model matter with
Wess-Zumino term. We have also given the scalar potentials of these
theories. Our method of constructing these theories relies on a
superfield method. Then we have shown that the Euclidean actions these
theories admit vortex type of bounds which generalise to higher
dimensions.

The gauge theories that we have constructed are not the most general
ones. It is known for example that the (1,1)-supersymmetric sigma
model admits a scalar potential which is the length of a killing
vector field \cite{Alvarez-Gaume:1983ab}. Our superfield method cannot describe such
a term. There are also other possibilities, for example the sigma
models with almost complex manifolds as a target space as well as
those associated with $(p,0)$ fermionic multiplets for which the
supersymmetry algebra closes on-shell \cite{Howe:1988cj,Hull:1985jv}. Other models
of interest that we have not described here are those with $(p, 2)$,
$p=2,4$, and (4,4) supersymmetry. All the above models can be
described using (1,0) superfields. This method has been used before,
see \cite{Papadopoulos:1994mf,Papadopoulos:1994tn,Papadopoulos:1995kj}. This means that the action of such models
can be written in terms of (1,0) superfields and the additional
supersymmetries can be implemented by requiring invariance of the
action under additional suitable transformations. The (2,2) and (4,4)
supersymmetric gauge theories have been described using other methods
in \cite{Witten:1993yc} and \cite{Gates:2000gu}.

The gauge theories coupled to sigma models which we have described
with $(p,1)$ supersymmetry have soliton type of bounds in addition to
the vortex type of bounds that we have described.  For the former
bounds the energy of these models can be written as a sum of squares
and a topological term. This is very similar to bounds of (ungauged)
sigma models \cite{Abraham:1992vb} and so we have not described them here. It
would be of interest to investigate the solutions of the vortex
equations we have presented for different types of moment maps. It may
be that for a suitable choice, the vortex equations can be solved
exactly.

\providecommand{\href}[2]{#2}\begingroup\raggedright\endgroup

\end{document}